%% file: main.tex
\documentclass[10pt,twocolumn,letterpaper]{article}
\usepackage[pagenumbers]{cvpr}
\input{preamble}
\definecolor{cvprblue}{rgb}{0.21,0.49,0.74}
\usepackage[pagebackref,hypertexnames=false,breaklinks,colorlinks,allcolors=cvprblue]{hyperref}
\title{\ours{}: Atomic Skills for Evidence-Grounded Video Reasoning}
\author{Xiyang Wu\textsuperscript{1}, Zongxia Li\textsuperscript{1}, Shengxin Zhang\textsuperscript{2}, Zhichao Liu\textsuperscript{1}, Dinesh Manocha\textsuperscript{1}\\[3pt]
{\normalsize \textsuperscript{1}University of Maryland, College Park\qquad \textsuperscript{2}Google}}
\hypersetup{pdftitle={ASSEMBLE: Atomic Skills for Evidence-Grounded Video Reasoning},pdfauthor={Xiyang Wu, Zongxia Li, Shengxin Zhang, Zhichao Liu, Dinesh Manocha}}
\begin{document}
\maketitle
\input{sections/0-Abstract}
\input{sections/1-Intro}
\input{sections/2-Related_Work}
\input{sections/3-Problem}
\input{sections/4-Method}
\input{sections/5-Results}
\input{sections/6-Conclusion}
{\small
\bibliographystyle{ieeenat_fullname}
\bibliography{references}
}
\clearpage
\appendix
\maketitlesupplementary
\input{sections/7-Appendix}

\end{document}

%% file: preamble.tex
\usepackage[T1]{fontenc}
\input{math_commands}

\usepackage{makecell}
\usepackage{algorithm}
\usepackage{algpseudocode}
\usepackage{array}
\usepackage{tabularx}
\input{showcase/showcase_style}
\newcommand{\ours}{ASSEMBLE}

%% file: math_commands.tex
\usepackage{amsmath,amsfonts,bm}

\def\eqref#1{equation~\ref{#1}}

\def\1{\bm{1}}

\DeclareMathAlphabet{\mathsfit}{\encodingdefault}{\sfdefault}{m}{sl}
\SetMathAlphabet{\mathsfit}{bold}{\encodingdefault}{\sfdefault}{bx}{n}

%% file: showcase/showcase_style.tex
\usepackage{xcolor}
\usepackage{tcolorbox}
\usepackage{url}
\usepackage{graphicx}
\newcommand{\ASfont}{\fontencoding{T1}\fontfamily{ptm}\selectfont}
\definecolor{ASblue}{HTML}{234A8A}
\definecolor{AScream}{HTML}{FFFCED}
\definecolor{ASgreen}{HTML}{226746}
\definecolor{ASred}{HTML}{A33333}
\definecolor{AScontext}{HTML}{85531D}
\newtcolorbox{AScase}[1]{colback=AScream,colframe=black!75,
  colbacktitle=black!85,coltitle=white,fonttitle=\ASfont\small\bfseries,
  title={#1},boxrule=0.5pt,arc=2pt,left=8pt,right=8pt,top=7pt,bottom=7pt,
  before skip=5pt,after skip=4pt}
\newcommand{\ASfield}[1]{\textcolor{ASblue}{\textbf{#1}}}
\newcommand{\AScode}[1]{{\ttfamily\footnotesize\nolinkurl{#1}}}
\newcommand{\ASgood}[1]{\textcolor{ASgreen}{\textbf{#1}}}
\newcommand{\ASbad}[1]{\textcolor{ASred}{\textbf{#1}}}
\newcommand{\ASbar}[1]{\par\vspace{5pt}\noindent
  \colorbox{ASblue!9}{\parbox{\dimexpr\linewidth-2\fboxsep\relax}{\centering\bfseries #1}}\par\vspace{4pt}}

\newcommand{\ASrefs}[1]{\textsf{#1}}
\newenvironment{ASbody}{\ASfont\small\raggedright\setlength{\parindent}{0pt}\setlength{\parskip}{4pt}}{}
\newcommand{\ASframe}[4]{%
  \begin{minipage}[t]{#1\linewidth}\centering
  \includegraphics[width=\linewidth,keepaspectratio]{#2}\par
  \vspace{2pt}{\footnotesize\bfseries #3}\par
  {\footnotesize #4}\par
  \end{minipage}%
}

%% file: sections/0-Abstract.tex
\begin{abstract}

Complex video reasoning often depends on evidence scattered across distant moments, entities, and events, yet a correct answer alone does not reveal whether a model relied on the right parts of the video. We introduce \ours{}, a framework that makes supporting evidence explicit throughout long-video reasoning. It organizes local observations and cross-clip narratives into timestamped evidence catalogs traceable to the source video. A grounding-aware reader then composes question-specific atomic skills whose structured outputs include explicit evidence references and support assessments. Following teacher-supervised fine-tuning, we train the reader with Group Relative Policy Optimization (GRPO) using answer correctness and correctness-gated citation alignment. The resulting reasoning traces are inspectable, and final predictions remain linked to explicit supporting evidence.
With a 9B reader supervised by a 235B teacher and shared precomputed evidence catalogs, \ours{} achieves 59.2\% macro-averaged answer accuracy across three long-video reasoning benchmarks, compared with 58.3\% for Gemini-2.5-Pro, while improving macro-averaged overlap-based Grounded accuracy by 6.7 percentage points, with gains on all three benchmarks. Ablations show that, under the same post-trained reader and inference budget, structured skill inference improves overlap-based Grounded accuracy over free-form reasoning. Compared with generic structured reasoning, the benefit is more setting-dependent, with the clearest answer-accuracy gain on VRBench. These results show that explicit evidence grounding can improve long-video reasoning while keeping predictions tied to their supporting evidence.
\end{abstract}

%% file: sections/1-Intro.tex
\section{Introduction}

Answering complex questions about long videos requires integrating clues across distant moments, tracking entities and events, and connecting local observations to broader context~\citep{cheng2025video,chen2025cg,yu2025vrbench}.
Failures can arise from missing evidence, incorrect entity or temporal grounding, or faulty reasoning over relevant observations~\citep{yu2026longvidsearch,meng2026videozerobench,tsuchiya2026ec}.
Answer-level supervision provides little indication of where these failures occur: an incorrect answer does not reveal where the reasoning failed, while a correct answer does not show whether the model relied on the relevant evidence~\citep{xie2025video}.
This motivates making evidence use explicit and evaluating grounding alongside answer correctness.

Prior work has explored reusable operations for multimodal reasoning by extracting visually grounded skills from trajectories~\citep{jiang2026xskill,zhang2026mmskills}, jointly learning hierarchical skills and policies~\citep{li2026comfyclaw,zhang2026spyce}, and representing skills as executable structures with verification and local repair~\citep{xia2026grasp}.
These approaches motivate making reasoning operations and the evidence supporting them explicit.
For long-video reasoning, however, the challenge is not only whether an operation is valid, but whether it is applied to the right evidence across entities, events, and distant time intervals~\citep{di2024grounded,xiao2024can,wei2026seeing,zhang2026cast}.
Studies on grounded VideoQA further show that strong answer accuracy does not necessarily imply accurate localization of supporting evidence~\citep{xiao2024can,wei2026seeing}.
This raises a practical question: how can explicit reasoning operations be tied to supervision that encourages a reader to ground its answers in relevant video evidence?

To connect explicit reasoning operations with grounding supervision, we use correctness-gated citation alignment as a measurable evidence-grounding signal.
Atomic skills provide structured interfaces for reasoning operations, evidence references, and support assessments, while timestamped references allow cited evidence to be compared with annotated video intervals.
When an answer is correct, the reader is further rewarded for citing the moments that support it; conditioning this signal on correctness prevents aligned citations from compensating for an incorrect prediction.
Post-training encourages correct, evidence-grounded answers, while atomic skills structure evidence selection, reasoning, and integration at inference time.

To implement this idea, we introduce \ours{}, which combines a timestamped evidence catalog, a fixed atomic-skill vocabulary, and a grounding-aware reader.
The catalog pairs textual descriptions with source intervals, making evidence reusable across questions and citations traceable to specific moments.
Atomic skills define structured reasoning operations with explicit evidence references and support assessments, and their ordered traces expose intermediate outputs, cited evidence, and model-generated diagnostic signals.
We first fine-tune the reader on teacher-generated responses with supervised fine-tuning (SFT), then apply Group Relative Policy Optimization (GRPO)~\citep{shao2024deepseekmath} with rewards for answer correctness, correctness-gated citation alignment, and output validity.
At inference time, the same reader performs planning, ordered skill application, integration, and final answer generation through a multi-call pipeline, ending with an answer and supporting citations.
We evaluate answer correctness and the temporal alignment
of cited evidence across three long-video reasoning benchmarks. Our contributions include:

\begin{itemize}
\item We introduce \ours{}, an atomic-skill framework for complex video reasoning that makes reasoning operations and their supporting evidence explicit through timestamped evidence catalogs and structured skill traces.

\item We develop a grounding-aware reader with teacher-response supervision and GRPO using correctness-gated citation alignment. At inference time, a structured multi-call pipeline produces answers with explicit citations to supporting video evidence.

\item Across three long-video reasoning benchmarks, \ours{} achieves 59.2\% macro-averaged accuracy versus 58.3\% for Gemini-2.5-Pro and improves Grounded accuracy by 6.7 percentage points. Structured skill inference further improves Grounded accuracy over free-form reasoning under matched settings.
\end{itemize}

%% file: sections/2-Related_Work.tex
\section{Related Work}

\textbf{Skill Learning and Compositional Program Induction.}
Early visual-programming approaches showed that complex visual reasoning can be decomposed into executable compositions of reusable modules. Visual Programming~\citep{gupta2023visual} generates modular programs that invoke pretrained vision and language tools, while ViperGPT~\citep{suris2023vipergpt} uses generated Python programs to compose visual APIs for reasoning.
More recent agents learn and maintain reusable skills from interaction: SkillRL~\citep{xia2026skillrl} jointly evolves skills and policies through RL, XSkills~\citep{jiang2026xskill} distills multimodal experience into reusable skills, and COS-PLAY~\citep{wu2026co} continually updates a skill library from trajectories.
Other systems organize or refine skills for composition, including MemSkill~\citep{zhang2026memskill}, GraSP~\citep{xia2026grasp}, SkillGraph~\citep{li2026skillgraph}, Skill-Pro~\citep{mi2026skill}, and SkillOS~\citep{ouyang2026skillos}.
Our work brings this modular view to evidence-grounded video reasoning: a shared Reader applies structured atomic-skill interfaces over a reusable evidence catalog and returns answers with citations to the supporting video intervals.

\noindent
\textbf{Evidence-Grounded Video Reasoning.}
Evidence-grounded video reasoning requires models to identify the temporal support for an answer, not just predict the answer itself.
NExT-GQA~\citep{xiao2024can} and EG-VQA~\citep{huang2026eg} add answer-aligned temporal annotations, while VITED~\citep{lu2025vited}, Grounded-VideoLLM~\citep{wang2024grounded}, and TimeChat~\citep{ren2024timechat} improve joint reasoning and temporal grounding.
GroundVQA~\citep{di2024grounded} and MultiHop-EgoQA~\citep{chen2025grounded} extend this setting to long egocentric videos and evidence spread across multiple intervals, while EV$^2$-Bench~\citep{wei2026seeing} and Video-in-the-Loop~\citep{wang2025video} evaluate or predict explicit supporting evidence.
VideoMind~\citep{liu2026videomind} structures reasoning into planning, grounding, verification, and answering, while Video-Grounded Entailment Tree~\citep{liu2025commonsense} decomposes hypotheses and verifies them against video evidence.
Evidence-aware training has also advanced rapidly: Video-VER~\citep{luo2026thinking} uses correctness-gated rewards for visually grounded reasoning traces, Open-o3-Video~\citep{meng2025open} predicts explicit spatio-temporal evidence, SER~\citep{xia2026ser} uses semantic evidence verification, and EG-Reasoner~\citep{huang2026eg} trains for grounded answering.
TimeThink~\citep{li2026timethink} and LOVER~\citep{chen2026long} further introduce temporal process and grounding-aware rewards.
Our method trains a shared reader to align catalog citations with annotated evidence spans. At inference time, it composes structured skills over a reusable, question-independent catalog while linking predictions to cited video intervals.

\noindent
\textbf{Long-Video Understanding and Evidence Retrieval.}
Long-video understanding requires finding sparse, question-relevant evidence across extended temporal contexts, as highlighted by CG-Bench~\citep{chen2025cg} and Video-Holmes~\citep{cheng2025video}.
LLoVi~\citep{zhang2024simple} aggregates dense clip captions with an LLM, while VideoAgent~\citep{wang2024videoagent}, VideoTree~\citep{wang2025videotree}, and T$^*$~\citep{ye2025re} retrieve relevant visual evidence through iterative, hierarchical, or adaptive temporal search.
ReWind~\citep{diko2025rewind} and SEAL~\citep{wang2025seal} use compact memory or semantic entities to guide selection, while LVAgent~\citep{chen2025lvagent} and DrVideo~\citep{ma2025drvideo} reason over selected segments or searchable video representations.
In contrast, we build a reusable, question-independent timestamped evidence catalog once per video and use a post-trained reader to compose atomic skills over it, linking answers back to the supporting video intervals.

%% file: sections/3-Problem.tex
\section{Problem Formulation}

\noindent
\textbf{Complex Video Reasoning.}
We study complex video reasoning across social-reasoning and long-video understanding tasks.
Each instance consists of a video $V$, a question $q$ including its answer options, a reference answer $y$, and, when available, annotated supporting evidence
$\mathcal{E}=\{e_j\}_{j=1}^{M}$ with associated temporal spans.
Using a set of perception tools $\mathcal{U}$, we construct an evidence catalog $\mathcal{C}(V)$ whose entries contain textual observations and their temporal provenance.
Given $q$ and $\mathcal{C}(V)$, the reader predicts an answer $\hat{y}$ and supporting citations $\hat{Z}$, where each citation indexes a catalog entry.
Reference answers $y$ and evidence annotations $\mathcal{E}$ are used for training and evaluation, but are excluded from the reader's inference inputs.

\noindent
\textbf{Atomic Skills and Structured Planning.}
We define a fixed vocabulary
$\mathcal{S}=\{s_1,\ldots,s_K\}$ of atomic skills for
evidence retrieval and structured reasoning. Each skill has
an explicit interface:
\begin{equation}
s_k
=
\left\langle
\mathcal{I}_k,
\mathcal{O}_k,
\ell_k,
\mathcal{A}_k
\right\rangle,
\end{equation}
where $\mathcal{I}_k$ and $\mathcal{O}_k$ specify the input and
output schemas, $\ell_k$ is the stage-specific instruction, and
$\mathcal{A}_k$ specifies evidence-support assessment fields,
including a support score, a status label, and an optional
diagnostic label.

Given a question $q$ and an evidence catalog $\mathcal{C}(V)$,
the reader's planning call produces an ordered skill sequence:
\begin{equation}
\mathcal{P}
=
\mathrm{Plan}_{\pi_\theta}
\left(q,\mathcal{C}(V);\mathcal{S}\right)
=
(a_1,a_2), \qquad L=2.
\label{eq:skill_plan}
\end{equation}
Each planned skill application
$a_i=(s_{k_i},\boldsymbol{\alpha}_i)$ specifies a skill and its arguments, which may reference the question, catalog entries, or outputs from earlier calls. The resulting plan $\mathcal{P}$ is serialized in execution order rather than represented as an executable program or dependency graph. Each question selects two skills, with repetition allowed. Together with planning, integration, and final-answer generation, this yields five reader calls in total.

\noindent
\textbf{Skill Outputs and Support Assessment.}
For each planned skill, the reader resolves its arguments from the question, evidence catalog, and preceding outputs, then follows the corresponding instruction:
\begin{equation}
(o_i,\rho_i,u_i,d_i)
\sim
\pi_\theta\!\left(
\cdot\mid q,\mathcal{C}(V),\mathbf{o}_{<i},
s_{k_i},\boldsymbol{\alpha}_i,\ell_{k_i}
\right),
\end{equation}
where $o_i$ is the structured output, $\rho_i$ contains its evidence references, $u_i$ records the associated support assessment and status, and $d_i$ is an optional diagnostic label.
We retain the ordered records
$\mathcal{R}=\{(o_i,\rho_i,u_i,d_i)\}_{i=1}^{L}$
for trace inspection.
Support assessments are generated by the reader and are not independently verified.

\noindent
\textbf{Reader Inference.}
Given the question $q$, evidence catalog $\mathcal{C}(V)$, and skill
records $\mathcal{R}$, the reader $\pi_{\theta}$ first produces an
integration summary $h$ and then generates the final response:
\begin{align}
h
&\sim
\pi_{\theta}(\cdot\mid q,\mathcal{C}(V),\mathcal{R}),\\
z
&\sim
\pi_{\theta}(\cdot\mid q,\mathcal{C}(V),\mathcal{R},h),
\qquad
(\hat{y},\hat{Z})=\mathrm{Parse}(z),
\end{align}
where $z$ contains an answer rationale and label, and
$\hat{Z}$ identifies the catalog entries cited in the rationale.
Generating $z$ is the final call of the inference pipeline.
Reference answers and evidence annotations are excluded
from its inputs.

\noindent
\textbf{Learning Objective.}
We warm-start the reader $\pi_{\theta}$ with supervised
fine-tuning on teacher-generated rationales and answer labels,
using cross-entropy loss on response tokens only.
Starting from this initialization, we optimize the reader
with GRPO toward
\begin{equation}
J(\theta)
=
\mathbb{E}_{
\substack{
(V,q,y,\mathcal{E})\sim\mathcal{D}_{\mathrm{RL}}\\
z\sim\pi_{\theta}(\cdot\mid q,\mathcal{C}(V))
}}
\left[
R(z;y,\mathcal{E},\mathcal{C}(V))
\right].
\end{equation}
The reward evaluates answer correctness and correctness-gated citation
alignment, together with auxiliary output checks. Training optimizes
responses generated from the question and catalog; skill traces provide
context only at inference time. Both stages update the reader's LoRA
parameters. Training and reward details
are provided in Appendix~\ref{app:training_objective}.

%% file: sections/4-Method.tex
\section{Methodology}

\begin{figure*}[t]
    \centering
    \includegraphics[width=0.98\linewidth]{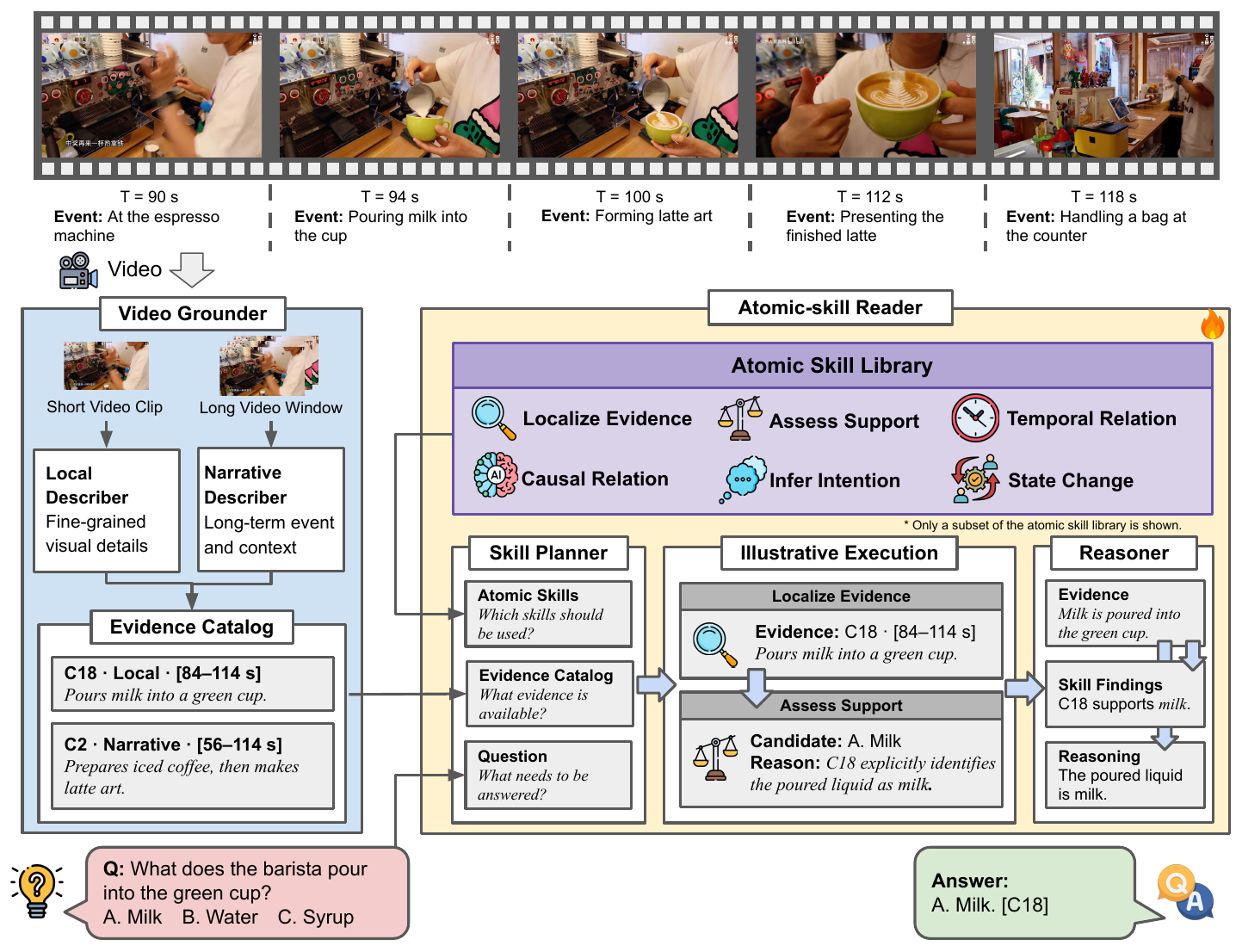}
    \vspace{-10pt}
    \caption{\textbf{Overview of \ours{}.}
    \ours{} combines a reusable timestamped evidence catalog with dynamically composed atomic skills for evidence-grounded long-video reasoning.
    The Video Grounder (\textbf{\textcolor{blue}{Blue Box}}) converts the input video into a timestamped catalog of local observations and, when available, longer-range narrative context.
    For each question, the Atomic-skill Reader (\textbf{\textcolor[rgb]{0.65,0.45,0.00}{Yellow Box}}) selects and composes structured operations from the Atomic Skill Library (\textbf{\textcolor{purple}{Purple Box}}) over this catalog.
    In the illustrated example, the selected skills localize relevant evidence and assess its support for a candidate answer.
    The reader then integrates the resulting evidence and intermediate findings to produce a final answer with explicit catalog citations (\textbf{\textcolor[rgb]{0.0,0.40,0.0}{Green Box}}).
    Only representative skills are shown.}
    \vspace{-10pt}
    \label{fig:atomic_skill_framework}
\end{figure*}

Figure~\ref{fig:atomic_skill_framework} provides an overview of \ours{}, which combines a reusable video evidence catalog, an atomic-skill reader, and grounding-aware post-training. Videos are converted into timestamped textual catalogs that preserve temporal provenance, while a fixed atomic skill vocabulary structures evidence selection, candidate assessment, and integration of intermediate findings. We train the reader with a supervised warm start on teacher-generated responses with explicit citations, followed by GRPO on answer correctness and correctness-gated citation alignment. At inference time, the same post-trained reader dynamically composes and applies atomic skills to gather evidence, integrate intermediate findings, and produce a grounded final answer.

\subsection{Video Evidence Catalog Construction}

The reader operates on a timestamped textual catalog
$\mathcal{C}(V)=\{c_i\}_{i=1}^{N}$, where each entry
$c_i=(x_i,I_i)$ pairs a textual observation $x_i$ with its source interval
$I_i$. Motivated by recent evidence that strong vision-language models can provide reliable grounded descriptions for downstream reasoning~\cite{li2026lenswalk}, we use a large vision-language model to generate grounded descriptions from sampled video segments, including observable scenes, entities, events, and interactions. Each catalog entry is assigned a unique identifier that maps back to its corresponding video interval, providing an explicit link between textual evidence and the source video. Catalog construction is question-independent, so the same evidence representation can be reused across different queries.

To capture evidence at different temporal scales, we construct both local descriptions and longer-range narratives. Local entries preserve fine-grained observations within short temporal windows, while narrative entries connect events across longer spans and help maintain consistent entities and relationships over time. Together, they form a multi-scale evidence representation that supports localized retrieval and reasoning over temporally distributed observations. The resulting catalog serves as the shared evidence interface for subsequent planning, skill application, and answer grounding. Construction settings, serialization details, and representative examples are provided in Appendix~\ref{app:catalog}.

\subsection{Atomic-Skill Reader}
\label{sec:dynamic_skill_execution}

\noindent
\textbf{Skill interfaces.}
We define six reusable skill interfaces for evidence retrieval, support assessment, and temporal, causal, intention, and state-change reasoning. Each specifies its inputs, instruction, structured output, and evidence references. All skills are executed by the same reader. The full registry appears in Appendix~\ref{app:atomic_skills}.

\noindent
\textbf{Question-dependent composition.}
Given a question $q$, answer options $\mathcal{O}$, and catalog $\mathcal{C}(V)$, the reader constructs an ordered plan
$\mathcal{P}=(a_1,\ldots,a_L)$, where
$a_i=(s_{k_i},\boldsymbol{\alpha}_i)$ specifies a skill and its arguments. The planning call also selects a catalog subset $\mathcal{C}_{\mathrm{sel}}\subseteq\mathcal{C}(V)$. This subset is distinct from the annotated evidence $\mathcal{E}$, which is not an inference input. The selected skills and their order depend on the question and available evidence.

\noindent
\textbf{Skill execution and integration.}
Each skill uses one reader call and returns a structured record $r_i$, yielding $\mathcal{R}=(r_1,r_2)$. The second skill receives the first result, and valid localization outputs update later evidence bindings. Integration then combines the plan and skill outputs without a separate verifier. Support fields are advisory. Stage instructions, schemas, and token limits are provided in Appendix~\ref{app:reader_stage_protocol}.
Algorithm~\ref{alg:dynamic_skills} summarizes the five-call Dynamic pipeline: planning, two skill calls, integration, and final answer generation. \textsc{Bind} and \textsc{UpdateRefs} are deterministic argument-binding operations. The historical verifier variant replaces integration with \texttt{verify\_claim\_support} while keeping the same reader and call budget. Other evaluation configurations are detailed in Appendix~\ref{app:experiment_configurations}.

\begin{algorithm}[t]
\caption{Five-call Dynamic inference (without verifier).}
\label{alg:dynamic_skills}
\begin{algorithmic}[1]
\Require Question $q$, options $\mathcal{O}$, precomputed catalog
$\mathcal{C}(V)$, skill library $\mathcal{S}$, reader $\pi_{\theta}$
\Ensure Answer $\hat{y}$, cited catalog identifiers $\hat{Z}$, intermediate records $\mathcal{R}$
\State $x \gets (q,\mathcal{O},\mathcal{C}(V))$
\State $p \gets \Call{Reader}{\pi_\theta,\textsc{Plan},x,\mathcal{S}}$ \Comment{1}
\State $((a_1,a_2),\mathcal{C}_{\mathrm{sel}},y_0) \gets \Call{ParsePlan}{p}$
\State $\boldsymbol{\alpha}_1 \gets \Call{Bind}{a_1,p,\mathcal{C}_{\mathrm{sel}},\varnothing}$
\State $r_1 \gets \Call{Reader}{\pi_\theta,s_{k_1},x,\boldsymbol{\alpha}_1}$ \Comment{2}
\State $\mathcal{C}_{\mathrm{sel}} \gets \Call{UpdateRefs}{\mathcal{C}_{\mathrm{sel}},r_1}$
\State $\boldsymbol{\alpha}_2 \gets \Call{Bind}{a_2,p,\mathcal{C}_{\mathrm{sel}},r_1}$
\State $r_2 \gets \Call{Reader}{\pi_\theta,s_{k_2},x,\boldsymbol{\alpha}_2}$ \Comment{3}
\State $\mathcal{R} \gets (r_1,r_2)$
\State $h \gets \Call{Reader}{\pi_\theta,\textsc{Integrate},x,p,\mathcal{R}}$ \Comment{4}
\State $z \gets \Call{Reader}{\pi_\theta,\textsc{Answer},x,p,\mathcal{R},h}$ \Comment{5}
\State $(\hat{y},\hat{Z}) \gets \Call{ParseAnswerAndCitations}{z}$
\State \Return $\hat{y},\hat{Z},\mathcal{R}$
\end{algorithmic}
\end{algorithm}

\subsection{Grounding-Aware Reader Training}

We train $\pi_\theta$ to produce a rationale, catalog citations, and a final answer from the question and evidence catalog. Training uses SFT followed by GRPO. Intermediate skill outputs are used only at inference time.

\noindent
\textbf{Supervised warm start.}
A teacher model generates cited rationales and answers from the question, answer choices, and catalog. We retain correct responses with valid rationales and, when evidence annotations are available, filter them by citation quality. The resulting responses are used for completion-only SFT.

\noindent
\textbf{Correctness-gated grounding reward.}
For a generated response $z$, let $A$ denote answer correctness and $P_{\mathrm{cite}}$ denote citation precision, measured by overlap between cited catalog intervals and annotated evidence. Our full reward is
\begin{equation}
R_{\mathrm{full}} = A + 0.5A P_{\mathrm{cite}} - 0.2T + 0.1F - 0.05L_{\mathrm{len}},
\end{equation}
where $T$ penalizes citations presented out of chronological
order on temporal questions, $F$ rewards valid response structure, and $L_{\mathrm{len}}$ penalizes excessive length. The product $A P_{\mathrm{cite}}$ makes citation reward conditional on a correct answer, encouraging grounding without rewarding well-cited incorrect predictions. Citation precision measures temporal alignment with annotated evidence rather than semantic entailment.

\noindent
\textbf{GRPO optimization.}
Starting from the SFT checkpoint, we apply Group Relative Policy Optimization (GRPO)~\citep{shao2024deepseekmath}. For each question, multiple sampled responses are scored with the reward above and optimized using group-relative advantages. Reference answers and evidence annotations are used only for reward computation and are not provided to the reader. Training details, data statistics, and reward ablations are given in Appendix~\ref{app:training_objective}.

%% file: sections/5-Results.tex
\section{Experiments}
\label{sec:experiments}

\input{tables/main}

\noindent
\textbf{Benchmarks.}
We evaluate on three complementary video reasoning benchmarks:
CG-Bench~\citep{chen2025cg} for clue-grounded long-video understanding,
VRBench~\citep{yu2025vrbench} for multi-step reasoning over long narrative videos,
and Video-Holmes~\citep{cheng2025video} for complex social reasoning that requires locating and integrating visual clues.
Dataset splits and evaluation statistics are provided in Appendix~\ref{app:data_split}.

\noindent
\textbf{Baselines and comparison protocol.}
For Table~\ref{tab:main_results}, external video-reasoning models retain their original training setups but are evaluated as readers of the same precomputed textual evidence catalogs used by our reader.
We compare \ours{} with three reinforcement-learning-based methods: GRPO-CARE~\citep{chen2026grpo}, which adds a reasoning--answer consistency reward to outcome-based training; VideoChat-R1~\citep{li2025videochat}, which applies GRPO to spatio-temporal video reasoning; and Video-R1-7B~\citep{feng2026video}, which combines a chain-of-thought cold start with temporal-aware reinforcement learning.
We also include Qwen2.5-VL-7B~\citep{bai2025qwen25vltechnicalreport} as a reference baseline.
The Qwen3.5-9B~\citep{qwen3.5} ablations in the same table provide controlled comparisons of training objectives.

\noindent
\textbf{Evaluation metrics.}
We report answer accuracy, Grounded accuracy, and generation length.
A response is Grounded if its answer is correct and at least half of its valid, distinct citations have positive-duration overlap with at least one annotated evidence span.
Duplicate citations are removed and invalid identifiers are ignored.
Responses with no valid citations or no evidence annotations are counted as ungrounded, and all evaluated questions remain in the Grounded-accuracy denominator.
Grounded accuracy therefore measures temporal alignment with annotated evidence rather than semantic entailment and does not require citations to cover every annotated span~\citep{gao2023enabling}.
Generation length is the average number of generated tokens per question, including both the rationale and final answer.
Reported gains are absolute differences between percentage-valued scores.
For each checkpoint, we generate one response per question and report 95\% bootstrap confidence intervals for answer and Grounded accuracy. 
Appendix~\ref{app:external_grounding} reports complementary benchmark-native temporal and semantic grounding metrics, while Appendix~\ref{app:evidence_dependence} tests whether predictions depend on cited catalog content through evidence interventions.

\noindent
\textbf{Training setup and controlled ablations.}
We train the Qwen3.5-9B reader in two stages: supervised fine-tuning (SFT) followed by Group Relative Policy Optimization (GRPO).
During SFT, the model learns from filtered teacher responses containing rationales, evidence citations, and final answers.
All GRPO variants start from the same SFT checkpoint and share the answer-correctness reward, format bonus, and excess-length penalty.
The citation variant adds a correctness-gated citation-alignment reward, the temporal variant adds a citation-order penalty, and the full model includes both.
Both stages update only the LoRA parameters.
Detailed reward definitions and optimization settings are provided in Appendix~\ref{app:training_objective}.

\subsection{Main Results}

\input{tables/inference}

\textbf{\ours{} maintains competitive answer accuracy while improving grounding.}
As shown in Table~\ref{tab:main_results}, \ours{} achieves macro-averaged answer accuracy of 59.2\% and Grounded accuracy of 25.4\% across the three benchmarks. Compared with Gemini-2.5-Pro, \ours{} improves answer accuracy by 0.9 points and overlap-based Grounded accuracy by 6.7 points, with higher Grounded accuracy on all three benchmarks. Relative to the Qwen3.5-9B base model, the gains are 8.8 and 7.2 points, respectively. With evidence catalogs shared across models, these results highlight the role of reader training and inference in using and citing available evidence. We further examine the contributions of training and inference through the Qwen3.5-9B ablations and inference-policy controls below. Table~\ref{tab:raw_video_baselines} reports complementary raw-video baselines using 16 uniformly sampled frames per video. In the shared-catalog setting, \ours{} achieves answer accuracy comparable to Gemini-2.5-Pro while attaining substantially higher overlap-based Grounded accuracy.

\input{tables/raw_video}

\noindent

\noindent
\textbf{Evidence grounding across benchmarks.}
Across the three benchmarks in Table~\ref{tab:main_results}, \ours{} improves overlap-based Grounded accuracy over Gemini-2.5-Pro by 9.7, 5.5, and 4.8 percentage points on Video-Holmes, VRBench, and CG-Bench, respectively. Table~\ref{tab:gemini_ours_grounding} shows higher Acc@GQA on all three benchmarks, with the largest gain on VRBench (16.77\% vs.\ 9.70\%), while Union mIoU and citation IoU-F1 vary across datasets. On EG-VQA, full-reward training improves EG-F1 from 38.78\% to 41.54\% over SFT under the same Dynamic inference pipeline (Appendix~\ref{app:external_grounding}), providing an additional measure of temporal and semantic evidence alignment.

\noindent
\textbf{Evidence dependence under catalog interventions.}
We rerun the full inference pipeline after modifying the evidence catalog (Appendix~\ref{app:evidence_dependence}). Removing cited entries reduces Accuracy more than matched random removal: 50.35\% vs.\ 52.53\% on Video-Holmes, 75.15\% vs.\ 77.78\% on VRBench, and 37.36\% vs.\ 42.94\% on CG-Bench. Conversely, keeping only cited entries largely preserves full-catalog Accuracy. This suggests that the cited evidence is behaviorally relevant to the model's predictions, although it does not by itself establish causal faithfulness or semantic support.

\input{tables/strict_gemini_ours}

\subsection{Ablation}

The ablation block of Table~\ref{tab:main_results} studies SFT, narrative evidence, and GRPO rewards. \textsc{Qwen3.5-9B} is the base reader, and \textsc{+ SFT} is trained on filtered teacher responses with rationales, citations, and answers. For Video-Holmes and CG-Bench, \textsc{w/o narrative} removes narrative entries at inference time; VRBench uses its default local catalog, with a separate narrative control in Appendix~\ref{app:catalog}. All GRPO variants start from the same SFT checkpoint and include answer correctness; citation and temporal rewards are added individually or together in the full model. Table~\ref{tab:inference_controls} separately compares inference policies.

\noindent
\textbf{Answer accuracy and citation alignment respond differently to training.}
Relative to Qwen3.5-9B, SFT improves answer accuracy by 8.6 and 5.9 percentage points on Video-Holmes and VRBench, respectively, while Grounded accuracy changes less consistently. Outcome-only GRPO further raises accuracy to 54.4\% and 76.4\%, again with limited Grounded gains. With Dynamic inference fixed, full-reward GRPO improves Grounded accuracy over outcome-only GRPO by 3.32, 0.82, and 3.52 percentage points on Video-Holmes, VRBench, and CG-Bench (Table~\ref{tab:training_inference_controls}). On VRBench, this grounding gain comes with lower answer accuracy, from 81.65\% to 78.14\%. Overall, answer correctness and citation alignment are related but distinct training objectives: improving one does not necessarily improve the other.

\begin{table}[t]
\centering
\footnotesize
\setlength{\tabcolsep}{2pt}
\renewcommand{\arraystretch}{1.08}
\caption{\textbf{Training objectives and inference policies under matched budgets.}
Cells report Accuracy / Grounded accuracy (\%) on Video-Holmes, VRBench, and CG-Bench. All settings use the same catalogs, five reader calls, and 4,096-token output budget. Outcome and Full denote outcome-only and full-reward GRPO; Generic uses structured reasoning without named skills, while Dynamic selects from six skills. Bold marks the best value per metric and benchmark.}
\vspace{-5pt}
\label{tab:training_inference_controls}
\resizebox{\columnwidth}{!}{%
\begin{tabular}{@{}l|r@{\,/\,}l|r@{\,/\,}l|r@{\,/\,}l@{}}
\toprule
& \multicolumn{2}{c|}{\textbf{VH}}
& \multicolumn{2}{c|}{\textbf{VRBench}}
& \multicolumn{2}{c}{\textbf{CG-Bench}} \\
\cmidrule(lr){2-3}\cmidrule(lr){4-5}\cmidrule(lr){6-7}
\textbf{Training + inference}
& \textbf{Acc.} $\uparrow$ & \textbf{Grd.} $\uparrow$
& \textbf{Acc.} $\uparrow$ & \textbf{Grd.} $\uparrow$
& \textbf{Acc.} $\uparrow$ & \textbf{Grd.} $\uparrow$ \\
\midrule
SFT + Generic
& 49.22 & 12.89 & 73.40 & 30.31 & 38.48 & 24.02 \\
SFT + Dynamic
& 50.20 & 12.11 & 78.56 & \textbf{35.05} & 42.97 & 22.66 \\
\midrule
Outcome + Dynamic
& \textbf{53.12} & 13.48 & \textbf{81.65} & 34.02 & 44.92 & 23.83 \\
\midrule
Full + Generic
& 50.59 & 16.02 & 72.16 & 30.72 & 42.97 & \textbf{27.34} \\
Full + Dynamic
& 51.56 & \textbf{16.80} & 78.14 & 34.85 & \textbf{45.12} & \textbf{27.34} \\
\midrule
\multicolumn{7}{c}{\textit{Full Dynamic $-$ Full Generic (percentage points)}} \\
\midrule
$\Delta$ Accuracy
& \multicolumn{2}{c|}{$+0.98 \pm 4.00$}
& \multicolumn{2}{c|}{$+5.98 \pm 3.36$}
& \multicolumn{2}{c}{$+2.15 \pm 3.93$} \\
$\Delta$ Grounded
& \multicolumn{2}{c|}{$+0.78 \pm 2.54$}
& \multicolumn{2}{c|}{$+4.12 \pm 3.18$}
& \multicolumn{2}{c}{$\phantom{+}0.00 \pm 3.42$} \\
\bottomrule
\end{tabular}%
}
\vspace{-10pt}
\end{table}

\noindent
\textbf{Narrative evidence is most useful for dispersed clues.}
We keep the reader checkpoint fixed and vary only the evidence catalog at inference time. On Video-Holmes and CG-Bench, the \emph{w/o narrative} rows remove narrative entries while retaining the local structured descriptions. On Video-Holmes, removing narratives from the SFT model reduces Accuracy and Grounded accuracy by 5.2 and 5.0 percentage points, respectively, consistent with the need to connect clues across distant parts of the video. On VRBench, adding narratives changes SFT Accuracy from 74.55\% to 72.32\% and Grounded accuracy from 25.25\% to 27.47\%; for the full-reward single-call reader, the corresponding changes are 75.96\% to 76.36\% and 30.10\% to 34.55\%. These catalog controls are separate from the main SFT and Dynamic Skills evaluations in Table~\ref{tab:main_results}. Overall, narrative augmentation can improve citation alignment without uniformly improving answer accuracy.

\noindent
\textbf{Citation and temporal rewards capture different aspects of supervision.}
Within the historical single-call reward ablations, citation supervision improves Grounded accuracy over outcome-only GRPO by 5.0 and 2.4 percentage points on Video-Holmes and VRBench, respectively, while the effect of the citation-order penalty varies across benchmarks. The complete \ours{} system combines full-reward training with five-call Dynamic inference; Table~\ref{tab:training_inference_controls} isolates the effect of the full objective by comparing it with outcome-only GRPO under the same inference pipeline.

\subsection{Structured Skill Inference}
\label{sec:structured_skill_inference}

\textbf{Inference controls.}
Table~\ref{tab:inference_controls} compares three inference policies using the same post-trained reader, catalogs, decoding settings, and five-call output budget. \textsc{Free Text} uses unconstrained reasoning. \textsc{Fixed Skills} applies localization and hypothesis-support assessment, with \texttt{verify\_claim\_support} in call 4. \textsc{Dynamic Skills} selects question-specific skills and uses integration in call 4 (Algorithm~\ref{alg:dynamic_skills}). Appendix~\ref{app:historical_dynamic_fixed} compares Fixed and Dynamic with integration in both pipelines. Failures are scored as incorrect. Skill definitions are provided in Appendix~\ref{app:atomic_skills}, and Appendix~\ref{app:qualitative_atomic_skills} presents qualitative traces from the historical verifier variant.

\noindent

\noindent
\textbf{Structured inference policies.}
Dynamic Skills achieves the highest Grounded accuracy in Table~\ref{tab:inference_controls}, averaging 5.5 percentage points above Free Text, with Accuracy similar to Fixed Skills. When Fixed and Dynamic use matched integration, Accuracy remains close while Grounded differences vary by dataset (Appendix~\ref{app:historical_dynamic_fixed}). Compared with Generic under the same reader and inference budget, Dynamic improves VRBench Accuracy from 72.16\% to 78.14\% with Full training and by 5.15 percentage points with SFT, while Grounded gains remain setting-dependent (Table~\ref{tab:training_inference_controls}). These comparisons evaluate the full Dynamic policy, including answer preselection, question-specific skills, and candidate-conditioned support.

\noindent
\textbf{Skill vocabulary under a fixed pipeline.}
Table~\ref{tab:skill_pool_ablation} varies only the candidate skill pool under the same reader and inference budget. The two-, three-, and six-skill pools progressively add temporal, causal, intention, and state-change reasoning. Smaller pools change answer accuracy by less than one point, but reduce Video-Holmes Grounded accuracy by 2.79 and 2.90 points. VRBench shows the same trend with wider intervals, while CG-Bench changes little. Thus, skill vocabulary can affect grounding even under a fixed pipeline.

\input{tables/skill_pool_ablation}

\noindent
\textbf{Initial answers and downstream evidence processing.}
On VRBench, the saved Full Dynamic planning calls achieve 80.62\% initial Accuracy. Downstream stages correct 5 answers but change 17 correct answers to wrong ones, reducing final Accuracy to 78.14\%, a drop of 2.47 percentage points. SFT similarly decreases from 80.62\% to 78.56\%. This distinction matters: the performance of the full inference policy should be separated from whether its later stages improve the initial answer. The intermediate traces make both useful corrections and harmful answer changes visible. Appendix~\ref{app:confirmation_bias} analyzes these transitions in more detail.

%% file: tables/main.tex
\begin{table*}[t]
\caption{\textbf{Answer accuracy and evidence grounding across three video-reasoning benchmarks.}
All methods are evaluated as catalog readers using the same default catalog within each benchmark's main comparison.
Base, SFT, reward-component, and catalog-control rows use single-call inference; \ours{} uses the full-reward reader with five-call Dynamic inference. These historical runs are summarized in Appendix~\ref{app:experiment_configurations}. Table~\ref{tab:training_inference_controls} compares training objectives under matched Dynamic inference.
The default VRBench catalog uses approximately 30-second local entries; narrative augmentation is evaluated separately.
\textbf{Grounded} requires a correct answer and citation precision of at least 0.5 against annotated evidence spans.
Values are percentages with 95\% bootstrap confidence intervals; bold indicates the best result within each large- or small-model group.
\textbf{Takeaway:} \ours{} achieves the highest overlap-based Grounded accuracy among the evaluated small models while maintaining competitive answer accuracy relative to large-model baselines.}
\vspace{-5pt}
    \label{tab:main_results}
    \centering
    \resizebox{\textwidth}{!}{%
    \setlength{\tabcolsep}{4pt}
    \renewcommand{\arraystretch}{0.98}
    \begin{tabular}{@{}l|cc|cc|cc@{}}
    \toprule
    \multicolumn{1}{c|}{}
    & \multicolumn{2}{c|}{\textbf{Video-Holmes}}
    & \multicolumn{2}{c|}{\textbf{VRBench}}
    & \multicolumn{2}{c}{\textbf{CGBench}}
    \\
    \cmidrule(lr){2-3}
    \cmidrule(lr){4-5}
    \cmidrule(lr){6-7}
    \textbf{Method}
    & \textbf{Accuracy} $\uparrow$
    & \textbf{Grounded} $\uparrow$
    & \textbf{Accuracy} $\uparrow$
    & \textbf{Grounded} $\uparrow$
    & \textbf{Accuracy} $\uparrow$
    & \textbf{Grounded} $\uparrow$  \\
    \midrule
    
    \multicolumn{7}{c}{\textit{Large-model baselines}} \\
    \midrule
    \textsc{Gemini-2.5-Pro}
    & $\bm{51.39 \pm 2.29}$ & $\bm{6.75 \pm 1.12}$ & $\bm{77.78 \pm 3.64}$ & $\bm{29.70 \pm 4.04}$ & $\bm{45.81 \pm 2.29}$ & $\bm{19.70 \pm 1.82}$ \\
    \textsc{Qwen3-235B-A22B} & $45.62 \pm 2.29$ & $4.35 \pm 0.93$ & $69.09 \pm 4.04$ & $23.64 \pm 3.74$ & $41.89 \pm 2.24$ & $13.74 \pm 1.57$ \\ %
    \textsc{DeepSeek-V4-Flash}
    & $41.97 \pm 2.26$ & $4.84 \pm 0.98$ & $61.01 \pm 4.24$ & $23.84 \pm 3.74$ & $26.93 \pm 2.04$ & $11.64 \pm 1.46$ \\
    
    \midrule
    \multicolumn{7}{c}{\textit{Video-reasoning models evaluated as catalog readers}} \\
    \midrule
    Qwen2.5-VL-7B
    & $31.30 \pm 2.12$ & $0.22 \pm 0.19$ & $37.78 \pm 4.24$ & $2.83 \pm 1.52$ & $19.43 \pm 1.79$ & $2.59 \pm 0.75$ \\ %
    \textsc{GRPO-CARE}~\citep{chen2026grpo}
    & $32.44 \pm 2.12$ & $0.93 \pm 0.44$ & $45.25 \pm 4.34$ & $4.24 \pm 1.72$ & $25.61 \pm 1.99$ & $3.70 \pm 0.86$ \\ %
    \textsc{VideoChat-R1}~\citep{li2025videochat}
    & $32.61 \pm 2.12$ & $0.27 \pm 0.24$ & $45.86 \pm 4.34$ & $2.22 \pm 1.31$ & $24.67 \pm 1.96$ & $3.70 \pm 0.86$ \\ %
    \textsc{Video-R1-7B}~\citep{feng2026video}
    & $28.91 \pm 2.07$ & $0.05 \pm 0.08$ & $37.17 \pm 4.24$ & $0.00 \pm 0.00$ & $21.41 \pm 1.85$ & $0.50 \pm 0.30$ \\ %
    
    \midrule
    \multicolumn{7}{c}{\textit{\ours{} and ablations (Qwen3.5-9B)}} \\
    \midrule
    \textsc{Qwen3.5-9B}
    & $43.44 \pm 2.26$ & $7.62 \pm 1.22$ & $69.09 \pm 4.04$ & $28.08 \pm 3.94$ & $38.85 \pm 2.24$ & $18.98 \pm 1.79$ \\
    \textsc{\quad + SFT}
    & $51.99 \pm 2.29$ & $7.19 \pm 1.17$ & $74.95 \pm 3.74$ & $23.43 \pm 3.74$ & $39.96 \pm 2.24$ & $20.75 \pm 1.85$ \\
    \quad + SFT w/o narrative & $46.82 \pm 2.29$ & $2.18 \pm 0.68$ & $74.55 \pm 3.84$ & $25.25 \pm 3.74$ & $39.46 \pm 2.25$ & $17.49 \pm 1.75$ \\
    \quad + GRPO (full reward; w/o narrative) & $51.06 \pm 2.29$ & $1.47 \pm 0.57$ & $75.96 \pm 3.74$ & $30.10 \pm 4.04$ & $39.01 \pm 2.25$ & $17.49 \pm 1.75$ \\
    \textsc{\quad + GRPO (outcome-only reward)}
    & $54.44 \pm 2.29$ & $6.64 \pm 1.12$ & $76.36 \pm 3.74$ & $24.04 \pm 3.74$ & $39.80 \pm 2.25$ & $22.46 \pm 1.90$ \\
    \textsc{\quad + GRPO (outcome + citation)}
    & $\bm{55.69 \pm 2.29}$ & $11.59 \pm 1.44$ & $77.98 \pm 3.64$ & $26.46 \pm 3.84$ & $42.83 \pm 2.26$ & $21.63 \pm 1.85$ \\
    \textsc{\quad + GRPO (outcome + temporal)}
    & $55.31 \pm 2.29$ & $5.50 \pm 1.03$ & $76.77 \pm 3.74$ & $23.43 \pm 3.74$ & $43.98 \pm 2.26$ & $17.94 \pm 1.77$ \\
    \midrule
    \textbf{\textsc{\ours{} (full)}} & $54.06 \pm 2.26$ & $\bm{16.49 \pm 1.69}$ & $\bm{78.99 \pm 3.64}$ & $\bm{35.15 \pm 4.24}$ & $\bm{44.59 \pm 2.29}$ & $\bm{24.50 \pm 1.99}$ \\
    \bottomrule
    \end{tabular}%
    }
    \vspace{-10pt}
    \end{table*}

%% file: tables/inference.tex
\begin{table*}[t]
    \caption{\textbf{Inference-time control study for atomic-skill application.}
All methods use the same reader checkpoint, evidence catalog, five-call budget, and 4,096-token output limit.
\textsc{Free Text} uses unconstrained reasoning; \textsc{Fixed Skills} follows a predefined two-skill sequence; and \textsc{Dynamic Skills} selects two skills based on the question.
Table~\ref{tab:dynamic_fixed_no_verifier} further compares Fixed and Dynamic when both use integration.
We report Accuracy, Grounded accuracy, and output tokens.
\textbf{Takeaway:} \textsc{Dynamic Skills} gives the highest Grounded accuracy across all three benchmarks, with similar answer accuracy to \textsc{Fixed Skills}.}
    \vspace{-5pt}
    \label{tab:inference_controls}
    \centering
    \resizebox{\textwidth}{!}{%
    \setlength{\tabcolsep}{4pt}
    \renewcommand{\arraystretch}{0.98}
    \begin{tabular}{@{}l|ccc|ccc|ccc@{}}
    \toprule
    \multicolumn{1}{c|}{}
    & \multicolumn{3}{c|}{\textbf{Video-Holmes}}
    & \multicolumn{3}{c|}{\textbf{VRBench}}
    & \multicolumn{3}{c}{\textbf{CGBench}} \\
    \cmidrule(lr){2-4}\cmidrule(lr){5-7}\cmidrule(lr){8-10}

    \textbf{Inference policy}
    & \textbf{Accuracy} $\uparrow$
    & \textbf{Grounded} $\uparrow$
    & \textbf{Tokens}
    & \textbf{Accuracy} $\uparrow$
    & \textbf{Grounded} $\uparrow$
    & \textbf{Tokens}
    & \textbf{Accuracy} $\uparrow$
    & \textbf{Grounded} $\uparrow$
    & \textbf{Tokens} \\
    \midrule

    \textsc{Free Text}
    & $53.40 \pm 2.29$
    & $11.76 \pm 1.47$
    & $662.7 \pm 7.0$
    & $72.73 \pm 3.94$
    & $26.67 \pm 3.94$
    & $765.2 \pm 28.3$
    & $38.91 \pm 2.21$
    & $21.14 \pm 1.88$
    & $722.0 \pm 17.2$ \\

    \textsc{Fixed Skills}
    & $53.78 \pm 2.29$
    & $12.63 \pm 1.53$
    & $715.0 \pm 12.4$
    & $76.57 \pm 3.74$
    & $31.31 \pm 4.04$
    & $731.9 \pm 19.8$
    & $44.54 \pm 2.29$
    & $23.90 \pm 1.99$
    & $760.8 \pm 19.1$ \\

    \midrule
    \textsc{Dynamic Skills} (\ours{})
    & $54.06 \pm 2.26$
    & $\bm{16.49 \pm 1.69}$
    & $670.5 \pm 10.1$
    & $78.99 \pm 3.64$
    & $\bm{35.15 \pm 4.24}$
    & $669.0 \pm 15.0$
    & $44.59 \pm 2.29$
    & $\bm{24.50 \pm 1.99}$
    & $723.5 \pm 16.1$ \\

    \bottomrule
    \end{tabular}%
    }
    \vspace{-10pt}
\end{table*}

%% file: tables/raw_video.tex
\begin{table}[t]
\centering
\small
\setlength{\tabcolsep}{3pt}
\caption{\textbf{Raw-video baselines.}
Accuracy (\%) with 16 uniformly sampled frames and a 768-token output cap.}
\vspace{-5pt}
\label{tab:raw_video_baselines}
\begin{tabular*}{\columnwidth}{@{\extracolsep{\fill}}lccc@{}}
\toprule
Model & VH & VRBench & CG-Bench \\
\midrule
Gemini-2.5-Pro & 54.10 & 81.82 & 44.26 \\
Qwen2.5-VL-7B & 34.78 & 59.60 & 34.71 \\
GRPO-CARE & 35.98 & 59.80 & 33.55 \\
VideoChat-R1 & 36.53 & 60.81 & 35.76 \\
Video-R1-7B & 37.78 & 58.79 & 35.43 \\
\bottomrule
\end{tabular*}
\vspace{-10pt}
\end{table}

%% file: tables/strict_gemini_ours.tex
\begin{table}[t]
\centering
\setlength{\abovecaptionskip}{0pt}
\setlength{\belowcaptionskip}{2pt}
\fontencoding{T1}\fontfamily{ptm}\selectfont
\caption{\textbf{Grounding diagnostics (\%).}
Each cell reports Gemini-2.5-Pro / \ours{};
the better result is bold.
\ours{} achieves higher Acc@GQA on all benchmarks,
with mixed temporal alignment results.}
\vspace{5pt}
\label{tab:gemini_ours_grounding}
\small
\setlength{\tabcolsep}{2pt}
\renewcommand{\arraystretch}{1.0}
\begin{tabular*}{\linewidth}{@{\extracolsep{\fill}}lccc@{}}
\toprule
\textbf{Metric} $\uparrow$
& \textbf{Video-Holmes} & \textbf{VRBench} & \textbf{CG-Bench} \\
\midrule
Union mIoU
& \textbf{1.41}\,/\,1.40
& \textbf{17.73}\,/\,11.76
& 7.92\,/\,\textbf{9.15} \\
IoU-F1
& \textbf{1.86}\,/\,1.42
& 6.34\,/\,\textbf{7.04}
& 7.87\,/\,\textbf{8.68} \\
Acc@GQA
& 0.11\,/\,\textbf{0.16}
& 9.70\,/\,\textbf{16.77}
& 0.99\,/\,\textbf{1.71} \\
\bottomrule
\end{tabular*}
\vspace{-10pt}
\end{table}

%% file: tables/skill_pool_ablation.tex
\begin{table}[t]
\centering
\small
\setlength{\tabcolsep}{4pt}
\renewcommand{\arraystretch}{1.12}
\caption{\textbf{Candidate-pool reduction under a fixed inference protocol.}
All settings select two skills within the same five-call, 4,096-token pipeline. Accuracy and Grounded are percentages. $\Delta$ denotes the change from the six-skill baseline in percentage points. Results use the full supplementary cohorts and a single training run and inference seed.}
\label{tab:skill_pool_ablation}
\resizebox{\columnwidth}{!}{%
\begin{tabular}{@{}llrrrll@{}}
\toprule
Benchmark & Pool & Accuracy & Grounded & & $\Delta$ Accuracy & $\Delta$ Grounded \\
\midrule
Video-Holmes & 6 skills & 53.86 & 16.97 & & --- & --- \\
 & 3 skills & 54.08 & 14.18 & & $+0.22 \pm 1.32$ & $-2.79 \pm 1.44$ \\
 & 2 skills & 54.30 & 14.07 & & $+0.44 \pm 1.34$ & $-2.90 \pm 1.54$ \\
\midrule
VRBench & 6 skills & 78.14 & 34.85 & & --- & --- \\
 & 3 skills & 78.35 & 32.78 & & $+0.21 \pm 2.16$ & $-2.06 \pm 2.95$ \\
 & 2 skills & 77.94 & 31.96 & & $-0.21 \pm 2.26$ & $-2.89 \pm 3.04$ \\
\midrule
CG-Bench & 6 skills & 44.01 & 24.31 & & --- & --- \\
 & 3 skills & 44.89 & 24.58 & & $+0.89 \pm 1.24$ & $+0.28 \pm 1.18$ \\
 & 2 skills & 43.84 & 23.86 & & $-0.17 \pm 1.27$ & $-0.44 \pm 1.22$ \\
\bottomrule
\end{tabular}
}
\end{table}

%% file: sections/6-Conclusion.tex
\section{Conclusion}

We introduced \ours{}, an evidence-grounded video reasoning approach that combines grounding-aware reader training with question-specific atomic-skill inference over timestamped evidence catalogs. Across three benchmarks, its 9B reader achieves competitive answer accuracy and higher overlap-based Grounded accuracy than Gemini-2.5-Pro under shared catalogs. Controlled experiments show that grounding-aware training improves citation alignment, while structured skill inference improves Grounded accuracy over free-form reasoning under matched budgets. The gains vary across tasks, and downstream evidence processing does not always improve the initial answer, but the intermediate records make the model's use of evidence explicit and inspectable.

\noindent
\textbf{Limitations.}
The video grounder remains an important bottleneck. Because the reader operates on precomputed evidence catalogs, information missed or distorted during catalog construction may be unavailable to downstream reasoning. Our experiments therefore evaluate reasoning over structured evidence rather than end-to-end retrieval from raw video. In addition, Grounded accuracy is based on temporal overlap and does not directly measure whether the cited content semantically supports the answer; this distinction is most apparent on Video-Holmes in the matched random-citation control.

\noindent
\textbf{Future work.}
We plan to extend atomic skills to video grounding for active evidence retrieval, refinement, and verification, and to make the skill library adaptive across new tasks and failure modes.
More broadly, this points toward long-video reasoning systems that can refine both their evidence and reasoning processes as they operate.

\subsection*{LLM Usage Statement}

We used generative AI tools, including Codex and Claude, to support coding and implementation, research brainstorming, literature search and related-work review, technical writing and proofreading, LaTeX formatting, and figure preparation. AI-assisted code was reviewed and tested by the authors. We did not use generative AI to formulate mathematical claims or proofs or to determine the scientific conclusions of this work. All AI-assisted content was reviewed and revised by the authors.

As part of the methodology, Qwen3.5-9B was used to construct local evidence descriptions, and Qwen3-VL-235B-A22B-Instruct was used to generate narrative entries and teacher responses for SFT. These uses are described in Appendices A and D.

%% file: sections/7-Appendix.tex
\input{sections/7-8-Evidence-Catalog}
\input{sections/7-1-Atomic-Skills}

\input{sections/7-2-Data-Splits}
\input{sections/7-3-Reader-Training}
\input{sections/7-4-Dynamic-Skill-Execution}
\input{sections/7-7-Temporal-Grounding}
\input{sections/7-5-External-Grounding}
\input{sections/7-6-Evidence-Dependence}
\input{sections/7-9-Prompt-Templates}
\input{sections/7-10-Qualitative-Examples}

%% file: sections/7-8-Evidence-Catalog.tex
\section{Evidence Catalog Construction Details}
\label{app:catalog}

\noindent
\textbf{Local descriptions.}
We use Qwen3.5-9B to describe sampled video windows and serialize the outputs as textual evidence entries.
Video-Holmes uses approximately 4-second windows, while CG-Bench and VRBench use approximately 30-second windows.

\noindent
\textbf{Narrative construction.}
For Video-Holmes and CG-Bench, we generate narrative entries with Qwen3-VL-235B-A22B-Instruct over approximately 30- and 60-second windows, respectively, using 16 uniformly sampled frames per window resized to a maximum width of 448 pixels.
Each request includes the sampled frames, overlapping timestamped dialogue, the preceding narrative (up to 1,500 characters), and a running character list (up to 20 entries).
Generation uses temperature 0 and a maximum of 1,200 output tokens; dialogue comes from Whisper transcripts for Video-Holmes and subtitle tracks for CG-Bench.
For VRBench, we use the same model, frame sampling, resolution, and generation settings with a target window duration of 58 seconds.
Source clips are grouped by temporal midpoint, with each narrative window spanning the first through last clip in the group, so actual durations may exceed 58 seconds and adjacent windows may overlap.
Each request includes the local descriptions (up to 900 characters per clip), the preceding narrative, and the running character list, without a separate ASR transcript or subtitle track.
Generated narratives are prepended to the local catalog and the combined entries are reindexed.
Video-Holmes and CG-Bench include narratives in their default catalogs, with \textsc{w/o narrative} removing these entries; VRBench uses approximately 30-second local entries by default and evaluates narrative augmentation separately.

\noindent
\textbf{Narrative controls.}
We evaluate VRBench narrative augmentation with single-call inference on 495 questions from 60 videos, comparing local-only and narrative-augmented catalogs for the SFT and full-reward GRPO readers.
Table~\ref{tab:vrbench_narrative_control} reports a supplementary experiment on the effect of adding narrative entries to the VRBench catalog.
The main SFT and Dynamic results in Table~\ref{tab:main_results} use the default local-only catalog:
SFT achieves 74.95\% Accuracy and 23.43\% Grounded accuracy, while five-call Dynamic \ours{} achieves 78.99\% and 35.15\%, respectively.

\begin{table}[t]
\centering
\caption{\textbf{VRBench narrative augmentation.} Single-call inference on 495 questions from 60 videos, with a fixed reader checkpoint within each pair. Accuracy (Acc.) and Grounded accuracy (Grounded) are percentages.}
\vspace{-5pt}
\label{tab:vrbench_narrative_control}
\footnotesize
\setlength{\tabcolsep}{3pt}
\renewcommand{\arraystretch}{1.1}
\begin{tabular}{@{}llrr@{}}
\toprule
Reader & Catalog & Acc. & Grounded \\
\midrule
SFT & Local & 74.55 & 25.25 \\
SFT & Local + narrative & 72.32 & 27.47 \\
\midrule
Full-reward GRPO & Local & 75.96 & 30.10 \\
Full-reward GRPO & Local + narrative & 76.36 & 34.55 \\
\bottomrule
\end{tabular}
\end{table}

\noindent
\textbf{Catalog serialization.}
Narrative entries precede local entries in the combined catalog.
Each entry retains its source interval and receives a one-based \texttt{rank} $i$.
The citation \texttt{(clip i)} and case-study notation $C_i$ refer to the same ranked entry and interval.
Descriptions are truncated to 1,200 characters, and narratives are cached per video and reused across questions.
Catalog construction does not use the questions or reference answers.

\begin{table}[t]
\caption{\textbf{Reader input and output format.}
The reader consumes a question, answer options, and indexed catalog entries,
then returns a citation-grounded rationale and answer label. The example is
excerpted from Figure~\ref{fig:showcase-phone}.}
\vspace{-5pt}
\label{tab:reader_io}
\centering
\footnotesize
\setlength{\tabcolsep}{5pt}
\renewcommand{\arraystretch}{1.16}
\begin{tabularx}{\linewidth}{@{}
    >{\raggedright\arraybackslash\bfseries}p{0.20\linewidth}
    >{\raggedright\arraybackslash\ttfamily}X@{}}
\toprule
\textrm{\textbf{Field}} & \textrm{\textbf{Serialized example}} \\
\midrule
Question &
What psychological state does the girl's action of gripping her phone tightly
in the car reflect? \\

Options &
A: Looking forward to news from friends.\newline
B: Afraid of being followed.\newline
C: Worry about the phone running out of power.\newline
D: Plan dinner menu.\newline
E: Recall childhood memories.\newline
F: Check time. \\

Catalog entry &
rank: 3\newline
time\_span: [63.0, 97.0] s\newline
description: The woman ... pulls out her phone and makes a call, her expression
anxious as she asks, ``Hello? Will you come get me, please?'' \\

\midrule
Rationale &
(clip 3) explicitly shows the woman making an anxious phone call. This supports
a psychological state of fear about being followed. (clip 4) provides
additional support. \\

Answer &
label: B \\
\bottomrule
\end{tabularx}
\vspace{-5pt}
\end{table}

\noindent
\textbf{Input and output format.}
Teacher supervision, reader training, and evaluation use the same catalog serialization, with citation indices mapping directly to the corresponding source intervals.

%% file: sections/7-1-Atomic-Skills.tex
\section{Atomic Skills}
\label{app:atomic_skills}

\textbf{Skill library.}
Long-video reasoning often involves recurring operations such as evidence localization, hypothesis assessment, temporal and causal reasoning, intention inference, and state tracking.
Following Section~\ref{sec:dynamic_skill_execution}, we represent these operations with six prompt-defined skill interfaces (Table~\ref{tab:skill_bank}).
Each interface specifies its inputs, instruction, structured outputs, and evidence references, together with model-generated support and diagnostic fields.

\noindent
\textbf{Connection to the reader.}
The skills are executed by the same post-trained Qwen3.5-9B reader rather than separate learned modules.
At inference time, the reader selects two skills for each question, executes them over the evidence catalog, and integrates the resulting records into the final prediction.
SFT and GRPO train this reader for answer generation and citation alignment; no separate planner or executor is trained.
The default Dynamic pool contains all six interfaces in Table~\ref{tab:skill_bank}, with repetition allowed across the two selections.
Dynamic inference uses five reader calls: planning, two skill calls, integration, and final-answer generation.
The \ours{} runs and supplementary Dynamic controls use integration in the fourth call, while a historical variant uses \texttt{verify\_claim\_support} in this position.
This operation is performed by the same reader and is not part of the six-skill candidate pool; it appears in the historical Fixed Skills runs and qualitative traces in Appendix~\ref{app:qualitative_atomic_skills}.
The broader implementation registry contains 23 interfaces, while all reported pool ablations use only the six interfaces listed here and their nested three- and two-skill subsets.
Appendix~\ref{app:experiment_configurations} maps these configurations to the corresponding results.

\input{tables/skills}

\subsection{Candidate-pool size and observed usage}
\label{app:skill_pool_ablation}

\textbf{Reduced-pool controls.}
We compare the full six-skill pool with nested three- and two-skill variants. The three-skill pool keeps localization, support assessment, and temporal reasoning, while the two-skill pool keeps localization and support assessment. All variants use the same full-reward checkpoint, evidence catalogs, decoding settings, two skill selections per question, and five-call budget, with repetition allowed. We evaluate them on the same supplementary cohorts as the main experiments and report the mean over three training runs, using a fixed inference seed across runs and conditions. As shown in Table~\ref{tab:skill_pool_ablation}, reducing the pool changes answer accuracy by less than one percentage point. On Video-Holmes, Grounded accuracy drops by 2.79 points with three skills and 2.90 points with two; VRBench shows smaller drops of 2.06 and 2.89 points, while CG-Bench changes little. The effect of the larger skill pool is most visible on Video-Holmes.

\noindent
\textbf{Observed selections.}
Table~\ref{tab:skill_selection_frequency} reports skill usage on a frozen 1,509-question diagnostic subset.
All six skills appear across the three benchmarks, with markedly different selection patterns.
Localization and support assessment account for 89.1\% of selections on CG-Bench, while causal and intention reasoning account for 52.0\% on Video-Holmes.
This variation suggests that the planner adjusts its skill choices to the benchmark.

\input{tables/skill_selection_frequency}

%% file: tables/skills.tex
\begin{table*}[t]
  \caption{\textbf{The six atomic-skill interfaces.}
    All interfaces use the same reader with the question, evidence catalog, bound arguments, and preceding results.
    For each question, the planner selects two interfaces, with repetition allowed.
    Evidence bindings are included in the structured outputs, while scores and confidences are model-generated.
    Appendix~\ref{app:reader_stage_protocol} provides the stage-level schemas.}
  \vspace{-5pt}
  \label{tab:skill_bank}
  \centering
  \small
  \setlength{\tabcolsep}{5pt}
  \renewcommand{\arraystretch}{1.12}
  \begin{tabular}{@{}p{0.275\textwidth}p{0.30\textwidth}p{0.365\textwidth}@{}}
  \toprule
  \textbf{Skill interface} & \textbf{Operation and bound arguments} & \textbf{Principal output fields} \\
  \midrule
  \texttt{localize\_clue}
    & Select relevant evidence for the question from the bound catalog entries.
    & \texttt{best\_refs}, \texttt{confidence}, \texttt{reasoning}. \\
  \texttt{score\_hypothesis\_support}
    & Assess a candidate hypothesis using supporting evidence and counterevidence.
    & \texttt{support\_score}, \texttt{contradiction\_score}, \texttt{reasoning}. \\
  \texttt{infer\_temporal\_relation}
    & Infer the temporal relation between the bound events.
    & \texttt{relation}, \texttt{confidence}, \texttt{reasoning}. \\
  \texttt{infer\_causal\_relation}
    & Assess a proposed cause--effect relation using the bound evidence.
    & \texttt{causal}, \texttt{confidence}, \texttt{reasoning}. \\
  \texttt{infer\_intention\_or\_motive}
    & Infer an agent's intention from actions and available context.
    & \texttt{intention}, \texttt{confidence}, \texttt{alternatives}. \\
  \texttt{infer\_state\_change}
    & Compare an entity's earlier and later states for the specified predicate.
    & \texttt{changed}, \texttt{before\_state}, \texttt{after\_state}, \texttt{confidence}. \\
  \bottomrule
  \end{tabular}
\end{table*}

%% file: tables/skill_selection_frequency.tex
\begin{table}[t]
\centering
\small
\setlength{\tabcolsep}{3pt}
\renewcommand{\arraystretch}{1.12}
\caption{\textbf{Observed selections from the six-skill pool.}
Percentages are computed over the \(2n\) selection slots, with repetition allowed.
The frozen diagnostic subsets contain 1,509 questions and are smaller than the full evaluation sets in Table~\ref{tab:skill_pool_ablation}.}
\vspace{-5pt}
\label{tab:skill_selection_frequency}
\resizebox{\columnwidth}{!}{%
\begin{tabular}{@{}lrrr@{}}
\toprule
Skill interface & Video-Holmes & VRBench & CG-Bench \\
Questions ($n$) & 512 & 485 & 512 \\
Selection slots ($2n$) & 1,024 & 970 & 1,024 \\
\midrule
\texttt{localize\_clue} & 5.27\% & 11.65\% & 46.48\% \\
\texttt{score\_hypothesis\_support} & 28.03\% & 34.85\% & 42.58\% \\
\texttt{infer\_temporal\_relation} & 7.32\% & 3.09\% & 4.59\% \\
\texttt{infer\_causal\_relation} & 25.59\% & 25.26\% & 3.22\% \\
\texttt{infer\_intention\_or\_motive} & 26.37\% & 18.25\% & 1.95\% \\
\texttt{infer\_state\_change} & 7.42\% & 6.91\% & 1.17\% \\
\bottomrule
\end{tabular}%
}
\end{table}

%% file: sections/7-2-Data-Splits.tex
\section{Data Splits and Training Statistics}
\label{app:data_split}

\textbf{Dataset splits.}
Video-Holmes uses its official train and test splits, with 1,551 training questions from 233 videos and 1,837 test questions from 270 disjoint videos. VRBench has no official split, so we use a video-level partition with 480 questions from 60 videos for the SFT candidate pool and 495 questions from a disjoint set of 60 videos for evaluation. For CG-Bench, the training pool contains 1,080 questions from 161 videos, and evaluation uses 1,812 questions from 823 videos in CG-Bench-mini. Training and evaluation sets are disjoint at both the question and source-video levels across all three benchmarks.

Reader prompts include the question, answer options, evidence catalog, and output-format instructions. During multi-call inference, later stages additionally receive the intermediate outputs from earlier calls, following the protocol in Appendix~\ref{app:reader_stage_protocol}. Gold answers and evidence annotations are used for training-example filtering, reward computation, and evaluation.

\noindent
\textbf{Training statistics.}
The combined SFT candidate pool contains 3,111 questions from 454 videos. Teacher-response filtering produces 4,520 response-level examples covering 1,879 unique questions. After length filtering, 4,416 responses covering 1,830 questions are used for SFT. All GRPO variants are trained on the official Video-Holmes training split of 1,551 questions from 233 videos.

\noindent
\textbf{Evaluation protocol.}
We follow the same evaluation setup as in the main experiments. Evaluation sets are fixed before scoring and are not filtered based on model outputs or correctness.

%% file: sections/7-3-Reader-Training.tex
\section{Reader Training and Reward Definitions}
\label{app:training_objective}

\noindent \textbf{SFT warm start.}
We fine-tune Qwen3.5-9B on responses generated by Qwen3-VL-235B-A22B-Instruct, each containing a rationale, evidence citations, and a final answer. For each of the 3,111 candidate questions, we generate responses using the original answer-option order and two shuffled orders. We retain responses with a correct final answer and a nonempty rationale; when gold evidence spans are available, we additionally require citation precision of at least $0.5$. This yields 4,520 response-level examples covering 1,879 unique questions. After length filtering, 4,416 examples covering 1,830 unique questions are used for SFT.

We train for one epoch with completion-only loss using LoRA with rank 16, scaling factor 32, and dropout 0.05. The learning rate is $10^{-4}$, with a per-device batch size of one, gradient accumulation of eight, and a maximum sequence length of 32,768. Thinking mode is disabled.

\noindent \textbf{Reward definitions.}
Let $A\in\{0,1\}$ indicate answer correctness, and let $F\in\{0,1\}$ indicate whether the response is successfully parsed and contains a nonempty rationale.

For citation precision $P_{\text{cite}}$, invalid catalog IDs are discarded and duplicate valid IDs are counted once. A valid citation is a hit if its temporal interval has positive overlap with any annotated evidence span. When annotations are available, $P_{\text{cite}}$ is the fraction of valid citations that are hits, with $P_{\text{cite}}=0$ if no valid citations are provided. For questions without annotated evidence spans, citation precision is undefined and the citation reward is set to zero.

Grounded accuracy requires both a correct answer and sufficiently precise evidence attribution. A question is counted as grounded-correct if $A=1$ and $P_{\text{cite}}\geq 0.5$. Questions without annotations or valid citations receive zero credit and remain in the denominator.

For a completion $z$, let $\ell_{\text{char}}(z)$ denote its character length. We define the length penalty as
\begin{align}
L_{\text{len}}
&=
\max\left(
0,\,
\frac{\ell_{\text{char}}(z)-2500}{2500}
\right).
\end{align}

For temporal-order questions (type prefix \texttt{TA}), we remove
duplicate and invalid citations while preserving their order
of first occurrence in the response.
We set $T=1$ if the start time decreases between any consecutive entries, and $T=0$ otherwise. If fewer than two valid entries remain, we set $T=0$.

The base outcome reward is
\begin{align}
R_{\text{out}}
&=
A + 0.1F - 0.05L_{\text{len}}.
\end{align}
The citation and temporal ablations are
\begin{align}
R_{\text{cite}}
&=
R_{\text{out}} + 0.5A P_{\text{cite}},\\
R_{\text{temp}}
&=
R_{\text{out}} - 0.2T,
\end{align}
and the full reward combines both terms:
\begin{align}
R_{\text{full}}
&=
R_{\text{out}}
+ 0.5A P_{\text{cite}}
- 0.2T.
\end{align}

\noindent \textbf{GRPO post-training.}
We train each GRPO configuration with three random seeds
and report the mean performance across the three runs.
All variants start from the same SFT checkpoint and train only on the 1,551 Video-Holmes training questions.
Each batch contains four questions, with eight sampled responses per question.
Rollouts use temperature 1.0 and a maximum of 1,024 generated tokens.
We train for 384 optimization steps and evaluate the checkpoint saved at step 384.

Rewards are standardized within each question group, and groups with near-zero reward variance are skipped. We optimize a token-level clipped policy-gradient objective with clipping parameter 0.2. LoRA parameters are updated with AdamW using a learning rate of $10^{-5}$, betas $(0.9,0.99)$, ten warm-up steps, and gradient-norm clipping at 1.0. We use neither a KL penalty nor an auxiliary SFT loss. Gold answers and evidence annotations are used only to compute rewards and are never included in the reader input. SFT and GRPO optimize complete generated responses for answer correctness and citation alignment; skill selections and intermediate outputs are not separately labeled or optimized. At inference time, the skill interfaces structure the operations performed by the same reader.

\paragraph{Experiment configurations.}
\label{app:experiment_configurations}

SFT, Outcome, and Full denote the reader training objective, while Free Text, Fixed, Generic, and Dynamic denote the inference policy. Supplementary experiments reuse the corresponding saved checkpoint.
Historical training ablations produce the answer in a single call. The complete \ours{} pipeline uses five calls for planning, skill execution, integration, and final answering. All supplementary inference uses deterministic decoding and the same output budget across methods. Stage prompts are provided in Appendix~\ref{app:reader_stage_protocol}.

\section{Reader Efficiency and Cost Scope}
\label{app:reader_efficiency}

\noindent\textbf{Measurement scope.}
Table~\ref{tab:reader_costs} reports reader usage for the supplementary runs with Qwen3.5-9B. Each question uses five reader calls with a total output budget of 4,096 tokens. Token counts reflect actual usage summed across all calls and averaged over questions. The cost analysis uses 512 Video-Holmes, 485 VRBench, and 512 CG-Bench questions, with Full Dynamic as the baseline within each cohort.

\input{tables/reader_costs}

\noindent\textbf{Latency and cost scope.}
Latency is measured per question across the hardware used for these runs, with paired conditions evaluated under the same hardware assignment. Reported costs cover reader inference only and exclude catalog construction, transcription, frame processing, and other preprocessing. For a video reused across $Q$ questions, the per-question cost can be written as
\begin{equation}
\overline{C}_{\mathrm{question}}
= \frac{C_{\mathrm{catalog}}}{Q}
+ \overline{C}_{\mathrm{reader}}.
\end{equation}

%% file: tables/reader_costs.tex
\begin{table}[t]
\caption{\textbf{Reader token usage and latency.}
We report the average total tokens used per question across five reader calls.
Latency is measured per question on mixed hardware and excludes catalog construction.
All Full settings in a cohort use the same full-reward checkpoint, catalogs, and output budget.
Full + Dynamic uses the six-skill baseline pool, and SFT + Dynamic follows the same call sequence.}
\label{tab:reader_costs}
\vspace{-5pt}
\centering
\small
\setlength{\tabcolsep}{4pt}
\renewcommand{\arraystretch}{0.98}
\resizebox{\columnwidth}{!}{%
\begin{tabular}{@{}l|cc|cc@{}}
\toprule
\multicolumn{1}{c|}{} & \multicolumn{2}{c|}{\textbf{Tokens}} & \multicolumn{2}{c@{}}{\textbf{Latency (s)}} \\
\cmidrule(lr){2-3}\cmidrule(lr){4-5}
\textbf{Condition} & \textbf{Input} & \textbf{Output} & \textbf{Mean} & \textbf{P95} \\
\midrule
\multicolumn{5}{c}{\textit{Video-Holmes ($n=512$)}} \\
\midrule
\textbf{Full + Dynamic (6 skills)} & 64,523.6 & 668.2 & 60.3 & 92.2 \\
Full + 3 skills & 66,440.4 & 713.2 & 62.1 & 97.6 \\
Full + No preselection & 64,461.7 & 743.6 & 63.8 & 105.3 \\
Full + 2 skills & 65,920.8 & 703.6 & 63.8 & 102.1 \\
Full + Generic & 61,915.5 & 559.4 & 52.7 & 86.0 \\
SFT + Dynamic & 64,570.0 & 652.5 & 58.3 & 93.8 \\
\midrule
\multicolumn{5}{c}{\textit{VRBench ($n=485$)}} \\
\midrule
\textbf{Full + Dynamic (6 skills)} & 72,398.2 & 663.5 & 69.1 & 136.8 \\
Full + 3 skills & 74,524.7 & 713.9 & 72.0 & 139.4 \\
Full + No preselection & 71,983.8 & 734.7 & 72.4 & 156.3 \\
Full + 2 skills & 74,542.8 & 725.4 & 73.8 & 132.6 \\
Full + Generic & 69,607.6 & 634.9 & 64.9 & 144.8 \\
SFT + Dynamic & 72,430.7 & 647.6 & 69.4 & 150.9 \\
Full + Symmetric bridge & 71,700.5 & 717.2 & 67.5 & 104.5 \\
\midrule
\multicolumn{5}{c}{\textit{CG-Bench ($n=512$)}} \\
\midrule
\textbf{Full + Dynamic (6 skills)} & 101,453.0 & 730.1 & 105.2 & 262.9 \\
Full + 3 skills & 101,441.6 & 710.4 & 100.4 & 240.2 \\
Full + No preselection & 100,807.2 & 837.2 & 113.1 & 263.0 \\
Full + 2 skills & 101,410.6 & 724.3 & 107.2 & 268.8 \\
Full + Generic & 96,661.3 & 629.3 & 93.6 & 256.8 \\
SFT + Dynamic & 101,506.0 & 695.8 & 106.4 & 281.9 \\
\bottomrule
\end{tabular}%
}
\vspace{-5pt}
\end{table}

%% file: sections/7-4-Dynamic-Skill-Execution.tex
\section{Inference Controls}
\label{app:dynamic_skill_execution}

Each policy uses five reader calls per question. Calls~1--4 are capped at 512 output tokens each. Call~5 uses the evidence catalog and the notes from the previous calls to produce the final answer and citations, with a limit of 2,048 tokens. This gives a total output budget of 4,096 tokens per question. For each comparison, we use the same checkpoint, evidence catalog, and decoding settings.

\begin{itemize}
    \item \textbf{Free Text.}
    The first four calls plan the approach, analyze the evidence, compare possible answers, and consider uncertainty or alternative interpretations. Each call produces a free-form reasoning note.

    \item \textbf{Fixed Skills.}
    Call~1 proposes a candidate answer and supporting evidence, call~2 localizes relevant clues, and call~3 evaluates how well those clues support the candidate. In Table~\ref{tab:inference_controls}, call~4 performs one additional support check using the same reader.

    \item \textbf{Dynamic Skills.}
    Call~1 proposes an initial answer and selects two ordered skills for the question (Algorithm~\ref{alg:dynamic_skills}). Calls~2 and~3 execute the two skills in sequence, with the output of the first passed to the second. Call~4 combines the plan and intermediate findings into a reasoning note. Call~5 then produces the final answer from the accumulated evidence and may revise the initial candidate.

    \item \textbf{Generic.}
    The first four calls plan, gather evidence, compare alternatives, and revisit the current reasoning. Each stage returns relevant catalog ranks and a reasoning note, without maintaining a designated candidate answer.
\end{itemize}

The Dynamic--Generic comparison in Table~\ref{tab:training_inference_controls} compares the full inference pipelines. Dynamic uses an initial answer, question-specific skill selection, and candidate-conditioned evidence assessment, while Generic follows the same fixed sequence of instructions for every question. The stage prompts are provided in Appendix~\ref{app:reader_stage_protocol}.

\noindent\textbf{Grounded accuracy by final-answer correctness.}
We pair Full + Dynamic and Full + Generic by question ID from Table~\ref{tab:training_inference_controls}. Each entry reports the number of predictions meeting the citation-overlap criterion within the corresponding answer-correctness group.

\begin{table}[t]
\caption{\textbf{Grounded accuracy by final-answer correctness.}
Each fraction reports the number of predictions meeting the citation-overlap criterion within the corresponding group. Under \emph{Only one correct}, each column contains questions answered correctly only by that policy.}
\label{tab:grounded_by_correctness}
\vspace{-5pt}
\centering
\small
\setlength{\tabcolsep}{2pt}
\begin{tabular*}{\columnwidth}{@{\extracolsep{\fill}}lrrrr@{}}
\toprule
& \multicolumn{2}{c}{\textbf{Both correct}} & \multicolumn{2}{c}{\textbf{Only one correct}} \\
\cmidrule(lr){2-3}\cmidrule(l){4-5}
\textbf{Benchmark} ($n$) & \textbf{Dynamic} & \textbf{Generic} & \textbf{Dynamic} & \textbf{Generic} \\
\midrule
Video-Holmes (512) & 66/212 & 65/212 & 20/52 & 17/47 \\
VRBench (485)      & 144/328 & 139/328 & 25/51 & 10/22 \\
CG-Bench (512)     & 110/172 & 114/172 & 30/59 & 26/48 \\
\bottomrule
\end{tabular*}
\vspace{-5pt}
\end{table}

On VRBench, most of Dynamic's Grounded gain comes from questions where only one policy answers correctly: 25 such predictions meet the grounding criterion for Dynamic, compared with 10 for Generic. Among questions both policies answer correctly, the difference is smaller (144 vs.\ 139). The corresponding jointly correct results are similar on Video-Holmes and CG-Bench. Appendix~\ref{app:grounding_robustness} evaluates the same predictions with stricter temporal metrics.

\subsection{Dynamic versus Fixed Skills}
\label{app:historical_dynamic_fixed}

Tables~\ref{tab:dynamic_fixed_no_verifier} and~\ref{tab:paired_dynamic_fixed_grounded} use the same set of questions and saved Dynamic predictions as Table~\ref{tab:inference_controls}. In Table~\ref{tab:dynamic_fixed_no_verifier}, both policies use integration in call~4. Table~\ref{tab:paired_dynamic_fixed_grounded} compares Dynamic with the Fixed support-assessment policy from Table~\ref{tab:inference_controls}.

\input{tables/dynamic_fixed_no_verifier}

\begin{table}[t]
\centering
\caption{\textbf{Grounded differences for the policies in Table~\ref{tab:inference_controls}.}
$n$ is the number of evaluated questions and $k$ the number of grounded-correct predictions.
$\Delta$ is Dynamic minus Fixed in percentage points.}
\vspace{-5pt}
\label{tab:paired_dynamic_fixed_grounded}
\small
\setlength{\tabcolsep}{2pt}
\renewcommand{\arraystretch}{1.08}
\begin{tabular*}{\columnwidth}{@{\extracolsep{\fill}}lrrr@{}}
\toprule
\textbf{Benchmark} ($n$) & \textbf{Dynamic} ($k$) & \textbf{Fixed} ($k$) & $\boldsymbol{\Delta}$ (\textbf{pp})\\
\midrule
Video-Holmes (1,837) & 303 & 232 & $+3.86$\\
VRBench (495)        & 174 & 155 & $+3.84$\\
CG-Bench (1,812)     & 444 & 433 & $+0.61$\\
\bottomrule
\end{tabular*}
\vspace{-5pt}
\end{table}

\subsection{Answer Preselection and Confirmation Bias}
\label{app:confirmation_bias}

Table~\ref{tab:preselection_controls} isolates the effect of answer preselection by comparing \textbf{Symmetric Bridge} and \textbf{No Preselection}. Both use the same downstream policy, which evaluates all answer options symmetrically; the only difference is whether the planning call proposes an initial answer. On VRBench, adding preselection increases final accuracy from 72.16\% to 77.73\%. \textbf{Original} includes both answer preselection and candidate-conditioned support.

\input{tables/preselection_controls}

\noindent\textbf{Planner-to-final evidence on VRBench.}
For Full + Dynamic, we compare the clips selected by the planner for skill execution (call~1, \texttt{evidence\_ranks}) with the final citations from call~5. On all questions answered correctly at both stages, union mIoU increases by 1.61 percentage points and mean cited duration by 59.09 seconds (Table~\ref{tab:planner-final-vrbench}). Citation IoU-F1 changes only slightly. The final citations therefore cover more of the video and show greater overlap with the annotated evidence union.

\begin{table}[t]
\caption{\textbf{Planner and final evidence on VRBench.}
We compare planner-selected clips with the final citations using the same clip catalog and temporal annotations. Union duration merges overlapping intervals, while citation IoU-F1 uses one-to-one interval matching. Changes are final minus planner, with overlap metrics in percentage points and duration in seconds. The $\pm$ term denotes half the width of the paired 95\% bootstrap interval over videos.}
\vspace{-5pt}
\label{tab:planner-final-vrbench}
\centering
\small
\setlength{\tabcolsep}{3pt}
\begin{tabular*}{\columnwidth}{@{\extracolsep{\fill}}lrrr@{}}
\toprule
\textbf{Metric} & \textbf{Planner} & \textbf{Final} & \boldmath$\Delta$ \\
\midrule
\multicolumn{4}{@{}c}{\textit{Both stages correct}} \\
\midrule
Union mIoU (\%)      & 9.03  & 10.64  & $+1.61 \pm 0.36$ \\
Citation IoU-F1 (\%) & 6.55  & 6.81   & $+0.26 \pm 0.17$ \\
Union duration (s)   & 98.08 & 157.17 & $+59.09 \pm 9.58$ \\
\midrule
\multicolumn{4}{@{}c}{\textit{Both stages correct, with temporal gold}} \\
\midrule
Union mIoU (\%)      & 11.18 & 13.17  & $+2.00 \pm 0.45$ \\
Citation IoU-F1 (\%) & 8.12  & 8.44   & $+0.32 \pm 0.22$ \\
Union duration (s)   & 96.60 & 144.88 & $+48.28 \pm 10.77$ \\
\midrule
\multicolumn{4}{@{}c}{\textit{All questions}} \\
\midrule
Union mIoU (\%)      & 9.50  & 11.56  & $+2.06 \pm 0.33$ \\
Citation IoU-F1 (\%) & 6.90  & 7.23   & $+0.33 \pm 0.19$ \\
Union duration (s)   & 96.39 & 160.02 & $+63.64 \pm 8.37$ \\
\bottomrule
\end{tabular*}
\vspace{-5pt}
\end{table}

%% file: tables/dynamic_fixed_no_verifier.tex
\begin{table}[t]
\caption{\textbf{Dynamic versus Fixed Skills without a verifier.}
Paired predictions use the same full-reward checkpoint, evidence catalogs, five-call inference budget, and 4,096-token output limit.
Scores are reported as percentages, and $\Delta$ is Dynamic minus Fixed in percentage points.
Grounded accuracy follows the overlap criterion defined in Section~\ref{sec:experiments}.}
\label{tab:dynamic_fixed_no_verifier}
\vspace{-5pt}
\centering
\small
\setlength{\tabcolsep}{2pt}
\begin{tabular*}{\columnwidth}{@{\extracolsep{\fill}}llrrr@{}}
\toprule
\textbf{Benchmark} & \textbf{Metric} $\uparrow$
& \textbf{Dynamic} & \textbf{Fixed} & $\boldsymbol{\Delta}$ \\
\midrule
Video-Holmes & Accuracy & 54.06 & 53.84 & $+0.22$ \\
 & Grounded & 16.49 & 13.39 & $+3.10$ \\
\midrule
VRBench & Accuracy & 78.99 & 77.37 & $+1.62$ \\
 & Grounded & 35.15 & 34.34 & $+0.81$ \\
\midrule
CG-Bench & Accuracy & 44.59 & 44.15 & $+0.44$ \\
 & Grounded & 24.50 & 24.72 & $-0.22$ \\
\bottomrule
\end{tabular*}
\vspace{-5pt}
\end{table}

%% file: tables/preselection_controls.tex
\begin{table}[t]
\caption{\textbf{Effect of answer preselection and symmetric support.}
All conditions use the full-reward reader on the same questions. Original uses candidate-conditioned support after preselection. No preselection evaluates all options symmetrically, while Symmetric bridge retains preselection but applies the same downstream policy symmetrically. Accuracy, overlap-based Grounded, and transition rates are percentages. W$\to$R is the percentage of initially incorrect answers that become correct.
R$\to$W is the percentage of initially correct answers that become incorrect.
Dashes indicate that no initial candidate is available.}
\label{tab:preselection_controls}
\vspace{-5pt}
\centering
\small
\setlength{\tabcolsep}{2pt}
\begin{tabular*}{\columnwidth}{@{\extracolsep{\fill}}llrrr@{}}
\toprule
& & & \multicolumn{2}{c@{}}{\textbf{Symmetric support}} \\
\cmidrule(l){4-5}
\makecell[bl]{\textbf{Benchmark}\\$(n)$} & \makecell[bl]{\textbf{Metric}\\(\%)}
& \textbf{Original} & \makecell[bc]{\textbf{No}\\\textbf{preselection}}
& \makecell[bc]{\textbf{Symmetric}\\\textbf{bridge}} \\
\midrule
Video-Holmes & Accuracy & 53.12 & 52.34 & 53.91 \\
$(256)$      & Grounded & 16.41 & 13.28 & 18.36 \\
             & W$\to$R  &  7.32 &   --- &  7.32 \\
             & R$\to$W  &  4.51 &   --- &  3.01 \\
\midrule
VRBench      & Accuracy & 78.14 & 72.16 & 77.73 \\
$(485)$      & Grounded & 34.85 & 33.20 & 36.29 \\
             & W$\to$R  &  5.32 &   --- &  5.26 \\
             & R$\to$W  &  4.35 &   --- &  4.62 \\
\midrule
CG-Bench     & Accuracy & 46.09 & 43.75 & 48.05 \\
$(256)$      & Grounded & 26.56 & 23.44 & 25.39 \\
             & W$\to$R  &  4.51 &   --- &  6.72 \\
             & R$\to$W  &  8.94 &   --- &  6.56 \\
\bottomrule
\end{tabular*}
\vspace{-5pt}
\end{table}

%% file: sections/7-7-Temporal-Grounding.tex
\section{Temporal Grounding and Citation Analysis}
\label{app:grounding_robustness}

\noindent\textbf{Full Dynamic versus Full Generic.}
On VRBench, Full Dynamic improves Grounded accuracy by 4.12 percentage points. Union mIoU is slightly lower and citation IoU-F1 slightly higher, with both confidence intervals including zero. On CG-Bench, the two policies have the same Grounded accuracy, while Dynamic is 1.17 points lower on Strict Grounded, which requires both a correct answer and citation IoU-F1 of at least 0.5.

\begin{table}[t]
\caption{\textbf{Paired differences in temporal grounding.}
Each entry reports Full Dynamic minus Full Generic in percentage points. The $\pm$ term gives half the width of the paired 95\% bootstrap interval clustered by video.}
\label{tab:paired_temporal_grounding}
\vspace{-5pt}
\centering
\small
\setlength{\tabcolsep}{2pt}
\begin{tabular*}{\columnwidth}{@{\extracolsep{\fill}}llr@{}}
\toprule
\textbf{Benchmark} ($n$) & \textbf{Metric} & \textbf{Difference (pp)} \\
\midrule
VRBench (485) & Grounded & $+4.12 \pm 3.18$ \\
 & Union mIoU & $-0.71 \pm 1.17$ \\
 & Citation IoU-F1 & $+0.55 \pm 0.55$ \\
\midrule
CG-Bench (512) & Grounded & $0.00 \pm 3.42$ \\
 & Strict Grounded & $-1.17 \pm 1.08$ \\
\bottomrule
\end{tabular*}
\vspace{-5pt}
\end{table}

\noindent Differences are Full Dynamic minus Full Generic, with paired bootstrap intervals clustered by video. Union mIoU and citation IoU-F1 measure temporal alignment with annotated evidence intervals. EG-VQA additionally evaluates semantic support using predicted and reference evidence descriptions (Appendix~\ref{app:external_grounding}).

Table~\ref{tab:gemini_ours_grounding} compares the saved shared-catalog predictions from Gemini-2.5-Pro and \ours{} (Dynamic Skills, without verifier) used in the main evaluation. Table~\ref{tab:grounding_robustness} compares Free Text, Fixed Skills, and Dynamic Skills on the full evaluation sets: 1,837 Video-Holmes, 495 VRBench, and 1,812 CG-Bench questions. The three policies use the same reader checkpoint and inference budget. The citation-type and duration-bin analyses use the Dynamic Skills predictions from this comparison.

\noindent\textbf{Temporal grounding metrics.}
Let $P$ and $G$ denote the valid predicted and annotated intervals after removing exact duplicates, and let $n_{\mathrm{invalid}}$ be the number of invalid citation IDs.
We compute a maximum-weight one-to-one matching $M$ with interval IoU as the edge weight:
\begin{align}
W
&=
\sum_{(p,g)\in M}
\operatorname{IoU}(p,g),
\end{align}
where $M$ is the matched set of predicted and annotated intervals. We then compute
\begin{align}
\mathrm{IoU\text{-}P}
&=
\frac{W}{|P|+n_{\mathrm{invalid}}},
&
\mathrm{IoU\text{-}R}
&=
\frac{W}{|G|},
\\
\mathrm{IoU\text{-}F1}
&=
\frac{
2\,\mathrm{IoU\text{-}P}\,\mathrm{IoU\text{-}R}
}{
\mathrm{IoU\text{-}P}+\mathrm{IoU\text{-}R}
}.
\end{align}
Invalid citation IDs receive zero matching weight and remain in the IoU-P denominator, while the IoU-R denominator includes only annotated intervals. Undefined precision, recall, and F1 are set to zero, and all metrics are averaged over the full evaluation set.

For Grounded accuracy, invalid IDs are removed and duplicate valid IDs are counted once. A correct answer is considered grounded when at least half of its valid citations overlap an annotated interval for a positive duration. Questions without valid citations or usable annotations receive zero credit.

Union mIoU compares the temporal unions of the predicted and annotated intervals, preserving gaps within each union. For the adapted Acc@GQA metric in Table~\ref{tab:gemini_ours_grounding}, we instead form an enclosing interval $H$ from the earliest valid citation start to the latest valid citation end. A correct answer satisfies the grounding criterion when
\begin{align}
\max_{g\in G}\frac{|H\cap g|}{|H|} \geq 0.5,
\end{align}
where $|\cdot|$ denotes temporal duration and each annotated interval $g\in G$ is compared with $H$ independently.

\begin{table*}[t]
\caption{\textbf{Temporal grounding and citation duration.}
Grounding metrics are percentages; \ours{} denotes Dynamic Skills. Sum and Union are mean citation durations before and after merging overlapping intervals. Video is the mean fraction of video duration covered by the citation union. Questions without valid citations remain in the denominator.}
\vspace{-5pt}
\label{tab:grounding_robustness}
\centering
\small
\setlength{\tabcolsep}{3pt}
\begin{tabular}{@{}llrrrrrrr@{}}
\toprule
Benchmark & Inference policy & IoU-P & IoU-R & IoU-F1 & Union mIoU & Sum (s) & Union (s) & Video (\%) \\
\midrule
Video-Holmes & Free Text & 1.13 & 2.46 & 1.37 & 1.33 & 95.33 & 92.07 & 53.13 \\
 & Fixed Skills & 1.22 & 2.86 & 1.57 & 1.40 & 108.56 & 104.88 & 60.20 \\
 & \ours{} & 1.19 & 2.32 & 1.42 & 1.40 & 93.12 & 90.71 & 52.44 \\
\midrule
VRBench & Free Text & 5.63 & 9.22 & 5.94 & 12.17 & 236.05 & 225.50 & 12.45 \\
 & Fixed Skills & 6.25 & 10.69 & 7.17 & 14.07 & 220.00 & 211.24 & 11.74 \\
 & \ours{} & 7.11 & 8.98 & 7.04 & 11.76 & 171.27 & 164.72 & 9.08 \\
\midrule
CG-Bench & Free Text & 7.39 & 11.75 & 8.38 & 8.80 & 126.86 & 119.33 & 8.77 \\
 & Fixed Skills & 6.59 & 15.45 & 8.74 & 8.97 & 167.59 & 151.38 & 10.67 \\
 & \ours{} & 6.63 & 14.84 & 8.68 & 9.15 & 165.29 & 149.97 & 10.65 \\
\bottomrule
\end{tabular}
\vspace{-5pt}
\end{table*}

\noindent\textbf{Count- and duration-matched random citations.}
We keep each answer fixed and replace its citations with entries from the same video catalog. Citation counts are preserved within duration strata, matching the summed duration of cited entries at the stratum resolution. For each replacement, we sample 64 candidate sets and select the one whose citation-union duration is closest to the original, repeating this 64 times per question. Summed duration counts entries separately, whereas union duration counts overlapping time only once.

\input{tables/citation_null}

On Video-Holmes, the observed and matched-random Grounded scores are similar for both Full and SFT (Table~\ref{tab:citation_null}), differing by only 0.78--1.13 percentage points. The differences are much larger on VRBench (13.98--16.25 points) and CG-Bench (18.04--22.56 points).

This difference partly reflects the catalog structure. On Video-Holmes, long, overlapping entries make random replacements more likely to cover the same evidence, and 49.95--51.62\% reproduce the original citation set. The resulting gap therefore reflects Grounded accuracy beyond the matched-random baseline. Appendix~\ref{app:evidence_dependence} further examines what happens when cited evidence is removed.

\noindent\textbf{Citation duration and evidence type.}
The inference policies also differ in their citation patterns. Fixed Skills tends to cite broader temporal regions, while Dynamic Skills produces shorter citation unions on Video-Holmes and VRBench.

We further separate local and narrative citations by catalog type. Type-specific scores use the full benchmark denominator, with zero assigned when a question contains no citation of that type. We also group questions by total citation-union duration to study how grounding quality changes with citation breadth.

\begin{table}[t]
\caption{\textbf{Local and narrative citation analysis for \ours{}.} $n$ counts questions with at least one valid citation of the indicated type.
Grounding scores (\%) use the full benchmark denominator. Questions may cite both entry types, whose intervals can overlap. VRBench uses local entries by default; the catalog ablation adds narrative entries.}
\vspace{-5pt}
\label{tab:grounding_types}
\centering
\small
\setlength{\tabcolsep}{2pt}
\begin{tabular*}{\columnwidth}{@{\extracolsep{\fill}}llrrrr@{}}
\toprule
\textbf{Benchmark} & \makecell[bl]{\textbf{Citation}\\\textbf{type}}
& $n$
& \makecell[br]{\textbf{Union}\\\textbf{mIoU}}
& \textbf{IoU-F1} & \makecell[br]{\textbf{Union}\\\textbf{(s)}} \\
\midrule
Video-Holmes & Local & 209 & 0.36 & 0.35 & 0.92 \\
 & Narrative & 1787 & 1.38 & 1.26 & 90.47 \\
\midrule
VRBench & Local & 485 & 11.76 & 7.04 & 164.72 \\
\midrule
CG-Bench & Local & 710 & 5.31 & 4.90 & 19.30 \\
 & Narrative & 1650 & 8.95 & 8.60 & 143.23 \\
\bottomrule
\end{tabular*}
\vspace{-5pt}
\end{table}

\begin{table}[t]
\caption{\textbf{Citation-duration stratification for \ours{}.}
IoU-F1 is reported as a percentage. Each question is assigned to one bin
according to its total citation-union duration.}
\label{tab:grounding_duration_bins}
\vspace{-5pt}
\centering
\small
\setlength{\tabcolsep}{2pt}
\begin{tabular*}{\columnwidth}{@{\extracolsep{\fill}}lrrrrrr@{}}
\toprule
Citation union (s) & \multicolumn{2}{c}{Video-Holmes}
                   & \multicolumn{2}{c}{VRBench}
                   & \multicolumn{2}{c}{CG-Bench} \\
\cmidrule(lr){2-3}\cmidrule(lr){4-5}\cmidrule(lr){6-7}
         & $n$ & IoU-F1 & $n$ & IoU-F1 & $n$ & IoU-F1 \\
\midrule
$0$       &   38 & 0.00 &  10 &  0.00 &  106 &  0.00 \\
$(0,30]$  &   39 & 3.00 &   5 & 10.03 &   17 &  9.48 \\
$(30,60]$ &  391 & 1.73 & 179 &  8.28 &  408 & 13.04 \\
$>60$     & 1369 & 1.33 & 301 &  6.49 & 1281 &  8.00 \\
\bottomrule
\end{tabular*}
\vspace{-5pt}
\end{table}

\noindent\textbf{Results and interpretation.}
These analyses separate citation quality from answer correctness. Dynamic Skills improves citation IoU-F1 over Free Text on all three benchmarks, while Fixed Skills achieves slightly higher IoU-F1 with broader citation coverage. Union mIoU shows no consistent ranking across policies.

Across all three benchmarks, mean IoU-F1 is higher when citation unions fall in $(30,60]$ seconds than when they exceed 60 seconds. On Video-Holmes and CG-Bench, narrative entries contribute most of the cited duration.

%% file: tables/citation_null.tex
\begin{table}[t]
\caption{\textbf{Random-citation control.}
Full (full-reward GRPO) and SFT reuse their Dynamic predictions from Table~\ref{tab:training_inference_controls} on the same $n$ questions.
With answers fixed, we compare observed Grounded accuracy (Obs.) with the average over 64 replacement sets per question (Rand.), matched in count and duration.
Union error is the mean absolute relative change in citation-union duration; Same is the fraction of unchanged sets.
All values are percentages except $\Delta$ (Obs.\ minus Rand.), in percentage points.}
\label{tab:citation_null}
\vspace{-5pt}
\centering
\small
\setlength{\tabcolsep}{2pt}
\begin{tabular*}{\columnwidth}{@{\extracolsep{\fill}}llrrrrr@{}}
\toprule
\makecell[bl]{\textbf{Benchmark}\\$(n)$} & \textbf{Reader}
& \textbf{Obs.} & \textbf{Rand.} & $\boldsymbol{\Delta}$
& \makecell[br]{\textbf{Union}\\\textbf{error}} & \textbf{Same} \\
\midrule
Video-Holmes & Full & 16.80 & 15.66 & $+1.13$ & 0.063 & 51.62 \\
$(512)$      & SFT  & 12.11 & 11.33 & $+0.78$ & 0.237 & 49.95 \\
\midrule
VRBench & Full & 34.85 & 20.87 & $+13.98$ & 0.850 & 1.87 \\
$(485)$ & SFT  & 35.05 & 18.80 & $+16.25$ & 0.853 & 12.01 \\
\midrule
CG-Bench & Full & 27.34 & 4.79 & $+22.56$ & 3.279 & 11.42 \\
$(512)$  & SFT  & 22.66 & 4.62 & $+18.04$ & 3.043 & 20.89 \\
\bottomrule
\end{tabular*}
\vspace{-5pt}
\end{table}

%% file: sections/7-5-External-Grounding.tex
\section{Additional Evidence-Grounding Benchmarks}
\label{app:external_grounding}

We further evaluate whether the reader's grounding behavior transfers beyond the three main benchmarks.
These experiments cover different answer formats, grounding annotations, and video domains.
We compare the Base, SFT, and \ours{} readers to examine the effect of post-training, with Gemini-2.5-Pro included as a shared-catalog large-model reference.

\input{tables/addtional_results}

\noindent\textbf{Benchmark coverage and evaluation setup.}
We evaluate on three additional evidence-grounding benchmarks.
NExT-GQA~\citep{xiao2024can} combines multiple-choice video QA with temporal localization in untrimmed videos; EG-VQA~\citep{huang2026eg} evaluates open-ended answers with fine-grained temporal and semantic evidence; and GroundVQA~\citep{di2024grounded} evaluates grounding in long egocentric videos.

We use 5,553 questions from 990 videos on NExT-GQA, 2,889 questions from 336 videos on EG-VQA, and 500 questions from 148 videos under the GroundVQA CloseQA protocol.
Each GroundVQA question is evaluated under five answer-option permutations.
Base, SFT, \ours{}, and Gemini-2.5-Pro use the same frozen catalogs and output budget.
Base, SFT, and \ours{} additionally share the same Dynamic Skills inference procedure and decoding settings, differing only in the reader checkpoint.
Prediction failures remain in the evaluation denominator.

\noindent\textbf{Evaluation metrics.}
We use each benchmark's native metrics rather than the Grounded criterion from the main experiments.

\textbf{NExT-GQA.}
We report answer accuracy, temporal mIoU, and Acc@GQA.
Acc@GQA requires a correct answer and temporal intersection-over-prediction of at least $0.5$, where intersection-over-prediction is the fraction of the predicted interval covered by the annotated evidence.

\textbf{EG-VQA inference.}
The fifth \ours{} call returns a free-form answer and supporting evidence in JSON. The single-call Gemini baseline returns an answer and up to five evidence items, each with a catalog rank and a description. The rank identifies the catalog entry and its corresponding start and end times.

\textbf{EG-VQA scoring.}
Gemini-2.5-Pro judges each predicted answer against the question and reference answer using the official EG-VQA prompt. Relaxed accuracy assigns 1 to fully correct answers, 0.5 to partially correct answers, and 0 otherwise; strict accuracy counts only fully correct answers.

For EG-F1, the official scorer uses BGE-M3 to compare predicted and reference evidence descriptions. A pair is eligible when temporal IoU is at least $0.3$ and semantic similarity is at least $0.5$. Eligible pairs are matched one-to-one with the Hungarian algorithm, and F1 is averaged over questions.

\textbf{GroundVQA.}
Answer accuracy is averaged over five answer-option permutations per question.
For temporal mIoU, cited intervals are merged into their enclosing interval and scored with the benchmark evaluator.
We additionally report Citation IoU-F1 over individual cited intervals.
Let $M$ denote the total IoU under maximum-weight one-to-one matching between predicted and annotated intervals, with $n_p$ predicted and $n_g$ annotated intervals.
We define citation precision and recall as $P=M/n_p$ and $R=M/n_g$, and report their harmonic mean, averaged over questions and option permutations.
Unlike the thresholded EG-VQA evidence F1, this metric retains the continuous IoU of matched intervals.

\noindent\textbf{Results.}
Post-training changes grounding behavior beyond the three main benchmarks, although the effects vary across datasets and metrics.
On NExT-GQA, \ours{} improves temporal grounding and joint grounded-QA performance over SFT, with a small decrease in answer accuracy.
On EG-VQA, \ours{} improves relaxed and strict answer accuracy over SFT by 1.73 and 1.32 percentage points, respectively, and improves EG-F1 by 2.76 points.
On GroundVQA, changes relative to SFT are small and mixed across metrics.

The results also show that SFT and GRPO do not improve every metric monotonically.
GRPO is trained only on Video-Holmes, while these benchmarks use different answer formats and grounding criteria.
The observed transfer is therefore strongest for some forms of answer--evidence alignment rather than for temporal localization or answer accuracy uniformly.

\noindent\textbf{Comparison with Gemini-2.5-Pro.}
The comparison with Gemini-2.5-Pro also depends on the evaluation metric.
Our reader performs better on some grounding measures, while Gemini retains advantages on others.
Because both use the same evidence catalogs and output budget, these differences reflect how the readers use the available evidence under the shared evaluation protocol.

%% file: tables/addtional_results.tex
\begin{table*}[t]
    \caption{\textbf{Answer accuracy and evidence grounding on three additional benchmarks.}
    We compare Base, SFT, \ours{}, and Gemini-2.5-Pro on NExT-GQA, EG-VQA, and GroundVQA under the same evidence catalogs and the same output budget. Base, SFT, and \ours{} additionally share the inference procedure and differ only in the reader checkpoint. \textbf{Takeaway:} Relative to SFT, \ours{} shows stronger grounding on NExT-GQA and improves relaxed accuracy and EG-F1 on EG-VQA, while differences on GroundVQA are small.}
    \vspace{-5pt}
    \label{tab:external_grounding}
    \centering
    \resizebox{\textwidth}{!}{%
    \setlength{\tabcolsep}{4pt}
    \renewcommand{\arraystretch}{0.98}
    \begin{tabular}{@{}l|ccc|ccc|ccc@{}}
    \toprule
    \multicolumn{1}{c|}{}
    &
    \multicolumn{3}{c|}{\textbf{NExT-GQA}}
    &
    \multicolumn{3}{c|}{\textbf{EG-VQA}}
    &
    \multicolumn{3}{c}{\textbf{GroundVQA}}
    \\
    \cmidrule(lr){2-4}
    \cmidrule(lr){5-7}
    \cmidrule(lr){8-10}
    \textbf{Method}
    & \textbf{Accuracy} $\uparrow$
    & \textbf{mIoU} $\uparrow$
    & \textbf{Acc@GQA} $\uparrow$
    & \makecell{\textbf{Relaxed}\\\textbf{Acc.} $\uparrow$}
    & \makecell{\textbf{Strict}\\\textbf{Acc.} $\uparrow$}
    & \textbf{EG-F1} $\uparrow$
    & \textbf{Accuracy} $\uparrow$
    & \textbf{mIoU} $\uparrow$
    & \makecell{\textbf{Citation}\\\textbf{IoU-F1} $\uparrow$}
    \\
    \midrule

    \multicolumn{10}{c}{\textit{Large-model baselines}} \\
    \midrule
    \textsc{Gemini-2.5-Pro}
    & $\bm{67.24 \pm 1.42}$
    & $\bm{25.78 \pm 0.98}$
    & $\bm{11.94 \pm 1.19}$
    & $\bm{47.53 \pm 2.49}$
    & $\bm{35.31 \pm 2.36}$
    & $\bm{42.20 \pm 1.81}$
    & $\bm{51.32 \pm 4.18}$
    & $\bm{4.95 \pm 0.95}$
    & $\bm{6.01 \pm 0.81}$
    \\

    \midrule
    \multicolumn{10}{c}{\textit{\ours{} and ablations (Qwen3.5-9B)}} \\
    \midrule
    \textsc{Qwen3.5-9B}
    & $64.81 \pm 1.35$
    & $24.50 \pm 0.89$
    & $\bm{12.08 \pm 1.08}$
    & $45.50 \pm 2.24$
    & $31.29 \pm 2.25$
    & $38.53 \pm 1.70$
    & $50.20 \pm 3.63$
    & $5.49 \pm 0.99$
    & $6.03 \pm 0.86$
    \\

    \textsc{\quad + SFT}
    & $\bm{66.50 \pm 1.35}$
    & $21.16 \pm 0.80$
    & $9.51 \pm 0.99$
    & $46.47 \pm 2.26$
    & $32.88 \pm 2.24$
    & $38.78 \pm 1.71$
    & $52.56 \pm 3.64$
    & $\bm{5.58 \pm 1.08}$
    & $6.11 \pm 0.99$
    \\

    \midrule
    \textbf{\textsc{\ours{} (full)}}
    & $65.48 \pm 1.37$
    & $\bm{25.95 \pm 0.93}$
    & $11.63 \pm 1.16$
    & $\bm{48.20 \pm 2.27}$
    & $\bm{34.20 \pm 2.28}$
    & $\bm{41.54 \pm 1.81}$
    & $\bm{53.24 \pm 3.32}$
    & $5.22 \pm 1.00$
    & $\bm{6.43 \pm 0.97}$
    \\
    \bottomrule
    \end{tabular}%
    }
    \vspace{-10pt}
\end{table*}

%% file: sections/7-6-Evidence-Dependence.tex
\section{Evidence Dependence under Catalog Interventions}
\label{app:evidence_dependence}

We test how predictions change when cited catalog evidence is retained or removed~\citep{deyoung2020eraser}. The analysis uses the full evaluation sets: 1,837 Video-Holmes, 495 VRBench, and 1,812 CG-Bench questions. Full-catalog predictions are taken from the saved five-call Dynamic Skills runs in Table~\ref{tab:inference_controls}, corresponding to the \ours{} (full) results in Table~\ref{tab:main_results}.

\noindent\textbf{Catalog interventions.}
\textbf{Cited-only} keeps only the entries cited in the full-catalog answer, while \textbf{Remove-cited} removes them. \textbf{Matched random removal} removes the same number of entries, selecting from 256 seeded random subsets the one closest to the cited set in union duration, total duration, and text length.
After each intervention, the remaining catalog is renumbered and all five reader calls are rerun with the same checkpoint, decoding settings, and output budget. All conditions are evaluated on the same question set.

\begin{table}[t]
\centering
\caption{\textbf{Evidence dependence.} Answer accuracy (\%) on the full evaluation sets.}
\vspace{-5pt}
\label{tab:evidence_dependence}
\small
\setlength{\tabcolsep}{2pt}
\begin{tabular*}{\columnwidth}{@{\extracolsep{\fill}}lrrr@{}}
\toprule
\textbf{Catalog} & \textbf{Video-Holmes} & \textbf{VRBench} & \textbf{CG-Bench} \\
\midrule
Full catalog           & 54.06 & 78.99 & 44.59 \\
\midrule
Cited-only             & 52.64 & 79.39 & 45.81 \\
Remove-cited           & 50.35 & 75.15 & 37.36 \\
Matched random removal & 52.53 & 77.78 & 42.94 \\
\bottomrule
\end{tabular*}
\vspace{-5pt}
\end{table}

\noindent\textbf{Results and interpretation.}
Keeping only the cited entries preserves most of the full-catalog accuracy, while removing them causes a larger drop than removing a matched amount of other catalog content. These interventions suggest that the cited entries are behaviorally relevant to the reader's predictions. The remaining gap may reflect overlapping information across catalog entries, which can provide alternative evidence after the catalog is modified.

%% file: sections/7-9-Prompt-Templates.tex
\section{Prompt Templates}
\label{app:prompt_templates}

\subsection{Five-Call Reader Protocol for Multiple-Choice QA}
\label{app:reader_stage_protocol}

Table~\ref{tab:reader_protocol_scope} distinguishes training targets from inference outputs. The five calls below apply to multiple-choice QA.

\begin{table}[!t]
\centering
\caption{\textbf{Reader training targets and inference outputs.}
Training conditions on the question, answer options, and evidence catalog. Gold answers and annotated evidence spans are used for supervision and evaluation as described in Appendix~\ref{app:training_objective}.}
\vspace{-5pt}
\label{tab:reader_protocol_scope}
\footnotesize
\setlength{\tabcolsep}{4pt}
\begin{tabularx}{\columnwidth}{@{}>{\raggedright\arraybackslash}p{0.23\columnwidth}>{\raggedright\arraybackslash}X@{}}
\toprule
\textbf{Setting} & \textbf{Target or output} \\
\midrule
\multicolumn{2}{@{}l}{\textit{Training}} \\
SFT &
Teacher completion with rationale, catalog citations, and answer. Loss applies to the completion. \\
\addlinespace
GRPO &
Responses sampled from the SFT-initialized reader, updated using within-question relative rewards. \\
\addlinespace
\multicolumn{2}{@{}l}{\textit{Inference}} \\
Multiple-choice QA &
Proposed \texttt{hypothesis\_label}, then final \texttt{label} and \texttt{reasoning} with clip citations. \\
\addlinespace
Open-ended EG-VQA &
Free-form answer with evidence. All three readers (Base, SFT, full-reward) share the Dynamic protocol and budget (Appendix~\ref{app:external_grounding}). \\
\bottomrule
\end{tabularx}
\vspace{-5pt}
\end{table}

\begin{table}[!t]
\centering
\caption{\textbf{From a skill call to its logged record.}
Each completed skill call is stored as a normalized record $(o_i,\rho_i,u_i,d_i)$ in \texttt{skill\_result}. Planner and skill fields may take the empty values listed below.}
\vspace{-5pt}
\label{tab:reader_trace_fields}
\footnotesize
\setlength{\tabcolsep}{3pt}
\begin{tabularx}{\columnwidth}{@{}>{\raggedright\arraybackslash}p{0.25\columnwidth}>{\raggedright\arraybackslash}X@{}}
\toprule
\textbf{Record component} & \textbf{Fields and interpretation} \\
\midrule
Reader response &
\texttt{raw}, \texttt{model\_json}\enspace
Original response and parsed skill JSON. Localization may return empty \texttt{best\_refs}, and intention reasoning may return empty \texttt{alternatives}. \\
\addlinespace
Input binding &
\texttt{skill\_args}\enspace
Valid planner ranks map to clip IDs in \texttt{evidence\_refs}, \texttt{evidence\_chain.\allowbreak evidence\_refs}, or \texttt{support\_evidence}. Empty \texttt{evidence\_ranks} yield an empty list. Unused strings and \texttt{actions} may also be empty. \\
\midrule
Finding $o_i$ and references $\rho_i$ &
\texttt{outputs}, \texttt{evidence\_refs}\enspace
The normalized finding and the references the skill returns. A skill can receive evidence and return none. Localization references carry forward. \\
\addlinespace
Score and status $u_i$ &
\texttt{confidence}, \texttt{status}\enspace
Normalized confidence and execution status. The skill-specific rule sets \texttt{verified} or \texttt{failed}. For example, \texttt{causal=false} yields a failed causal result. \\
\addlinespace
Diagnostic $d_i$ &
\texttt{failure\_code}\enspace
Optional executor diagnostic. A failed result may have a \texttt{null} failure code. \\
\bottomrule
\end{tabularx}
\vspace{-5pt}
\end{table}

\begin{table}[!t]
\caption{\textbf{Representative prompts for evidence-catalog construction.}
The local-describer excerpt retains the original rule numbers. The narrative prompt is reproduced in full.}
\vspace{-5pt}
\label{tab:catalog_prompt_templates}
\centering
\footnotesize
\begin{tabularx}{\columnwidth}{@{}>{\raggedright\arraybackslash}X@{}}
\toprule
\textbf{Local describer}\par
{\raggedright
You convert one video clip span into a structured perception record.\par
1. Use only information supported by the clip frames or provided subtitle/context text.\newline
2. Do not invent characters, objects, or events.\newline
3. Prefer clue-oriented noun phrases over generic captions: objects, colors, place, repeated-looking props, screen text, spoken clues, temporal changes.\newline
5. For social questions, record only visible social cues: facial expression, gaze direction, hesitation-like motion, distance, posture, gesture, or group interaction. Do not infer private motives unless the visual evidence is clear.\newline
7. If nothing is visible, return empty lists and a cautious \texttt{scene\_description}.} \\
\midrule
\textbf{Narrative describer}\par
{\raggedright
You are annotating one stretch of a short film for a detective-style question set. You are given evenly spaced frames from that stretch in time order, the dialogue heard in it (if any), the previous stretch's annotation and the running cast. Write ONE paragraph, the way a careful human annotator would, of what happens: who is present (stable descriptors reused from the cast, e.g. 'the man with the backpack'), what each person does and says, how they react, what changes, and what the sequence shows or implies about intentions, relationships and cause-and-effect when the frames and dialogue make it clear. Quote or paraphrase dialogue in English. Describe events, not camera work. Do not invent what is not shown or heard. Reply with JSON only:\newline
{\ttfamily\footnotesize \{"narrative": "\textless{}paragraph\textgreater{}", "cast": ["\textless{}stable descriptor\textgreater{}", ...]\}.}} \\
\bottomrule
\end{tabularx}
\vspace{-10pt}
\end{table}

\begin{table}[!t]
\caption{\textbf{Representative prompts for evidence assessment and answer generation.}}
\label{tab:representative_prompts}
\centering
\footnotesize
\begin{tabularx}{\columnwidth}{@{}>{\raggedright\arraybackslash}X@{}}
\toprule
\textbf{Evidence support}\par
{\raggedright
How well does the evidence support this hypothesis?\newline
Hypothesis: '\texttt{\{hypothesis\}}'\newline
Supporting evidence: \texttt{\{support\}}\newline
Counterevidence: \texttt{\{counter\}}\newline
Answer with JSON:\newline
{\ttfamily\footnotesize \{\{"support\_\allowbreak score": 0.0-1.0, "contradiction\_\allowbreak score": 0.0-1.0, "reasoning": "..."\}\}}} \\
\midrule
\textbf{Final answer}\par
{\raggedright
Answer the original multiple-choice question using the catalog and the accumulated notes. Notes are fallible; use the evidence to decide. Return only JSON\newline
{\ttfamily\footnotesize \{"reasoning":"reason step by step, citing evidence as clip 1, clip 2, etc.","label":"option letter"\}.}} \\
\bottomrule
\end{tabularx}
\vspace{-10pt}
\end{table}

\begingroup\small
Dynamic and Generic use the same checkpoint, question, options, catalog, decoding settings, final prompt, and parser. Calls~1--4 request at most 350 tokens each (512-token cap). Call~5 allows 2,048 tokens, giving a total cap of 4,096. Usage sums all five outputs. Decoding uses temperature 0 and seed 7 with native thinking disabled.

\noindent\textbf{Dynamic planning (call 1).}
The reader proposes an option, selects two ordered skills (repeats allowed), and identifies up to eight catalog entries. It returns \texttt{skills}, \texttt{hypothesis\_label}, \texttt{evidence\_ranks}, \texttt{cause}, \texttt{effect}, \texttt{agent}, \texttt{actions}, \texttt{state\_entity}, \texttt{state\_predicate}, and \texttt{reasoning}. Unused arguments are empty. Later calls bind valid ranks to catalog entries. Reduced-pool controls vary only the available skills.

\noindent\textbf{Dynamic skill applications (calls 2 and 3).}
The same reader executes each selected skill, including repeats, in a separate call. Skill selection and records are produced at inference time and have no separate training targets. Call~3 receives the full call~2 output. Valid localization references carry forward. Each skill returns the following fields:
\begin{itemize}\raggedright\setlength{\itemsep}{0pt}\setlength{\parskip}{0pt}
\item Localization: \texttt{best\_refs}, \texttt{confidence}, \texttt{reasoning}.
\item Hypothesis assessment: \texttt{support\_score}, \texttt{contradiction\_score}, \texttt{reasoning}.
\item Temporal: \texttt{relation} (\texttt{before}, \texttt{after}, \texttt{overlap}, or \texttt{simultaneous}), \texttt{confidence}, \texttt{reasoning}.
\item Causal: \texttt{causal}, \texttt{confidence}, \texttt{reasoning}.
\item Intention: \texttt{intention}, \texttt{confidence}, \texttt{alternatives}.
\item State change: \texttt{changed}, \texttt{before\_state}, \texttt{after\_state}, \texttt{confidence}.
\end{itemize}
Scores and confidences lie in $[0,1]$. Both \texttt{causal} and \texttt{changed} are Boolean.

\noindent\textbf{Dynamic integration (call 4).}
The reader combines the plan, both skill outputs, and the catalog into \texttt{\{"reasoning":"..."\}}. The historical variant uses \texttt{verify\_claim\_support} in this slot and returns \texttt{supported}, \texttt{target\_aligned}, a score in $[0,1]$, and \texttt{reasoning}.

\noindent\textbf{Generic (calls 1--4).}
The four prompts ask the reader to plan, gather evidence, compare options, and reconsider its answer. Each returns \texttt{evidence\_ranks} and \texttt{reasoning}. The historical Free Text control requests reasoning-only JSON.

\noindent\textbf{Shared final answer (call 5).}
Given the original question context, full catalog, and all intermediate notes, the reader returns \texttt{reasoning} and \texttt{label}. It may revise the planner's answer. Citations map catalog ranks to source intervals (Table~\ref{tab:representative_prompts}).

\par\endgroup

\subsection{Raw JSON and Normalized Skill Records}
\label{app:skill_record_fields}

Table~\ref{tab:reader_trace_fields} maps skill responses to logged records. In Figure~\ref{fig:showcase-game}, localization passes \texttt{clip:9} to the causal skill, which returns empty \texttt{evidence\_refs}. The final answer cites clips 9 and 48.

\subsection{Representative Prompt Text}

Tables~\ref{tab:catalog_prompt_templates} and~\ref{tab:representative_prompts} give the catalog and reader prompts. Braced placeholders are filled at runtime.

%% file: sections/7-10-Qualitative-Examples.tex
\section{Qualitative Atomic-Skill Traces}
\label{app:qualitative_atomic_skills}

Figures~\ref{fig:showcase-coffee}--\ref{fig:showcase-tires} show six successful cases from CG-Bench, Video-Holmes, and VRBench, covering evidence localization, causal and intention reasoning, and support assessment. In Figure~\ref{fig:showcase-stickers}, the cited entries do not establish the sticker count, and the final response reaches its length limit.
In Figure~\ref{fig:showcase-extortion}, the final answer retains an unsupported hypothesis.
Figure~\ref{fig:showcase-citation-gap} shows a correct answer with incomplete citation coverage that still satisfies the overlap-based Grounded criterion. These examples use the historical verifier variant, with \texttt{verify\_claim\_support} in the fourth reader call instead of the integration step used in the default Dynamic pipeline.

\input{showcase/showcase_coffee}
\clearpage
\input{showcase/showcase_game}
\clearpage
\input{showcase/showcase_kate}
\clearpage
\input{showcase/showcase_phone}
\clearpage
\input{showcase/showcase_choir}
\clearpage
\input{showcase/showcase_tires}
\clearpage
\input{showcase/showcase_stickers}
\clearpage
\input{showcase/showcase_extortion}
\clearpage
\input{showcase/showcase_citation_gap_entry}

%% file: showcase/showcase_coffee.tex
\begin{figure*}[!htp]
\ASfont
\centering
\begin{AScase}{CG-Bench: Is the coffee shop open?}
\begin{ASbody}
\ASfield{Question:} \textbf{In the video, what happened when the protagonist came
to the ``BEAR POND'' store?}

\ASbar{Selected video frames}
\noindent
\ASframe{.24}{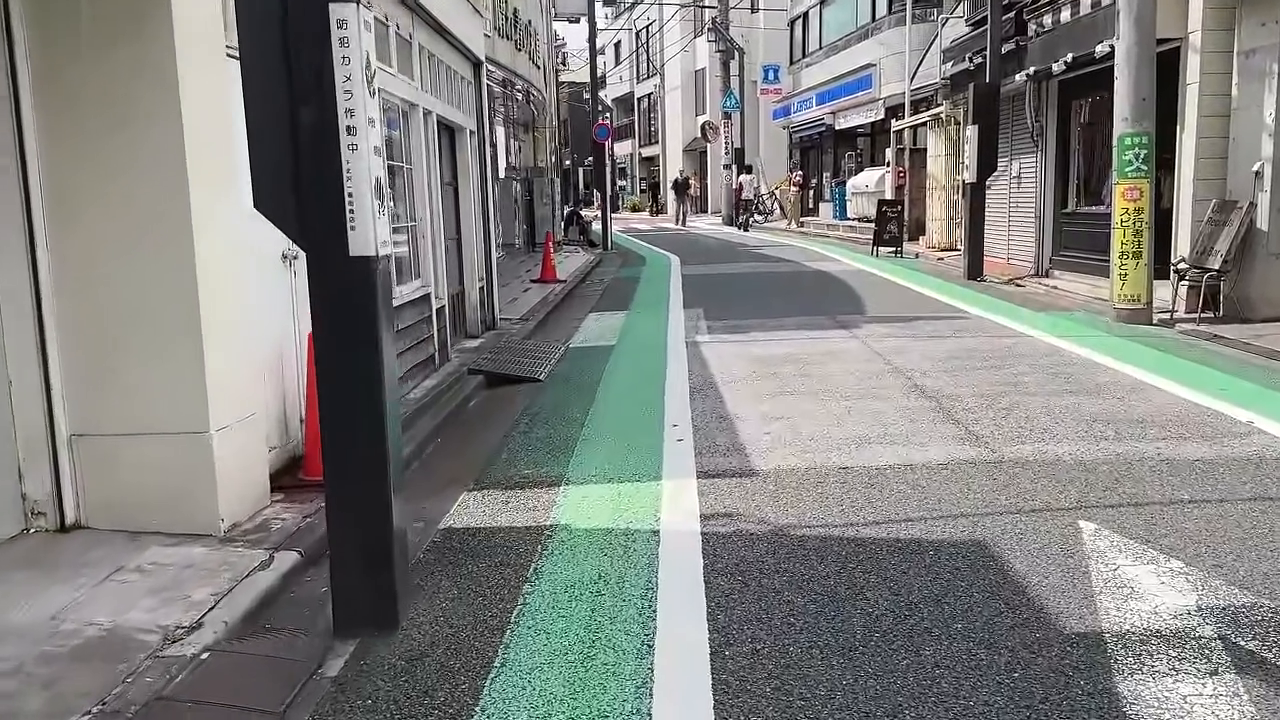}{C32 / 1808.0 s}{Approaching the coffee shop.}\hfill
\ASframe{.24}{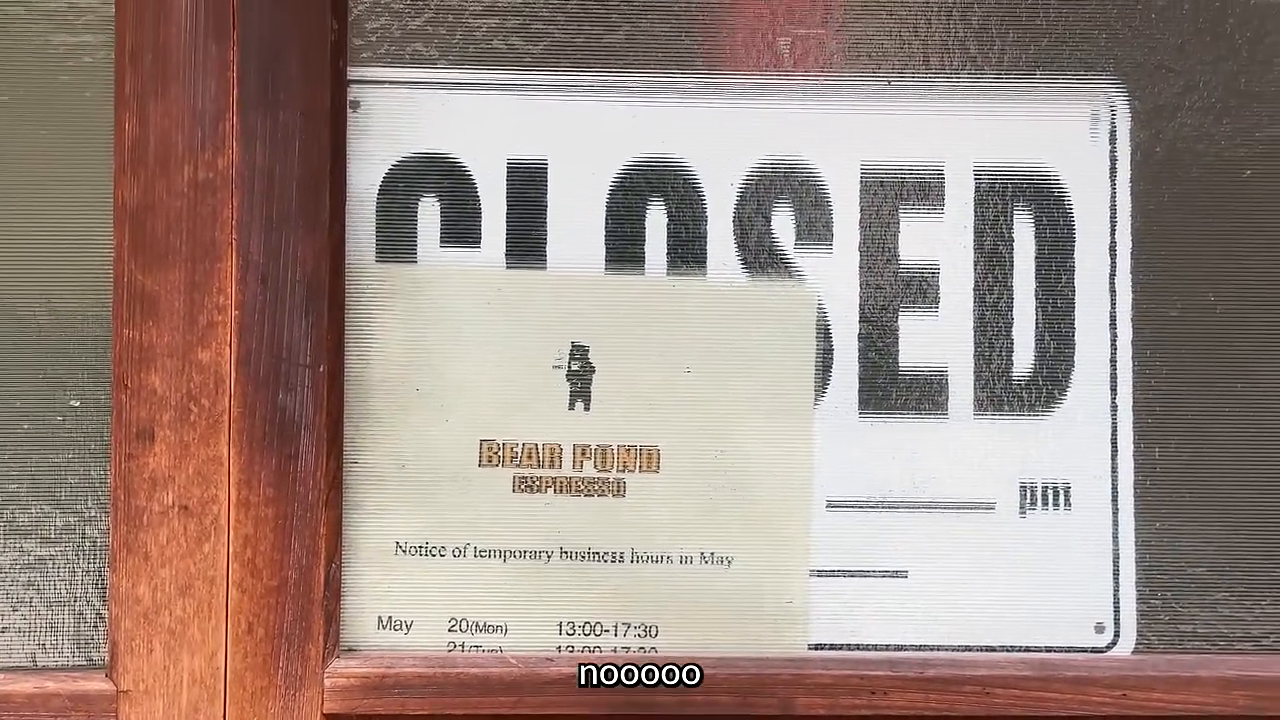}{C32 / 1814.0 s}{The door displays a CLOSED sign.}\hfill
\ASframe{.24}{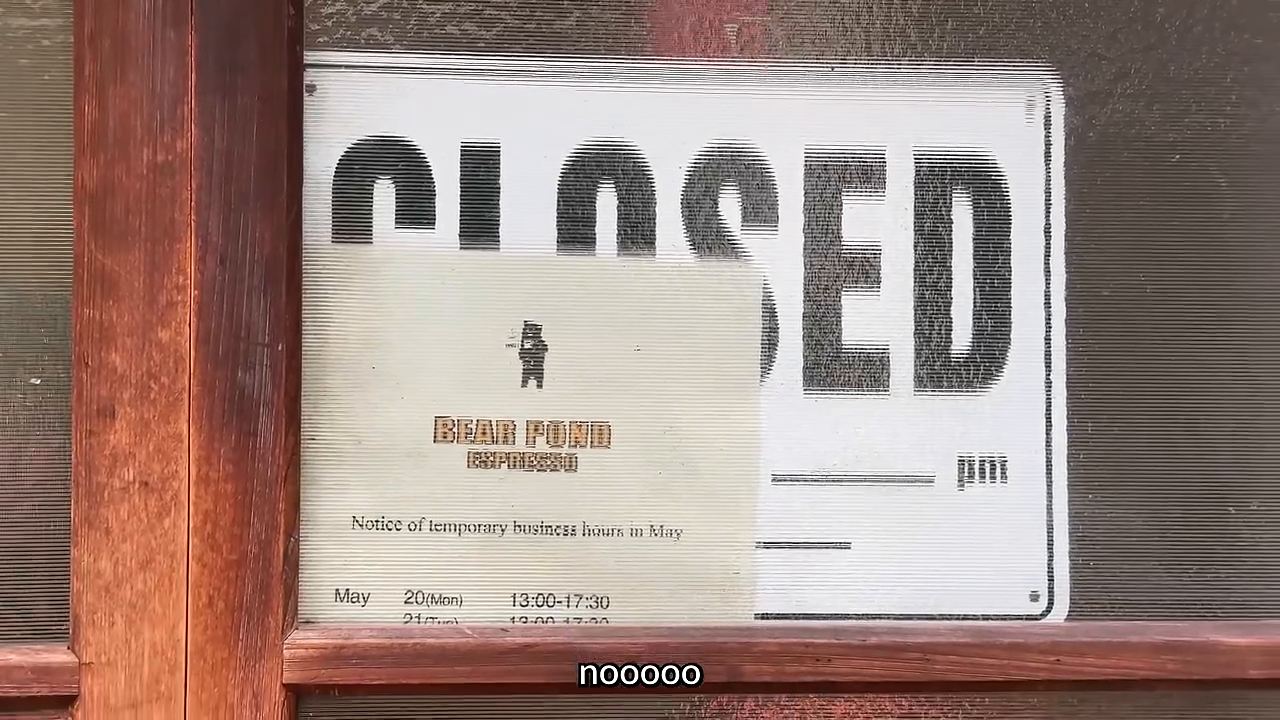}{C105 / 1816.0 s}{The sign and the shop's notice.}\hfill
\ASframe{.24}{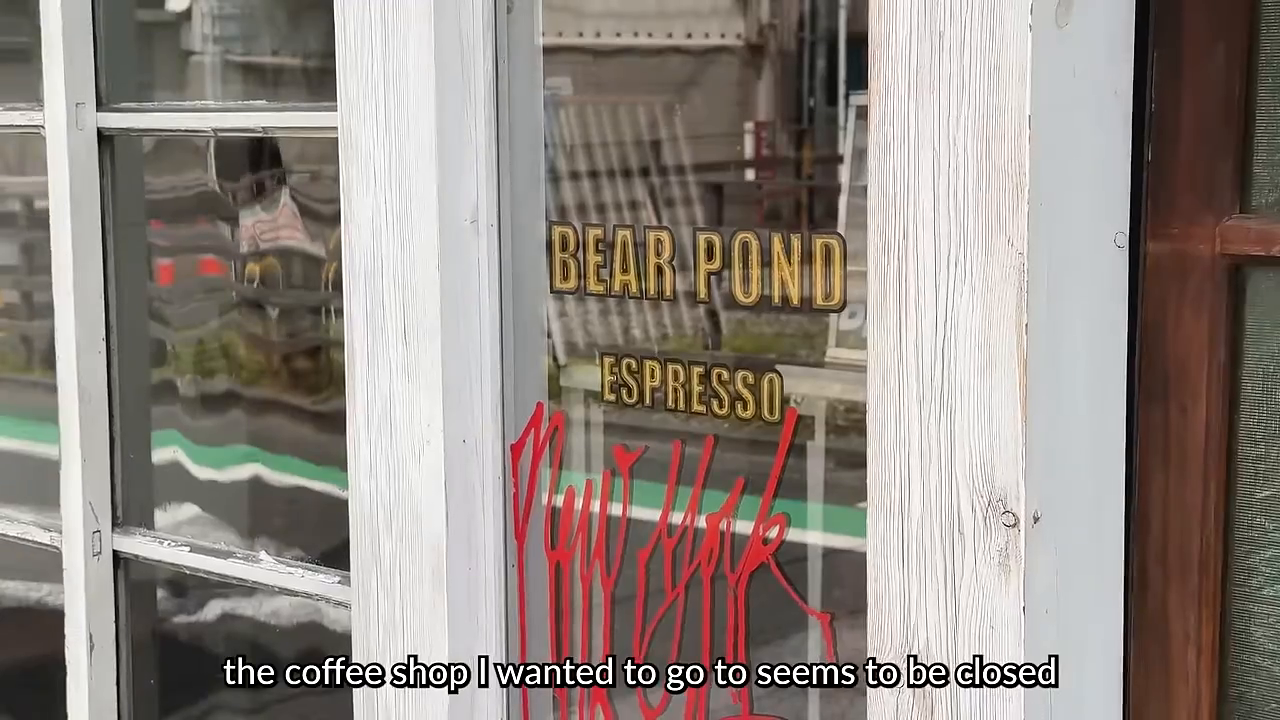}{C105 / 1818.0 s}{The subtitle says the shop is closed.}
\par

\ASbar{Evidence from the catalog}
\begin{minipage}[t]{0.48\linewidth}
\raggedright
\ASfield{C32 / [1792, 1850] s}\par
The narrator arrives at Bear Pond Espresso, finds it closed, and goes
to another coffee shop.
\end{minipage}\hfill
\begin{minipage}[t]{0.48\linewidth}
\raggedright
\ASfield{C105 / [1792, 1822] s}\par
The entry describes the shop name, a ``CLOSED'' sign on the door,
and a subtitle saying that the shop seems closed.
\end{minipage}

\ASbar{Planner-selected skills}
\ASfield{Plan:} \AScode{localize_clue} $\rightarrow$
\AScode{score_hypothesis_support}\par

\ASfield{Why these skills:}
\textbf{Localization} finds the entry for this particular visit.
\textbf{Support assessment} then evaluates whether it establishes that the shop
is closed. This matches the question's request for an observed state.

\ASfield{Candidate:} F. The shop is not open.\quad
\ASfield{Starting references:} \ASrefs{C32, C105}.

\ASfield{1. Localize the clue.}
Given \ASrefs{C32} and \ASrefs{C105}, the localizer selects only
\ASrefs{C32}, with confidence \textbf{0.95}
and status \ASgood{assessed as supported}.

\ASfield{2. Score the hypothesis.}
Using \ASrefs{C32} and the localizer's result, the scorer assigns
``The shop is not open'' a support score of \textbf{1.00} and a
contradiction score of \textbf{0.00}. Its status is \ASgood{assessed as supported}.

\ASbar{Support assessment and final answer}
\ASfield{3. Support-assessment output:}
\AScode{verify_claim_support} checks \ASrefs{C32} and the scoring result.
It returns \ASgood{assessed as supported}, score \textbf{1.00},
\AScode{passed=true}, and \AScode{supported_by_refs} = [\ASrefs{C32}].

\ASfield{4. Final answer:}
\ASgood{F. The shop is not open. [C32, C105]}\par
\ASfield{Gold answer:} F.\quad
\ASfield{Logged result:} \AScode{ok=true}, \AScode{grounded=true}.

\ASfield{Discussion:}
\textbf{Localization changes what the next calls receive.}
The reference set shrinks from \ASrefs{C32, C105} to \ASrefs{C32}, and
both scoring and support assessment use that selected entry. The resulting
support score is tied to a specific description of the closure.
The final-answer call still sees the full catalog and cites both entries.
Recording these stages separately lets us trace which evidence was used
to check the answer and which evidence appeared in the final response.
\end{ASbody}
\end{AScase}
\caption{\textbf{CG-Bench: Is the coffee shop open?} Localization narrows the evidence passed to both downstream checks. The intermediate trace distinguishes those inputs from the citations in the final answer.}
\label{fig:showcase-coffee}
\end{figure*}

%% file: showcase/showcase_game.tex
\begin{figure*}[!htp]
\ASfont
\centering
\begin{AScase}{CG-Bench: Why is the player stuck?}
\begin{ASbody}
\ASfield{Question:} \textbf{Why couldn't the player move the next day in the video?}

\ASbar{Selected video frames}
\noindent
\ASframe{.24}{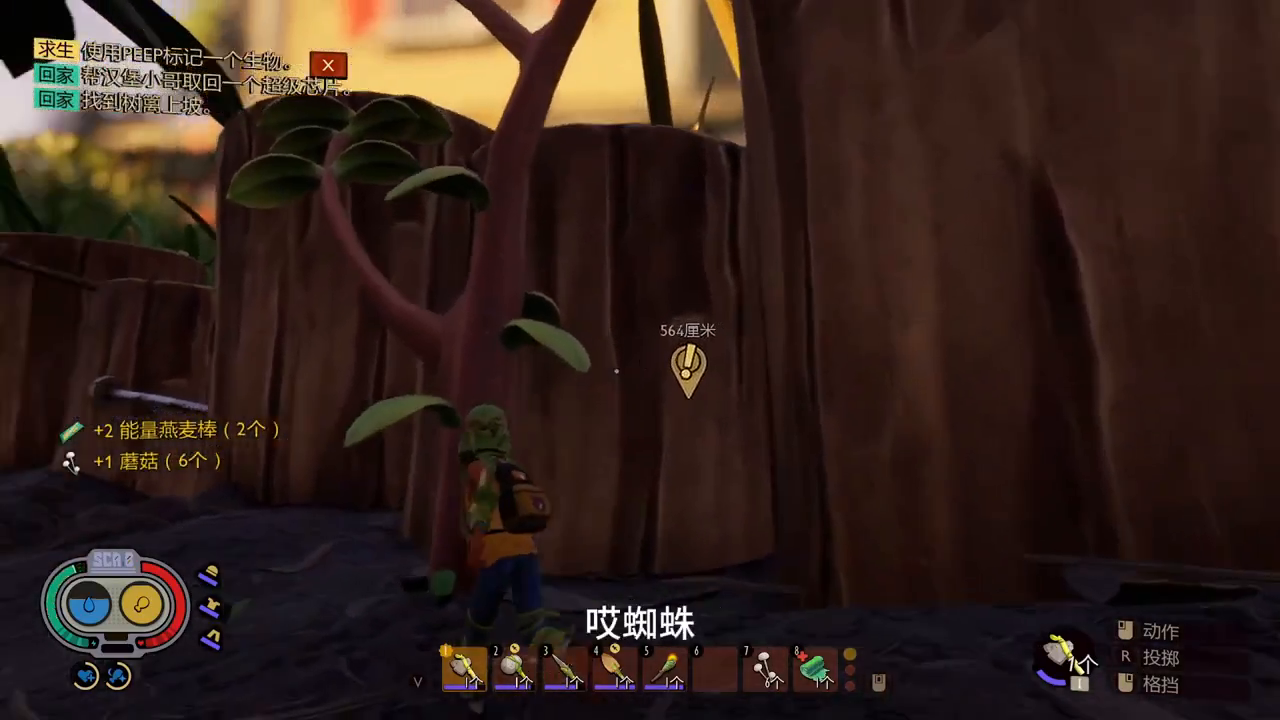}{C9 / 516.0 s}{Entering the narrow passage.}\hfill
\ASframe{.24}{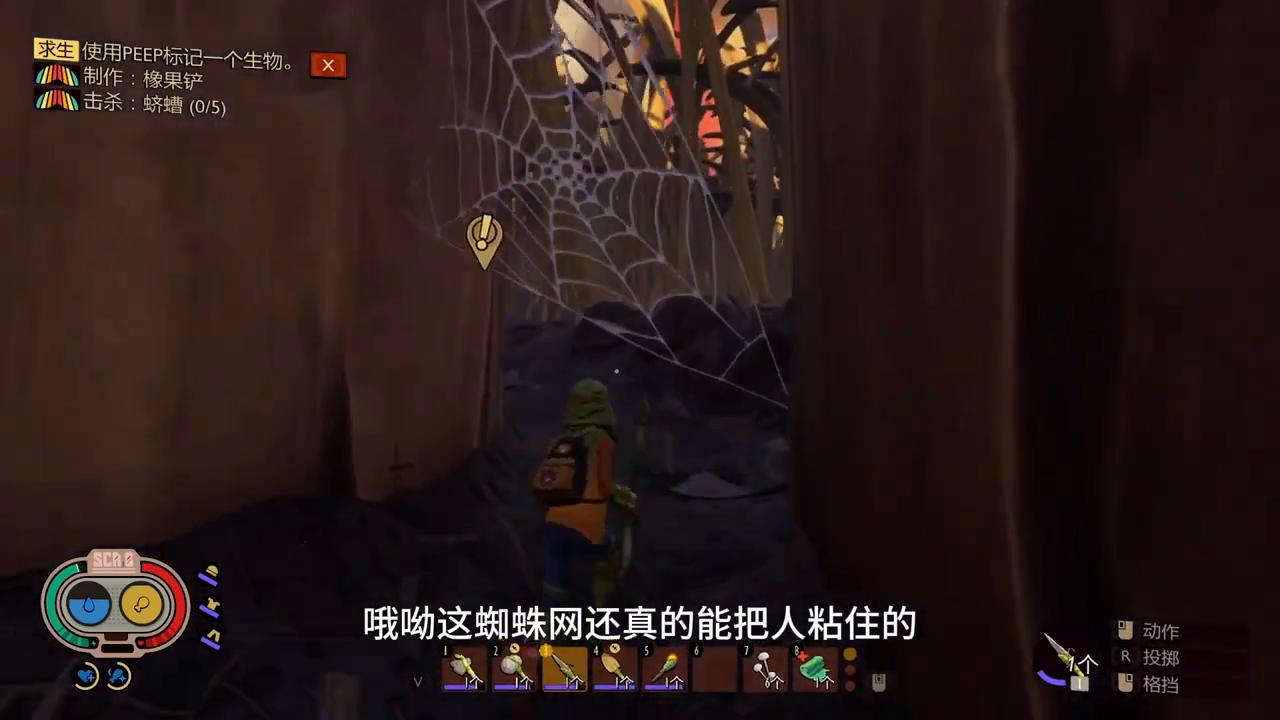}{C9 / 524.0 s}{The subtitle says the web can trap people.}\hfill
\ASframe{.24}{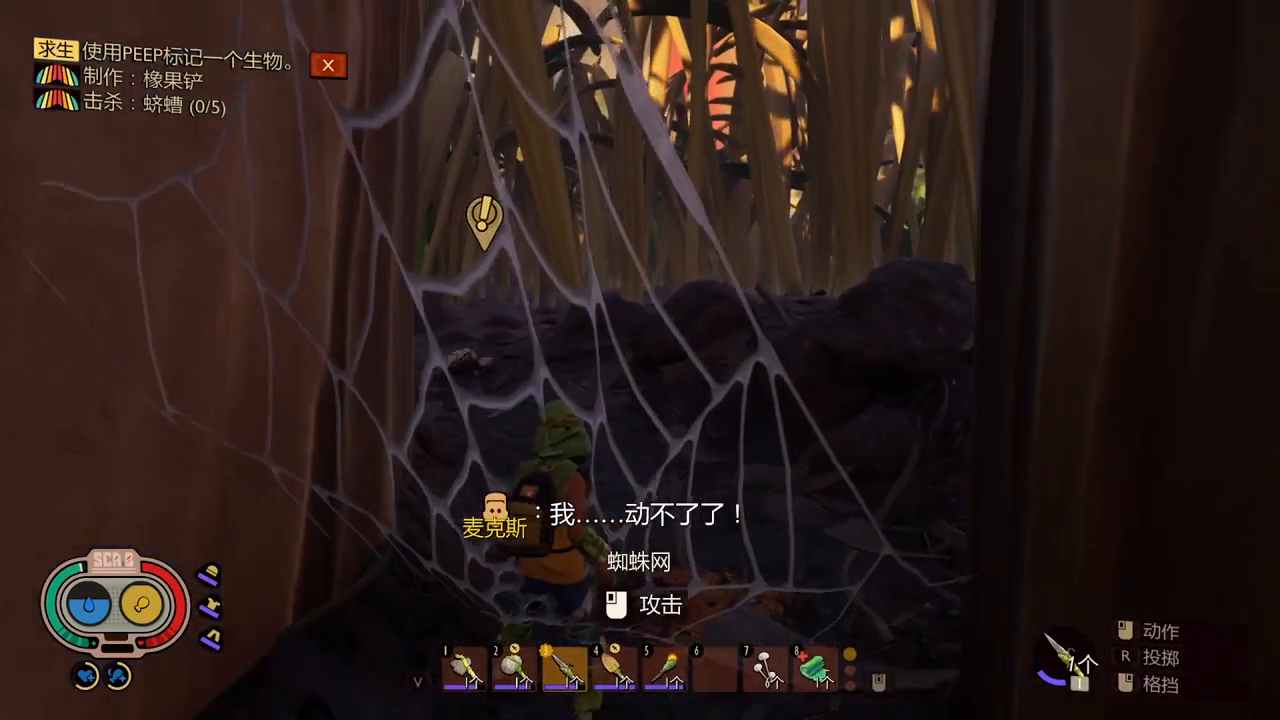}{C48 / 526.0 s}{The player says he cannot move.}\hfill
\ASframe{.24}{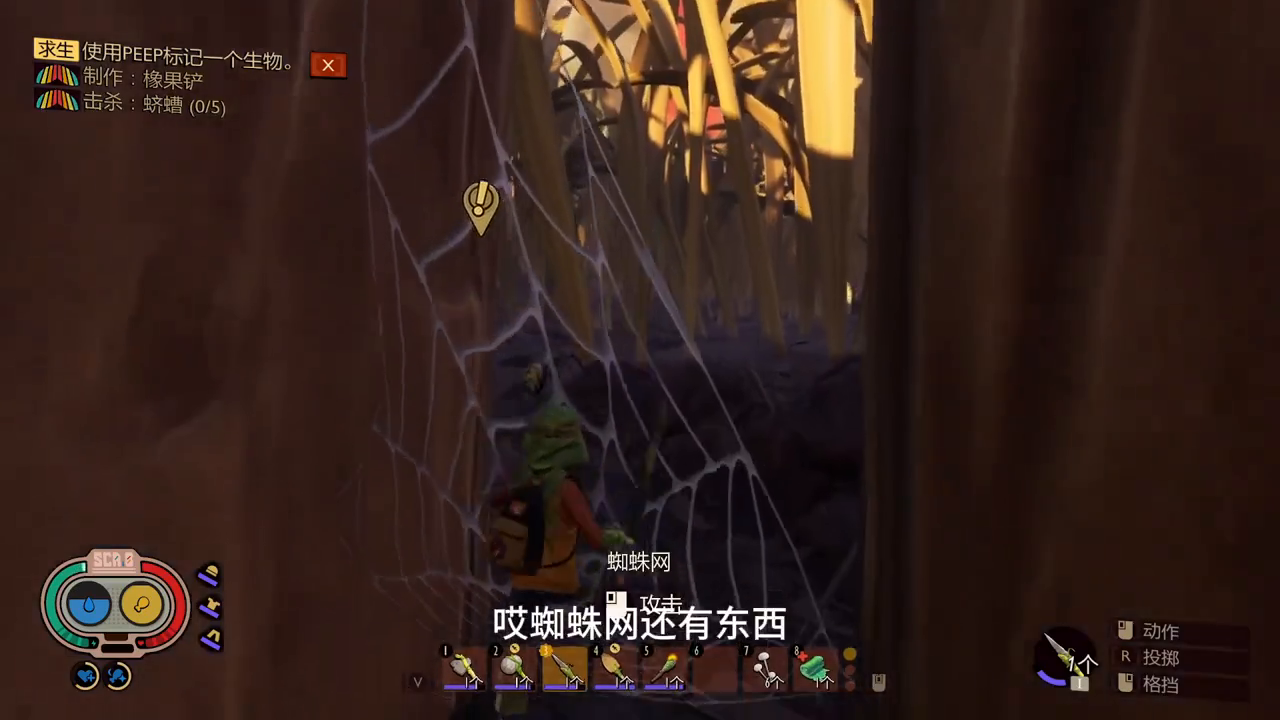}{C48 / 527.0 s}{A web spans the passage.}
\par

\ASbar{Evidence from the catalog}
\textbf{C9 [476, 534] s} describes Max entering a narrow passage with
a spiderweb and saying that he cannot move. \textbf{C48 [504, 534] s}
also describes the web across the passage and a subtitle mentioning it.

\ASbar{Planner-selected skills}
\ASfield{Plan:} \AScode{localize_clue} $\rightarrow$
\AScode{infer_causal_relation}

\ASfield{Why these skills:}
The question asks for a cause. \textbf{Localization} selects
the encounter in which the player becomes stuck; \textbf{causal inference}
connects the web to his inability to move. Finding the scene and explaining
the event are separate tasks.

\ASfield{Skill outputs:}
The planner proposes \textbf{B} using \ASrefs{C9, C48}.
The localizer keeps \ASrefs{C9} (confidence \textbf{0.95}).
The causal skill then uses \ASrefs{C9} and the localizer's result to
infer that the web prevents the player from moving (confidence
\textbf{0.98}). Both skills return \ASgood{assessed as supported}.

\ASbar{Support assessment and final answer}
\ASfield{Support-assessment output:} \ASgood{assessed as supported}; score \textbf{1.00};
supported references: \ASrefs{C9}.

\ASfield{Final answer:} \ASgood{B. Caught in a spider's web. [C9, C48]}\par
\ASfield{Gold answer:} B.\quad \ASfield{Logged grounding:} \AScode{true}.

\ASfield{Discussion:}
\textbf{The same localization skill supports a different next step.}
For the shop question, the planner checks a state claim; here it selects
causal inference to explain a physical consequence. \ASrefs{C9} contains
both the web encounter and the player's complaint. Passing this entry
and the localizer's result to the next call ties the explanation to the
selected event. The causal output has empty \AScode{evidence_refs}, but
the input and support-assessment output still identify \ASrefs{C9}, so the evidence used
for the causal judgment remains traceable.
\end{ASbody}
\end{AScase}
\caption{\textbf{CG-Bench: Why is the player stuck?} The planner reuses localization, then selects causal inference to explain why the player is stuck. The selected event supplies both the obstacle and its effect.}
\label{fig:showcase-game}
\end{figure*}

%% file: showcase/showcase_kate.tex
\begin{figure*}[!htp]
\ASfont
\centering
\begin{AScase}{Video-Holmes: How was Kate's location exposed?}
\begin{ASbody}
\ASfield{Question:} \textbf{What is the direct reason for Kate's position being
ultimately exposed?}

\ASbar{Selected video frames}
\noindent
\ASframe{.24}{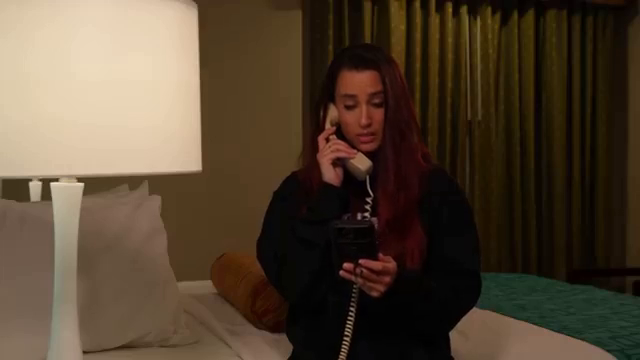}{C3 / 84.0 s}{Kate holds a receiver and a mobile phone.}\hfill
\ASframe{.24}{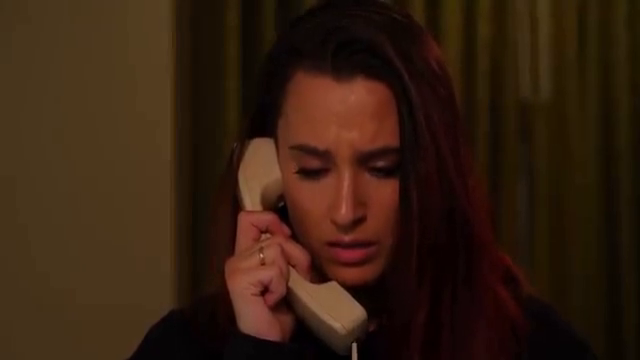}{C37 / 93.0 s}{Close-up during the phone call.}\hfill
\ASframe{.24}{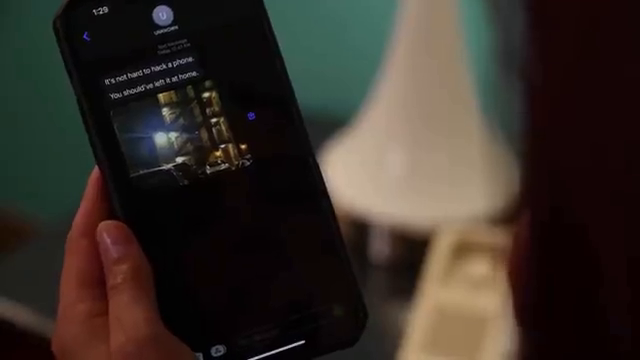}{\textcolor{AScontext}{Supplement / 95.0 s}}{A message says phones are easy to hack.}\hfill
\ASframe{.24}{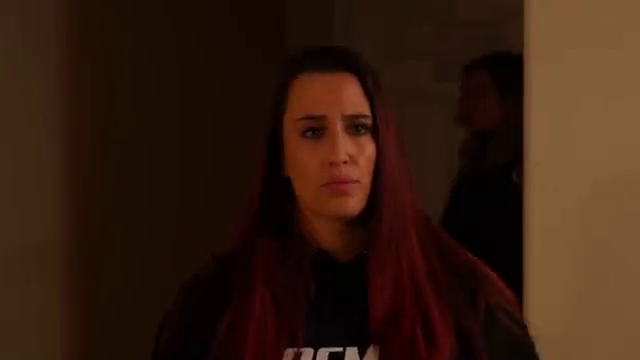}{\textcolor{AScontext}{Supplement / 145.0 s}}{A visitor appears at the doorway.}
\par

{\footnotesize\textcolor{AScontext}{\textbf{Supplementary context:}}
The 95 s and 145 s frames are outside C3 and C37, which both end
at 94 s. Neither frame belongs to the intervals cited in the original run.}\par

\ASbar{Evidence from the catalog}
\textbf{C3 [63, 94] s} and \textbf{C37 [90, 94] s} describe a
threatening message: a phone is easy to hack and should have been left
at home. The entries disagree about the attached image, so we retain
only the message they both describe.

\ASbar{Planner-selected skills}
\ASfield{Plan:} \AScode{infer_causal_relation} $\rightarrow$
\AScode{score_hypothesis_support}

\ASfield{Why these skills:}
\textbf{Causal inference} turns the message about hacking into
an explanation of the exposed location. \textbf{Support assessment} evaluates
whether \ASrefs{C3, C37} substantiate that explanation. The sequence
separates proposing a cause from assessing its evidence.

\ASfield{Skill outputs:}
The planner proposes \textbf{C} using \ASrefs{C3, C37}.
The causal skill infers that a phone hack revealed Kate's location
(confidence \textbf{0.95}). Using that result and the same references,
the scorer assigns support \textbf{0.95} and contradiction
\textbf{0.00}. Both skills return \ASgood{assessed as supported}.

\ASbar{Support assessment and final answer}
\ASfield{Support-assessment output:} \ASgood{assessed as supported}; score \textbf{0.95};
supported references: \ASrefs{C3, C37}.

\ASfield{Final answer (condensed):}
\ASgood{C. Her location was exposed through a phone hack. [C3, C37]}\par
\ASfield{Gold answer:} C.\quad \ASfield{Logged grounding:} \AScode{true}.

\ASfield{Discussion:}
\textbf{The proposed cause becomes an explicit claim to check.}
The first skill attributes the exposure to a phone hack; the second
checks that claim against the same catalog entries, reporting support
\textbf{0.95} and contradiction \textbf{0.00}. This exposes the inference
being evaluated and the evidence used to evaluate it, beyond the final
choice of \textbf{C}. The causal output itself has empty
\AScode{evidence_refs}; the scoring and support-assessment outputs record both references.
Because the intervals overlap, their agreement should not be read as
two independent observations.
\end{ASbody}
\end{AScase}
\caption{\textbf{Video-Holmes: How was Kate's location exposed?} Causal inference proposes the phone-hacking explanation, and the support assessment evaluates it against the cited message. The supplementary frames do not alter the recorded evidence.}
\label{fig:showcase-kate}
\end{figure*}

%% file: showcase/showcase_phone.tex
\begin{figure*}[!htp]
\ASfont
\centering
\begin{AScase}{Video-Holmes: Why does the girl seem afraid?}
\begin{ASbody}
\ASfield{Question:} \textbf{What psychological state does the girl's action of
gripping her phone tightly in the car reflect?}

\ASbar{Selected video frames}
\noindent
\ASframe{.24}{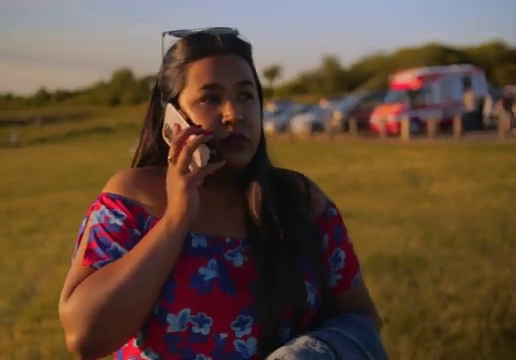}{C3 / 76.0 s}{The girl talks on a phone outdoors.}\hfill
\ASframe{.24}{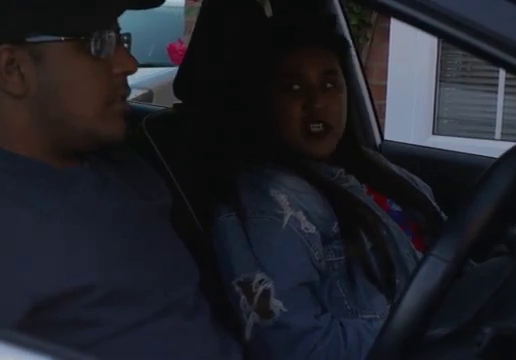}{C3 / 90.0 s}{She turns toward the driver.}\hfill
\ASframe{.24}{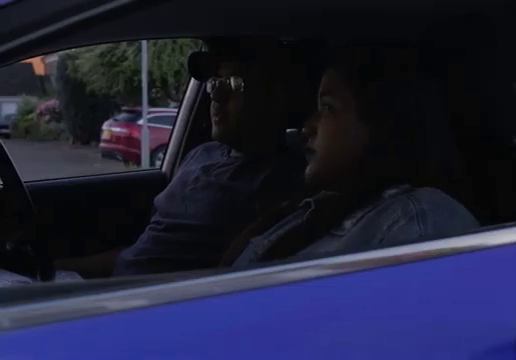}{C4 / 99.0 s}{Both people are seated in the car.}\hfill
\ASframe{.24}{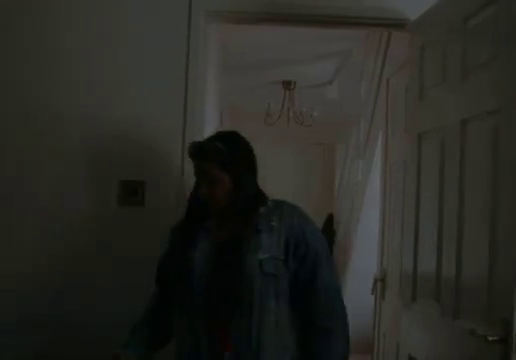}{C4 / 115.0 s}{She looks back after entering.}
\par

\ASbar{Evidence from the catalog}
In \textbf{C3 [63, 97] s}, the girl asks to be picked up and says
that she feels she is being followed. \textbf{C4 [96, 127.98] s}
describes her conversation with the driver about seeking help if she
feels unsafe. She remains nervous and is startled during the ride.

\ASbar{Planner-selected skills}
\ASfield{Plan:} \AScode{infer_intention_or_motive}
$\rightarrow$ \AScode{score_hypothesis_support}

\ASfield{Why these skills:}
The grip described in the question could have several meanings.
\textbf{Motive inference} uses the call and conversation to infer the
girl's state of mind; \textbf{support assessment} evaluates the proposed fear
of being followed against those entries.

\ASfield{Skill outputs:}
The planner proposes \textbf{B} using \ASrefs{C3, C4}.
The first skill infers that she is afraid of being followed
(confidence \textbf{0.95}). Using this result and the same references,
the scorer assigns support \textbf{0.95} and contradiction
\textbf{0.00}. Both skills return \ASgood{assessed as supported}.

\ASbar{Support assessment and final answer}
\ASfield{Support-assessment output:} \ASgood{assessed as supported}; score \textbf{0.95};
supported references: \ASrefs{C3, C4}.

\ASfield{Final answer:} \ASgood{B. Afraid of being followed. [C3, C4]}\par
\ASfield{Gold answer:} B.\quad \ASfield{Logged grounding:} \AScode{true}.

\ASfield{Discussion:}
\textbf{The answer draws on context across two entries.}
The request to be picked up and the feeling of being followed in
\ASrefs{C3} suggest a threat; the conversation in \ASrefs{C4} adds her
continued nervousness. The first skill also records alternatives:
expecting news from friends and worrying about battery life. The next
skill assigns support to the proposed fear explanation using the same
evidence. This makes the interpretation and its support separately
inspectable. The tight grip itself is not described in the catalog or
clearly shown in these frames; the answer is supported by the dialogue.
\end{ASbody}
\end{AScase}
\caption{\textbf{Video-Holmes: Why does the girl seem afraid?} Motive inference interprets the surrounding dialogue, while the support assessment evaluates the resulting fear explanation. The recorded evidence supports the emotional context, not the hand gesture itself.}
\label{fig:showcase-phone}
\end{figure*}

%% file: showcase/showcase_choir.tex
\begin{figure*}[!htp]
\ASfont
\centering
\begin{AScase}{VRBench: Why are some children told not to sing?}
\begin{ASbody}
\ASfield{Question:} \textbf{What is the reason that certain kids in the choir
have to be silent during the competition?}

\ASbar{Selected video frames}
\noindent
\ASframe{.24}{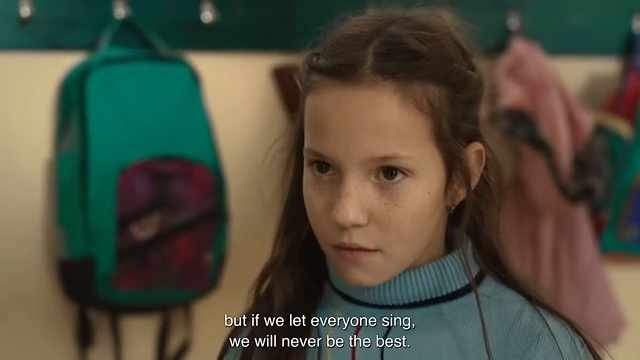}{C38 / 1046.0 s}{Teacher: we cannot be the best if everyone sings.}\hfill
\ASframe{.24}{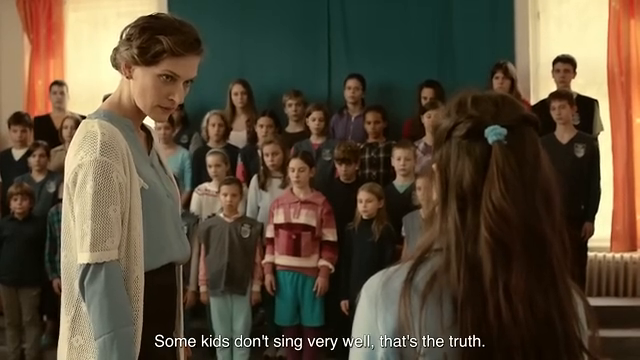}{C38 / 1052.0 s}{Teacher: some children do not sing well.}\hfill
\ASframe{.24}{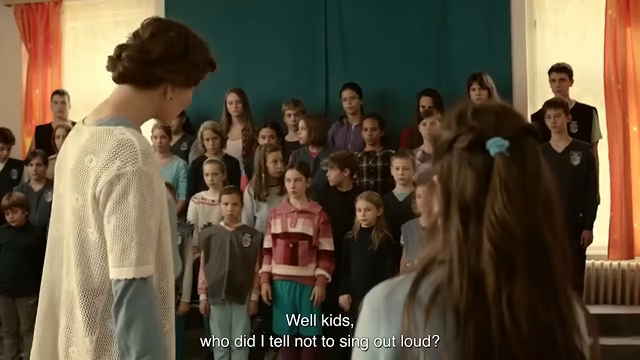}{C39 / 1066.0 s}{The teacher asks who was told not to sing.}\hfill
\ASframe{.24}{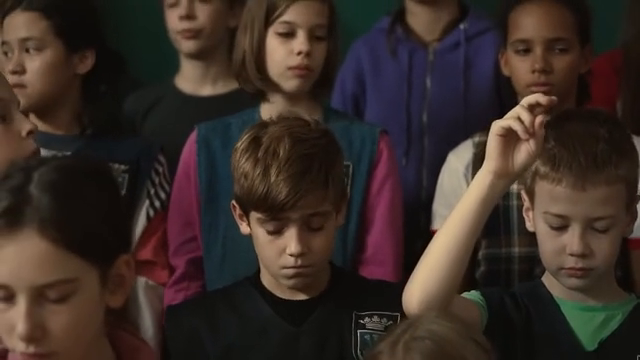}{C39 / 1074.0 s}{Children raise their hands.}
\par

\ASbar{Evidence from the catalog}
\ASfield{C38 / [1036, 1066] s:}
The teacher gives her reason for keeping some children silent:
``but if we let everyone sing, we will never be the best.''

\ASfield{C39 / [1064, 1094] s:}
She asks the group: ``Well kids, who did I tell not to sing out loud?''

\ASbar{Planner-selected skills}
\ASfield{Plan:} \AScode{infer_causal_relation} $\rightarrow$
\AScode{infer_intention_or_motive}\par

\ASfield{Why these skills:}
The question links a rule to the teacher's objective.
\textbf{Causal inference} connects silencing weaker singers to her
expectation of winning; \textbf{motive inference} then uses that relation
to explain what she hopes to achieve.

\ASfield{Candidate:} D. The teacher believes silencing weaker singers
is necessary to win.\par
\ASfield{References:} \ASrefs{C38, C39} for both skills.

\ASfield{1. Infer the cause.}
The causal skill links the instruction to the teacher's stated goal
of winning. It returns \AScode{causal=true}, confidence \textbf{0.95},
and status \ASgood{assessed as supported}.

\ASfield{2. Infer the motive.}
Using the first result and \ASrefs{C38, C39}, the motive skill reads
the rule as an effort to win by excluding weaker singers. It returns
confidence \textbf{0.95} and status \ASgood{assessed as supported}.

\ASbar{Support assessment and final answer}
\ASfield{3. Support-assessment output:}
\AScode{verify_claim_support} checks \ASrefs{C38, C39} and the second
skill's result. It accepts the candidate with score \textbf{0.95},
\AScode{passed=true}, and \AScode{supported_by_refs} = [\ASrefs{C38, C39}].

\ASfield{4. Final answer (condensed):}
\ASgood{D. The teacher believes silencing weaker singers is necessary
to win. [C38, C39]}\par
\ASfield{Gold answer:} D.\quad
\ASfield{Logged result:} \AScode{ok=true}, \AScode{grounded=true}.

\ASfield{Discussion:}
\textbf{The sequence connects an instruction, its expected effect,
and the teacher's goal.} \ASrefs{C39} identifies the instruction, while
\ASrefs{C38} supplies her stated reason. The motive skill receives the
causal result together with those references, forming an explicit
dependency that can be inspected. The resulting answer explains why she
imposes the rule, beyond identifying which children stay silent.
Both intermediate outputs have empty \AScode{evidence_refs}; the input
bindings and support-assessment output preserve \ASrefs{C38, C39}. The reasoning concerns
one conversation and the teacher's expectations about winning.
\end{ASbody}
\end{AScase}
\caption{\textbf{VRBench: Why are some children told not to sing?} The two skills connect the teacher's instruction to its expected benefit and her competitive goal. The second call uses the first result as well as the original evidence.}
\label{fig:showcase-choir}
\end{figure*}

%% file: showcase/showcase_tires.tex
\begin{figure*}[!htp]
\ASfont
\centering
\begin{AScase}{VRBench: Why end with a skidding car?}
\begin{ASbody}
\ASfield{Question:} \textbf{Why was a car skidding episode inserted at the end of
the short movie?}

\ASbar{Selected video frames}
\noindent
\ASframe{.24}{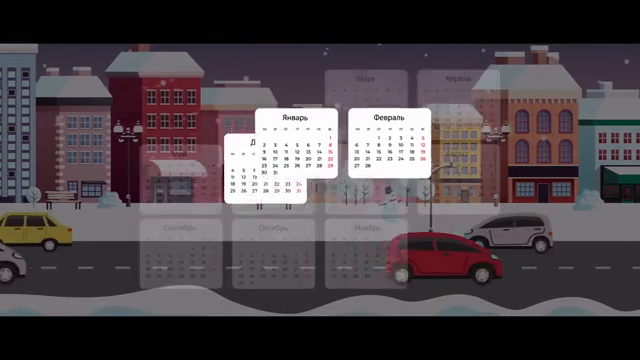}{C48 / 1335.0 s}{Calendars show winter months.}\hfill
\ASframe{.24}{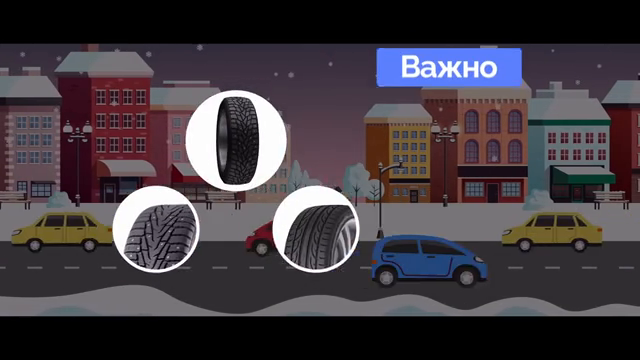}{C48 / 1343.0 s}{Tire types under the ``Important'' heading.}\hfill
\ASframe{.24}{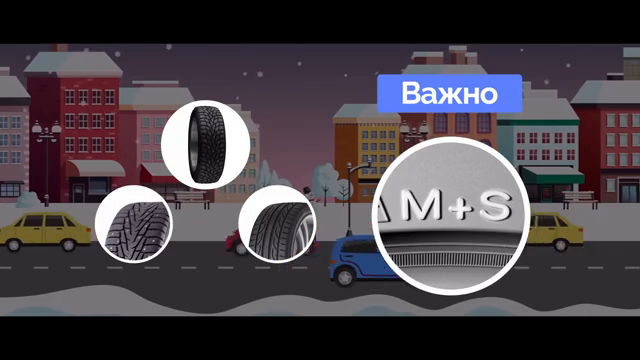}{C49 / 1348.0 s}{Close-up of the M+S marking.}\hfill
\ASframe{.24}{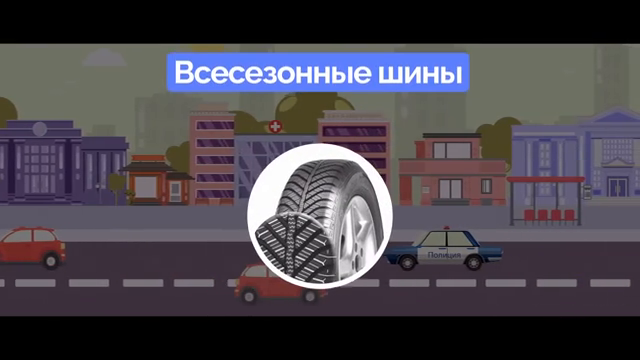}{C49 / 1360.0 s}{The heading reads ``All-season tires.''}
\par

\ASbar{Evidence from the catalog}
\textbf{C48 [1316, 1346] s} describes police on a snowy street,
followed by winter-tire graphics, a December--February calendar,
and an M+S marking. \textbf{C49 [1344, 1374] s} describes more
tire graphics, a June--August calendar, and text about all-season tires.

\ASbar{Planner-selected skills}
\ASfield{Plan:} \AScode{infer_intention_or_motive}
$\rightarrow$ \AScode{score_hypothesis_support}

\ASfield{Why these skills:}
The question concerns the filmmaker's purpose.
\textbf{Motive inference} explains why the skidding scene was included,
using the surrounding tire advice. \textbf{Support assessment} evaluates whether
that interpretation follows from the cited calendars and tire markings.

\ASfield{Skill outputs:}
The planner proposes \textbf{A} using \ASrefs{C48, C49}.
The motive skill reads the ending as advice about tire safety
(confidence \textbf{0.95}). Using this result and the same references,
the scorer assigns support \textbf{0.95} and contradiction
\textbf{0.00}. Both skills return \ASgood{assessed as supported}.

\ASbar{Support assessment and final answer}
\ASfield{Support-assessment output:} \ASgood{assessed as supported}; score \textbf{0.95};
supported references: \ASrefs{C48, C49}.

\ASfield{Final answer (condensed):}
\ASgood{A. Explain road safety and remind viewers to change to winter
tires in time. [C48, C49]}\par
\ASfield{Gold answer:} A.\quad \ASfield{Logged grounding:} \AScode{true}.

\ASfield{Discussion:}
\textbf{Skill selection follows the kind of explanation requested.}
A physical account of skidding would leave the reason for inserting the
scene unanswered. The planner instead asks what the ending communicates
to viewers. Winter months, the M+S marking, and all-season tire graphics
connect the scene to practical advice across \ASrefs{C48, C49}.
The scorer then checks the specific educational claim against those
details. The creator's purpose remains an inference, and the intermediate
records make clear which observations are used to support it.
\end{ASbody}
\end{AScase}
\caption{\textbf{VRBench: Why end with a skidding car?} The planner treats the question as one about communicative intent. Motive inference links the ending to tire advice, and the support assessment evaluates that interpretation.}
\label{fig:showcase-tires}
\end{figure*}

%% file: showcase/showcase_stickers.tex
\begin{figure*}[!htp]
\ASfont
\centering
\begin{AScase}{Failure / CG-Bench: An unsupported sticker count}
\begin{ASbody}
\ASfield{Question:} \textbf{How many green puppy stickers are there on the coffee
machine when the girl is making coffee in the video?}

\ASbar{Selected video frames}
\noindent
\ASframe{.24}{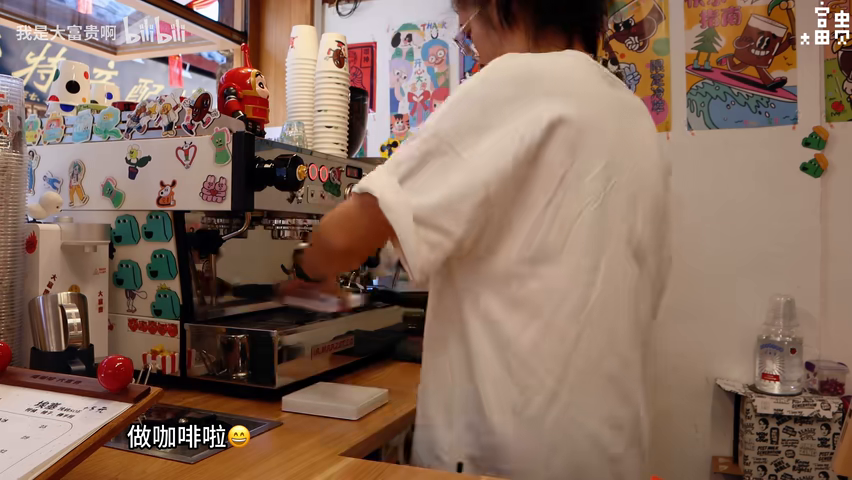}{C2 / 57.0 s}{The machine is partly obscured.}\hfill
\ASframe{.24}{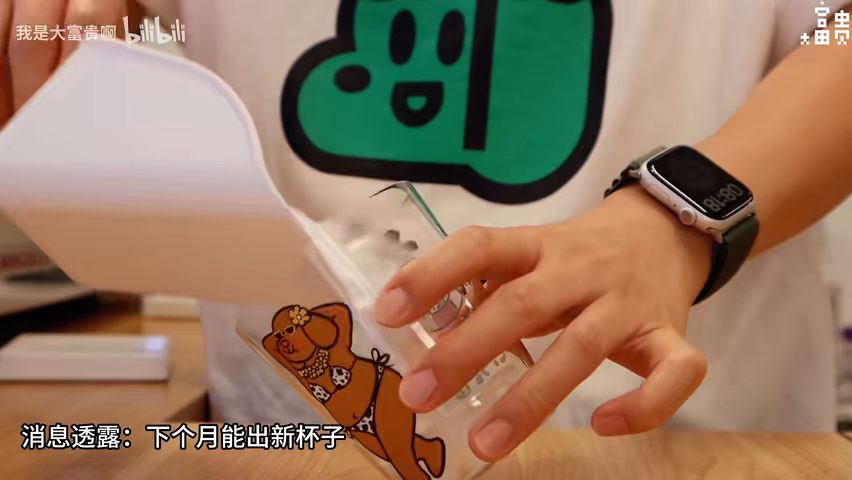}{C2 / 65.0 s}{A glass with a dog graphic.}\hfill
\ASframe{.24}{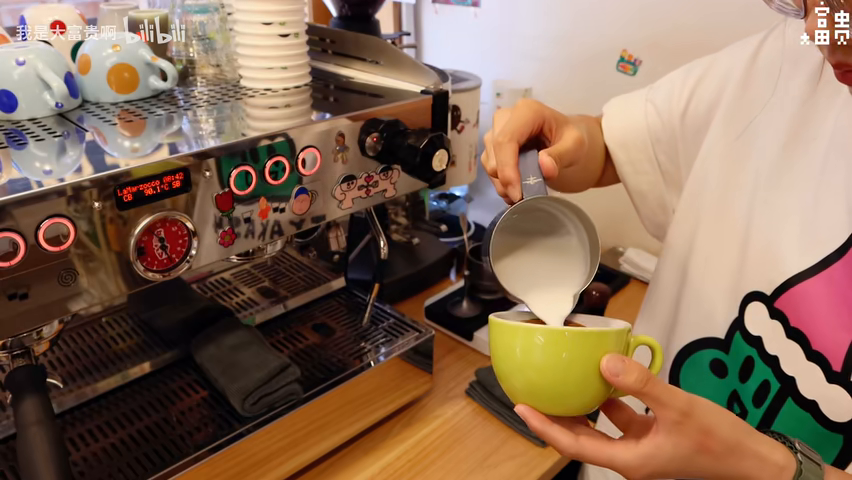}{C2 / 94.0 s}{Milk is poured beside the machine.}\hfill
\ASframe{.24}{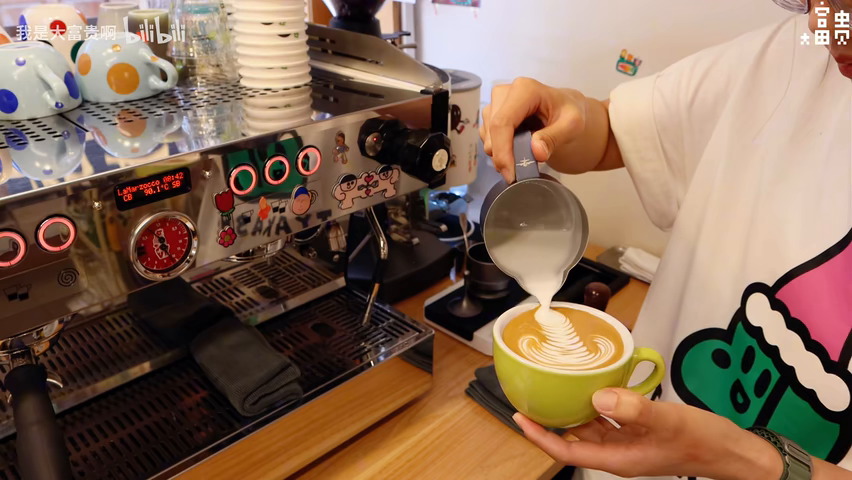}{C18 / 100.0 s}{Latte art in the green cup.}
\par

\ASbar{Evidence from the catalog}
\textbf{C2 [56, 114] s} mentions coffee preparation and a cartoon dog
on a glass; \textbf{C18 [84, 114] s} describes milk being poured into
a green cup. Neither entry states how many green puppy stickers are
on the coffee machine.

\ASbar{Planner-selected skills}
\ASfield{Plan:} \AScode{localize_clue} $\rightarrow$
\AScode{score_hypothesis_support}

\ASfield{Why these skills:}
Counting requires evidence about both the right object and its
quantity. \textbf{Localization} searches for entries about the coffee
machine; \textbf{support assessment} evaluates whether those entries justify
the proposed count of seven. A relevant scene alone cannot settle that claim.

\ASfield{Skill outputs:}
The planner proposes \textbf{C. Seven} using \ASrefs{C1, C2}.
The localizer returns \ASrefs{C1, C2, C17, C18, C26, C31, C42},
with status \ASgood{assessed as supported} and confidence \textbf{0.85}.
The scorer uses all seven references and the localizer's result.
It assigns support \textbf{0.00} and contradiction \textbf{1.00},
returning \ASbad{assessed as unsupported} with diagnostic label
\AScode{no_hypothesis_support}.

\ASbar{Support assessment and final answer}
\ASfield{Support-assessment output:}
\ASbad{assessed as unsupported}; score \textbf{0.00};
\AScode{insufficient_evidence};
no supported references. The catalog does not give the requested count.

\ASfield{Final outcome:}
\ASbad{No valid answer}. The response reaches its length limit
(\AScode{finish_reason=length}).\par
\ASfield{Gold answer:} \textbf{E. Five}.

\ASfield{Discussion:}
\textbf{The trace distinguishes successful retrieval from unsupported
count support.} The planner uses a dog graphic on a glass to support a
claim about stickers on the machine. The localizer returns seven related
entries with status \ASgood{assessed as supported}, yet the scorer assigns zero
support and the support-assessment output marks the claim as unsupported. Adding references has
not supplied the missing quantity. These records locate the problem in
the evidence for the count, before the final call reaches its length
limit. The assessment is advisory and triggers no abstention, replanning,
or repair.
\end{ASbody}
\end{AScase}
\caption{\textbf{Failure / CG-Bench: An unsupported sticker count.} Finding related scenes does not establish a count. Separate retrieval and support-assessment outputs reveal where the proposed answer loses evidential support.}
\label{fig:showcase-stickers}
\end{figure*}

%% file: showcase/showcase_extortion.tex
\begin{figure*}[!htp]
\ASfont
\centering
\begin{AScase}{Failure / VRBench: The final answer retains an unsupported claim}
\begin{ASbody}
\ASfield{Question:} \textbf{Why were the two boys in the video repeatedly extorted
for money?}

\ASbar{Selected video frames}
\noindent
\ASframe{.24}{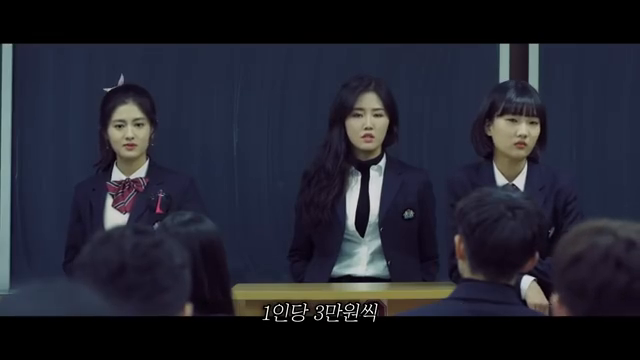}{C4 / 88.0 s}{Students face seated classmates.}\hfill
\ASframe{.24}{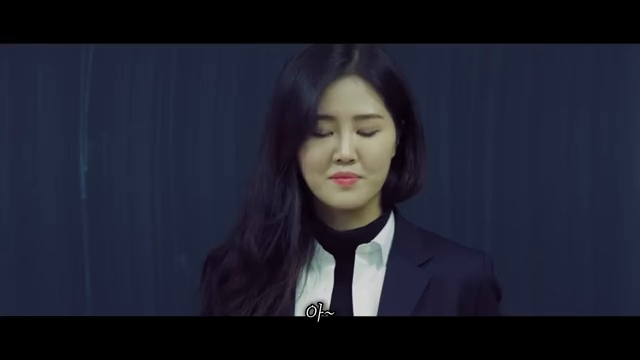}{C4 / 100.0 s}{A close-up during the exchange.}\hfill
\ASframe{.24}{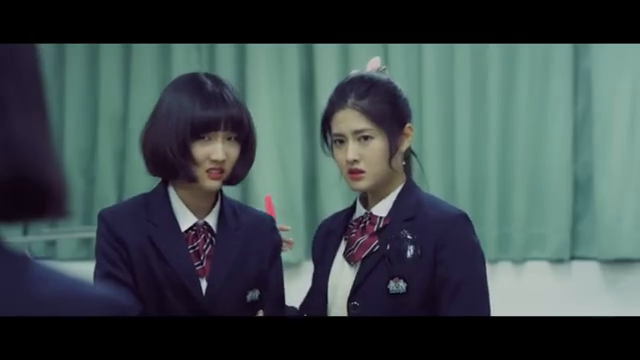}{C8 / 210.0 s}{Students confront one another.}\hfill
\ASframe{.24}{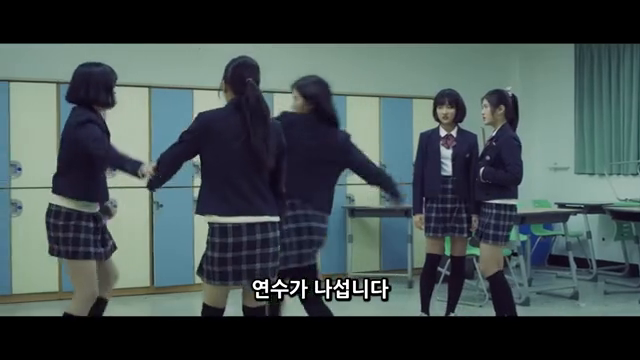}{C8 / 216.0 s}{One student steps between the others.}
\par

\ASbar{Evidence from the catalog}
\textbf{C4 [84, 114] s} describes classroom extortion and resistance.
\textbf{C8 [196, 226] s} describes a confrontation and physical restraint.
The two cited entries (C4 and C8) and the selected frames provide no support for the claim that the bullies expected the victims not to report them.

\ASbar{Planner-selected skills}
\ASfield{Plan:} \AScode{infer_causal_relation} $\rightarrow$
\AScode{infer_intention_or_motive}

\ASfield{Why these skills:}
Repeated extortion calls for an explanation of behavior over time.
\textbf{Causal inference} relates the confrontation to the repeated
targeting; \textbf{motive inference} examines the proposed belief that
the victims will not report it. Evidence of intimidation alone does
not establish that belief.

\ASfield{Skill outputs:}
The planner proposes \textbf{D}: the bullies believed the victims would
not report them. Both skills use \ASrefs{C4, C8}; the second also uses
the first result. Both return \ASgood{assessed as supported} with confidence
\textbf{0.85}, although their output \AScode{evidence_refs} lists are empty.

\ASbar{Support assessment and final answer}
\ASfield{Support-assessment output:}
\ASbad{assessed as unsupported}; score \textbf{0.00};
\AScode{insufficient_evidence};
no supported references. The support-assessment stage assesses the proposed motive
as unsupported.

\ASfield{Final answer (condensed):}
\ASbad{D. The bullies believed the victims would not report them. [C4, C8]}\par
\ASfield{Gold answer (condensed):} \textbf{B. The victim felt unable to fight back
after an earlier failed attempt.}\par
\ASfield{Logged result:} \AScode{ok=false}, \AScode{grounded=false}.

\ASfield{Discussion:}
\textbf{The trace exposes both an unsupported inference and a
failure to use the support assessment.} The selected skills retain the
same proposed motive, but the support-assessment output indicates insufficient evidence.
The final call nevertheless returns \textbf{D}, with
\AScode{grounded=false}. The wrong answer alone would not reveal that
the lack of support had already been detected. Here the separate assessment
provides a diagnostic signal, but that advisory signal does not gate the
final decision or trigger abstention, replanning, or repair.
\end{ASbody}
\end{AScase}
\caption{\textbf{Failure / VRBench: The final answer retains an unsupported claim.} The support-assessment output indicates insufficient evidence, but because this model-generated signal is advisory, the final generation may still retain the hypothesis. The intermediate trace exposes this system boundary.}
\label{fig:showcase-extortion}
\end{figure*}

%% file: showcase/showcase_citation_gap_entry.tex
\input{showcase/showcase_citation_gap}

%% file: showcase/showcase_citation_gap.tex
\begin{figure*}[!htp]
\ASfont
\centering
\tcbset{top=4pt,bottom=4pt}
\begin{AScase}{CG-Bench: Correct answer, incomplete citation coverage}
\begin{ASbody}
\ASfield{Question:} \textbf{In the video, when the subtitle ``Task 2'' appears,
what is the status of the phone screen?}

\ASbar{Selected video frames}
\noindent
\ASframe{.24}{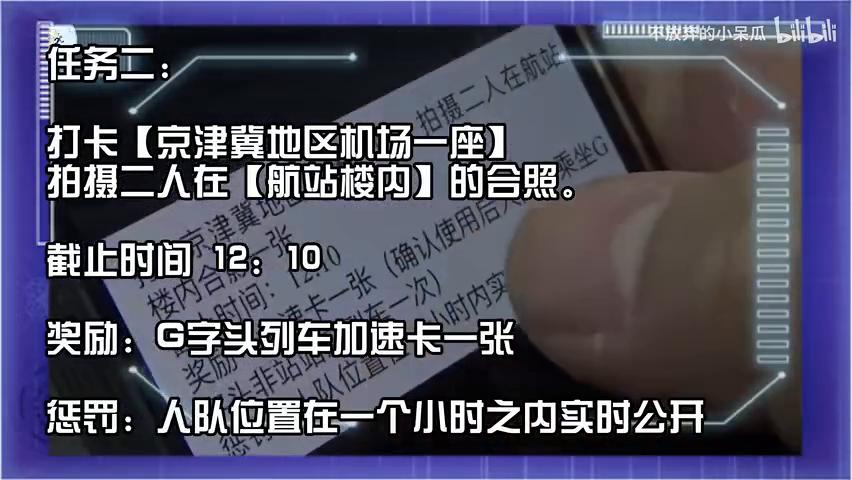}{Cited / 580 s}{``Task 2'' over a lit phone screen.}\hfill
\ASframe{.24}{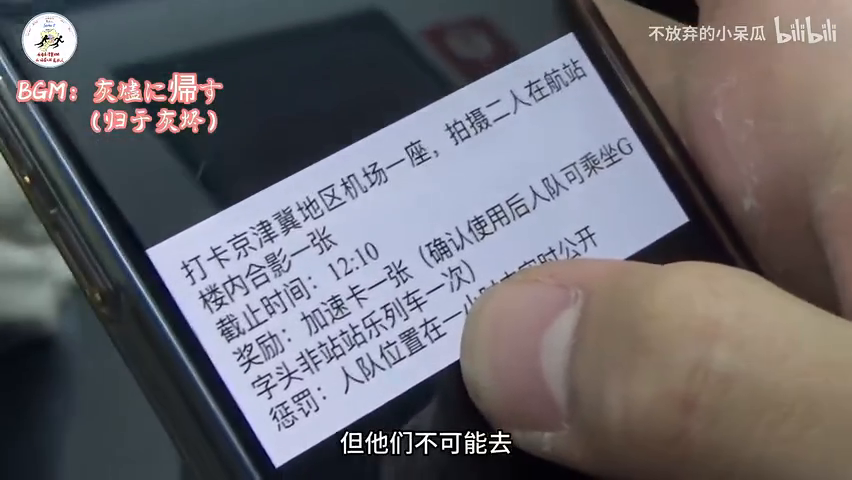}{Covered / 589 s}{Task text in the first annotated span.}\hfill
\ASframe{.24}{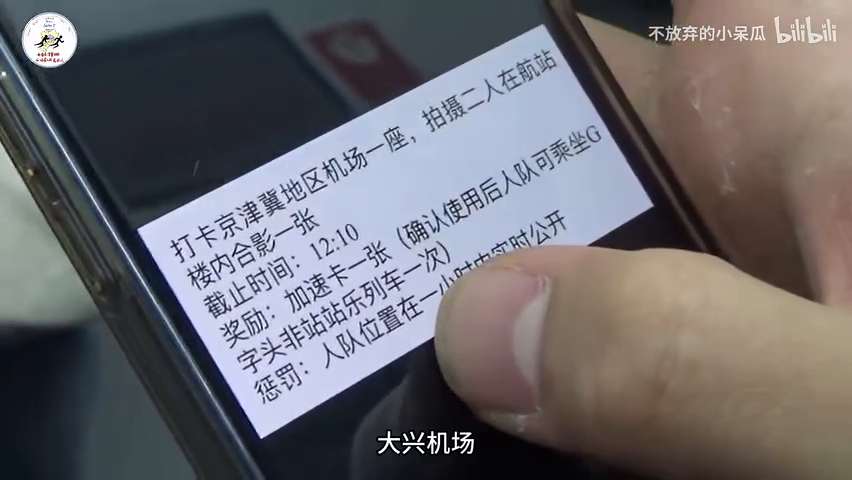}{\textcolor{ASred}{Missed / 592 s}}{The screen stays lit after the citations end.}\hfill
\ASframe{.24}{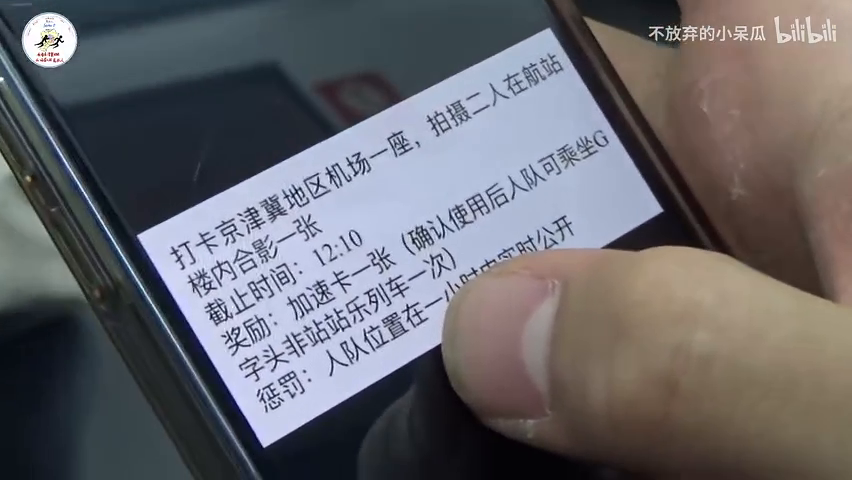}{\textcolor{ASred}{Missed / 593 s}}{The same task text remains visible.}
\par
{\footnotesize Frames are post-run illustrations; ``missed'' means outside the final cited intervals.}

\ASbar{Evidence and planner-selected skills}
\textbf{C10 [532, 590] s} describes the team reading Task 2 on a smartphone.
\textbf{C53 [560, 590] s} describes a close-up of the screen displaying task text.

\ASfield{Plan:} \AScode{localize_clue} $\rightarrow$ \AScode{score_hypothesis_support}

\ASfield{Why these skills:} \textbf{Localization} finds the Task 2 episode;
\textbf{support assessment} evaluates whether the screen description supports
``Bright screen status.'' This separates finding the event from checking its state.

\ASfield{Skill outputs:} The localizer keeps \ASrefs{C10} from \ASrefs{C10, C53}
(confidence \textbf{0.95}). Using \ASrefs{C10} and that result, the scorer assigns
support \textbf{0.90}, contradiction \textbf{0.10}, and overall score \textbf{0.80}.
Both return \ASgood{assessed as supported}.

\ASbar{Support assessment and final answer}
\ASfield{Support-assessment output:} \AScode{verify_claim_support} accepts \ASrefs{C10}
and the scoring result: \ASgood{assessed as supported}, score \textbf{0.90}, \AScode{passed=true}.

\ASfield{Final answer / Gold:} \ASgood{D. Bright screen status.}\quad
\ASfield{Final citations:} \ASrefs{C10, C53}.\par
\ASfield{Logged metrics:} \AScode{ok=true}; \AScode{citation_precision=1.00};
\AScode{grounded=true}.

\ASbar{Citation coverage against the saved annotations}
{\centering\includegraphics[width=0.85\linewidth]{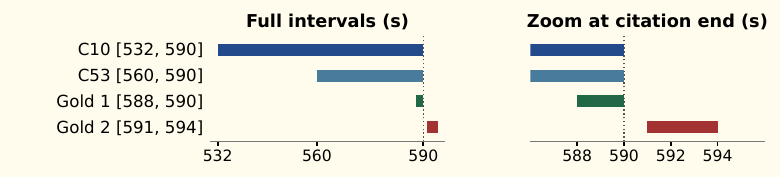}\par}
\ASfield{Post-hoc diagnostics:} \textbf{1/2 annotated spans (50\%)} and
\textbf{2/5 annotated seconds (40\%)} are covered. The cited union is \textbf{58 s};
its intersection with the annotations is \textbf{2 s}. Temporal precision is
\textbf{2/58 = 3.45\%}; temporal IoU is \textbf{2/61 = 3.28\%}.

\ASfield{Discussion:} \textbf{A citation hit does not guarantee complete or precise temporal coverage.}
Both references overlap [588, 590] s, so the logged citation precision is 1.00,
yet neither reaches [591, 594] s. The nested C53 adds no temporal coverage
beyond C10. This motivates checking coverage alongside citation precision.
It does not make the answer unsupported: the screen is already lit at 580 s,
and the annotations need not exhaust all valid evidence. The uncited
\ASrefs{C54} [588, 618] s covers both spans, but the advisory assessment does
not trigger citation revision.
\end{ASbody}
\end{AScase}
\caption{\textbf{Correct answer, incomplete citation coverage.} The final answer and logged grounding check pass, while both citations cover only the first annotated span. Coverage values are supplementary diagnostics.}
\label{fig:showcase-citation-gap}
\end{figure*}

%% file: main.bbl
\begin{thebibliography}{54}
\providecommand{\natexlab}[1]{#1}
\providecommand{\url}[1]{\texttt{#1}}
\expandafter\ifx\csname urlstyle\endcsname\relax
  \providecommand{\doi}[1]{doi: #1}\else
  \providecommand{\doi}{doi: \begingroup \urlstyle{rm}\Url}\fi

\bibitem[Bai et~al.(2025)Bai, Chen, Liu, Wang, Ge, Song, Dang, Wang, Wang,
  Tang, Zhong, Zhu, Yang, Li, Wan, Wang, Ding, Fu, Xu, Ye, Zhang, Xie, Cheng,
  Zhang, Yang, Xu, and Lin]{bai2025qwen25vltechnicalreport}
Shuai Bai, Keqin Chen, Xuejing Liu, Jialin Wang, Wenbin Ge, Sibo Song, Kai
  Dang, Peng Wang, Shijie Wang, Jun Tang, Humen Zhong, Yuanzhi Zhu, Mingkun
  Yang, Zhaohai Li, Jianqiang Wan, Pengfei Wang, Wei Ding, Zheren Fu, Yiheng
  Xu, Jiabo Ye, Xi Zhang, Tianbao Xie, Zesen Cheng, Hang Zhang, Zhibo Yang,
  Haiyang Xu, and Junyang Lin.
\newblock Qwen2.5-vl technical report, 2025.

\bibitem[Chen et~al.(2025{\natexlab{a}})Chen, Yue, Chen, Wang, Liu, Li, and
  Wang]{chen2025lvagent}
Boyu Chen, Zhengrong Yue, Siran Chen, Zikang Wang, Yang Liu, Peng Li, and Yali
  Wang.
\newblock Lvagent: Long video understanding by multi-round dynamical
  collaboration of mllm agents.
\newblock In \emph{2025 IEEE/CVF International Conference on Computer Vision
  (ICCV)}, pages 20237--20246. IEEE, 2025{\natexlab{a}}.

\bibitem[Chen et~al.(2025{\natexlab{b}})Chen, Liu, Huang, Pei, Xu, He, Lu,
  Wang, and Wang]{chen2025cg}
Guo Chen, Yicheng Liu, Yifei Huang, Baoqi Pei, Jilan Xu, Yuping He, Tong Lu,
  Yali Wang, and Limin Wang.
\newblock Cg-bench: Clue-grounded question answering benchmark for long video
  understanding.
\newblock In \emph{International Conference on Learning Representations}, pages
  45647--45682, 2025{\natexlab{b}}.

\bibitem[Chen et~al.(2026{\natexlab{a}})Chen, Xiao, and Yang]{chen2026long}
Kaiyan Chen, Junbin Xiao, and Xun Yang.
\newblock Long-to-short video evidence reasoning for grounded question
  answering.
\newblock \emph{arXiv preprint arXiv:2609.15224}, 2026{\natexlab{a}}.

\bibitem[Chen et~al.(2025{\natexlab{c}})Chen, Di, and Xie]{chen2025grounded}
Qirui Chen, Shangzhe Di, and Weidi Xie.
\newblock Grounded multi-hop videoqa in long-form egocentric videos.
\newblock In \emph{Proceedings of the AAAI Conference on Artificial
  Intelligence}, pages 2159--2167, 2025{\natexlab{c}}.

\bibitem[Chen et~al.(2026{\natexlab{b}})Chen, Ge, Wang, Ge, Cheng, Shan, and
  Liu]{chen2026grpo}
Yi Chen, Yuying Ge, Rui Wang, Yixiao Ge, Junhao Cheng, Ying Shan, and Xihui
  Liu.
\newblock Grpo-care: Consistency-aware reinforcement learning for multimodal
  reasoning.
\newblock In \emph{Findings of the Association for Computational Linguistics:
  ACL 2026}, pages 4305--4320, 2026{\natexlab{b}}.

\bibitem[Cheng et~al.(2025)Cheng, Ge, Wang, Ge, Liao, and Shan]{cheng2025video}
Junhao Cheng, Yuying Ge, Teng Wang, Yixiao Ge, Jing Liao, and Ying Shan.
\newblock Video-holmes: Can mllm think like holmes for complex video reasoning?
\newblock \emph{arXiv preprint arXiv:2505.21374}, 2025.

\bibitem[DeYoung et~al.(2020)DeYoung, Jain, Rajani, Lehman, Xiong, Socher, and
  Wallace]{deyoung2020eraser}
Jay DeYoung, Sarthak Jain, Nazneen~Fatema Rajani, Eric Lehman, Caiming Xiong,
  Richard Socher, and Byron~C Wallace.
\newblock Eraser: A benchmark to evaluate rationalized nlp models.
\newblock In \emph{Proceedings of the 58th annual meeting of the association
  for computational linguistics}, pages 4443--4458, 2020.

\bibitem[Di and Xie(2024)]{di2024grounded}
Shangzhe Di and Weidi Xie.
\newblock Grounded question-answering in long egocentric videos.
\newblock In \emph{2024 IEEE/CVF Conference on Computer Vision and Pattern
  Recognition (CVPR)}, pages 12934--12943. IEEE, 2024.

\bibitem[Diko et~al.(2025)Diko, Wang, Swaileh, Sun, and Patras]{diko2025rewind}
Anxhelo Diko, Tinghuai Wang, Wassim Swaileh, Shiyan Sun, and Ioannis Patras.
\newblock Rewind: Understanding long videos with instructed learnable memory.
\newblock In \emph{2025 IEEE/CVF Conference on Computer Vision and Pattern
  Recognition (CVPR)}, pages 13734--13743. IEEE, 2025.

\bibitem[Feng et~al.(2026)Feng, Gong, Li, Guo, Wang, Peng, Wu, Zhang, Wang, and
  Yue]{feng2026video}
Kaituo Feng, Kaixiong Gong, Bohao Li, Zonghao Guo, Yibing Wang, Tianshuo Peng,
  Junfei Wu, Xiaoying Zhang, Benyou Wang, and Xiangyu Yue.
\newblock Video-r1: Reinforcing video reasoning in mllms.
\newblock \emph{Advances in Neural Information Processing Systems},
  38:\penalty0 99114--99137, 2026.

\bibitem[Gao et~al.(2023)Gao, Yen, Yu, and Chen]{gao2023enabling}
Tianyu Gao, Howard Yen, Jiatong Yu, and Danqi Chen.
\newblock Enabling large language models to generate text with citations.
\newblock In \emph{Proceedings of the 2023 Conference on Empirical Methods in
  Natural Language Processing}, pages 6465--6488, 2023.

\bibitem[Gupta and Kembhavi(2023)]{gupta2023visual}
Tanmay Gupta and Aniruddha Kembhavi.
\newblock Visual programming: Compositional visual reasoning without training.
\newblock In \emph{2023 IEEE/CVF Conference on Computer Vision and Pattern
  Recognition (CVPR)}, pages 14953--14962. IEEE, 2023.

\bibitem[Huang et~al.(2026)Huang, Chen, Chen, Liu, and Lin]{huang2026eg}
Linpeng Huang, Weixing Chen, Zexin Chen, Yang Liu, and Liang Lin.
\newblock Eg-vqa: Benchmarking verifiable video question answering with
  grounded temporal evidence.
\newblock \emph{arXiv preprint arXiv:2606.24797}, 2026.

\bibitem[Jiang et~al.(2026)Jiang, Su, Qu, and Fung]{jiang2026xskill}
Guanyu Jiang, Zhaochen Su, Xiaoye Qu, and Yi~R Fung.
\newblock Xskill: Continual learning from experience and skills in multimodal
  agents.
\newblock \emph{arXiv preprint arXiv:2603.12056}, 2026.

\bibitem[Li et~al.(2026{\natexlab{a}})Li, Guo, Liu, Hao, Tang, Zhao, Jiang,
  Jin, Chen, Lu, et~al.]{li2026timethink}
Handong Li, Longteng Guo, Zikang Liu, Dongze Hao, Yepeng Tang, Zijia Zhao, Jie
  Jiang, Zhiwei Jin, Chen Chen, Haonan Lu, et~al.
\newblock Timethink: Reasoning with time for video llms.
\newblock \emph{arXiv preprint arXiv:2607.05089}, 2026{\natexlab{a}}.

\bibitem[Li et~al.(2026{\natexlab{b}})Li, Li, Shen, Liu, Chang, and
  Shan]{li2026lenswalk}
Keliang Li, Yansong Li, Hongze Shen, Mengdi Liu, Hong Chang, and Shiguang Shan.
\newblock Lenswalk: Agentic video understanding by planning how you see in
  videos.
\newblock \emph{arXiv preprint arXiv:2603.24558}, 2026{\natexlab{b}}.

\bibitem[Li et~al.(2025)Li, Yan, Meng, Dong, Zeng, He, Wang, Qiao, Wang, and
  Wang]{li2025videochat}
Xinhao Li, Ziang Yan, Desen Meng, Lu Dong, Xiangyu Zeng, Yinan He, Yali Wang,
  Yu Qiao, Yi Wang, and Limin Wang.
\newblock Videochat-r1: Enhancing spatio-temporal perception via reinforcement
  fine-tuning.
\newblock \emph{arXiv preprint arXiv:2504.06958}, 2025.

\bibitem[Li et~al.(2026{\natexlab{c}})Li, Li, Bao, Ma, Wang, Liu, and
  Feng]{li2026skillgraph}
Xiaoyuan Li, Moxin Li, Keqin Bao, Yubo Ma, Wenjie Wang, Dayiheng Liu, and Fuli
  Feng.
\newblock Skillgraph: Skill-augmented reinforcement learning for agents via
  evolving skill graphs.
\newblock \emph{arXiv preprint arXiv:2605.12039}, 2026{\natexlab{c}}.

\bibitem[Li et~al.(2026{\natexlab{d}})Li, Liu, Liu, Zhou, Wu, Chen, Xie, Wu,
  and Sun]{li2026comfyclaw}
Zongxia Li, Dawei Liu, Fuxiao Liu, Yuhang Zhou, Xiyang Wu, Jingxi Chen, Jing
  Xie, Xiaomin Wu, and Lichao Sun.
\newblock Comfyclaw: Self-evolving skill harnesses for image generation
  workflows.
\newblock \emph{arXiv preprint arXiv:2607.01709}, 2026{\natexlab{d}}.

\bibitem[Liu et~al.(2025)Liu, Ilievski, and Snoek]{liu2025commonsense}
Huabin Liu, Filip Ilievski, and Cees~GM Snoek.
\newblock Commonsense video question answering through video-grounded
  entailment tree reasoning.
\newblock In \emph{2025 IEEE/CVF Conference on Computer Vision and Pattern
  Recognition (CVPR)}, pages 3262--3271. IEEE, 2025.

\bibitem[Liu et~al.(2026)Liu, Lin, Chen, and Shou]{liu2026videomind}
Ye Liu, Kevin~Qinghong Lin, Chang-Wen Chen, and Mike~Zheng Shou.
\newblock Videomind: A chain-of-lora agent for temporal-grounded video
  reasoning.
\newblock In \emph{International Conference on Learning Representations}, pages
  57481--57506, 2026.

\bibitem[Lu et~al.(2025)Lu, Song, Wang, Torresani, and Nagarajan]{lu2025vited}
Yujie Lu, Yale Song, William Wang, Lorenzo Torresani, and Tushar Nagarajan.
\newblock Vited: Video temporal evidence distillation.
\newblock In \emph{2025 IEEE/CVF Conference on Computer Vision and Pattern
  Recognition (CVPR)}, pages 8501--8511. IEEE, 2025.

\bibitem[Luo et~al.(2026)Luo, Xue, Dimakis, and Grauman]{luo2026thinking}
Romy Luo, Zihui~Sherry Xue, Alex Dimakis, and Kristen Grauman.
\newblock When thinking drifts: Evidential grounding for robust video
  reasoning.
\newblock \emph{Advances in Neural Information Processing Systems},
  38:\penalty0 83696--83727, 2026.

\bibitem[Ma et~al.(2025)Ma, Gou, Shi, Sun, Li, Rezatofighi, and
  Cai]{ma2025drvideo}
Ziyu Ma, Chenhui Gou, Hengcan Shi, Bin Sun, Shutao Li, Hamid Rezatofighi, and
  Jianfei Cai.
\newblock Drvideo: Document retrieval based long video understanding.
\newblock In \emph{2025 IEEE/CVF Conference on Computer Vision and Pattern
  Recognition (CVPR)}, pages 18936--18946. IEEE, 2025.

\bibitem[Meng et~al.(2025)Meng, Li, Wang, Tan, Zhang, Kong, Tong, Wang, Teng,
  Wang, et~al.]{meng2025open}
Jiahao Meng, Xiangtai Li, Haochen Wang, Yue Tan, Tao Zhang, Lingdong Kong,
  Yunhai Tong, Anran Wang, Zhiyang Teng, Yujing Wang, et~al.
\newblock Open-o3-video: Grounded video reasoning with explicit spatio-temporal
  evidence.
\newblock \emph{arXiv preprint arXiv:2510.20579}, 2025.

\bibitem[Meng et~al.(2026)Meng, Yue, Xu, Wang, Ren, Liu, Wang, Zhang, Tong, and
  Duan]{meng2026videozerobench}
Jiahao Meng, Tan Yue, Qi Xu, Haochen Wang, Zhongwei Ren, Weisong Liu, Yuhao
  Wang, Renrui Zhang, Yunhai Tong, and Haodong Duan.
\newblock Videozerobench: Probing the limits of video mllms with
  spatio-temporal evidence verification.
\newblock \emph{arXiv preprint arXiv:2604.01569}, 2026.

\bibitem[Mi et~al.(2026)Mi, Ma, Yang, Li, Wang, Zhang, and Wang]{mi2026skill}
Qirui Mi, Zhijian Ma, Mengyue Yang, Haoxuan Li, Yisen Wang, Haifeng Zhang, and
  Jun Wang.
\newblock Skill-pro: Learning reusable skills from experience via
  non-parametric ppo for llm agents.
\newblock \emph{arXiv preprint arXiv:2602.01869}, 2026.

\bibitem[Ouyang et~al.(2026)Ouyang, Yan, Chen, Han, Wang, Mishra, Meng, Li,
  Jiao, Zha, et~al.]{ouyang2026skillos}
Siru Ouyang, Jun Yan, Yanfei Chen, Rujun Han, Zifeng Wang, Bhavana~Dalvi
  Mishra, Rui Meng, Chun-Liang Li, Yizhu Jiao, Kaiwen Zha, et~al.
\newblock Skillos: Learning skill curation for self-evolving agents.
\newblock \emph{arXiv preprint arXiv:2605.06614}, 2026.

\bibitem[{Qwen Team}(2026)]{qwen3.5}
{Qwen Team}.
\newblock {Qwen3.5}: Towards native multimodal agents, 2026.

\bibitem[Ren et~al.(2024)Ren, Yao, Li, Sun, and Hou]{ren2024timechat}
Shuhuai Ren, Linli Yao, Shicheng Li, Xu Sun, and Lu Hou.
\newblock Timechat: A time-sensitive multimodal large language model for long
  video understanding.
\newblock In \emph{2024 IEEE/CVF Conference on Computer Vision and Pattern
  Recognition (CVPR)}, pages 14313--14323. IEEE, 2024.

\bibitem[Shao et~al.(2024)Shao, Wang, Zhu, Xu, Song, Bi, Zhang, Zhang, Li, Wu,
  et~al.]{shao2024deepseekmath}
Zhihong Shao, Peiyi Wang, Qihao Zhu, Runxin Xu, Junxiao Song, Xiao Bi, Haowei
  Zhang, Mingchuan Zhang, YK Li, Yang Wu, et~al.
\newblock Deepseekmath: Pushing the limits of mathematical reasoning in open
  language models.
\newblock \emph{arXiv preprint arXiv:2402.03300}, 2024.

\bibitem[Sur{\'\i}s et~al.(2023)Sur{\'\i}s, Menon, and
  Vondrick]{suris2023vipergpt}
D{\'\i}dac Sur{\'\i}s, Sachit Menon, and Carl Vondrick.
\newblock Vipergpt: Visual inference via python execution for reasoning.
\newblock In \emph{2023 IEEE/CVF International Conference on Computer Vision
  (ICCV)}, pages 11854--11864. IEEE, 2023.

\bibitem[Tsuchiya et~al.(2026)Tsuchiya, Miyanishi, Ukai, Inoue, Kurita,
  Iwasawa, and Matsuo]{tsuchiya2026ec}
Fumihiko Tsuchiya, Taiki Miyanishi, Mahiro Ukai, Nakamasa Inoue, Shuhei Kurita,
  Yusuke Iwasawa, and Yutaka Matsuo.
\newblock Ec-bench: Enumeration and counting benchmark for ultra-long videos.
\newblock \emph{arXiv preprint arXiv:2603.29943}, 2026.

\bibitem[Wang et~al.(2025{\natexlab{a}})Wang, Bai, Yang, Jin, Zhang, Wang,
  Jiang, Yang, Wu, Dai, et~al.]{wang2025video}
Chendong Wang, Donglin Bai, Yifan Yang, Xiao Jin, Anlan Zhang, Rui Wang, Shiqi
  Jiang, Yuqing Yang, Hao Wu, Qi Dai, et~al.
\newblock Video-in-the-loop: Span-grounded long video qa with interleaved
  reasoning.
\newblock \emph{arXiv preprint arXiv:2510.04022}, 2025{\natexlab{a}}.

\bibitem[Wang et~al.(2024{\natexlab{a}})Wang, Xu, Cheng, Diao, Zhou, Cao, Wang,
  Ge, and Huang]{wang2024grounded}
Haibo Wang, Zhiyang Xu, Yu Cheng, Shizhe Diao, Yufan Zhou, Yixin Cao, Qifan
  Wang, Weifeng Ge, and Lifu Huang.
\newblock Grounded-videollm: Sharpening fine-grained temporal grounding in
  video large language models.
\newblock \emph{arXiv preprint arXiv:2410.03290}, 2024{\natexlab{a}}.

\bibitem[Wang et~al.(2025{\natexlab{b}})Wang, Chen, Tran, Boddeti, and
  Chu]{wang2025seal}
Lan Wang, Yujia Chen, Du Tran, Vishnu~Naresh Boddeti, and Wen-Sheng Chu.
\newblock Seal: Semantic attention learning for long video representation.
\newblock In \emph{2025 IEEE/CVF Conference on Computer Vision and Pattern
  Recognition (CVPR)}, pages 26192--26201. IEEE, 2025{\natexlab{b}}.

\bibitem[Wang et~al.(2024{\natexlab{b}})Wang, Zhang, Zohar, and
  Yeung-Levy]{wang2024videoagent}
Xiaohan Wang, Yuhui Zhang, Orr Zohar, and Serena Yeung-Levy.
\newblock Videoagent: Long-form video understanding with large language model
  as agent.
\newblock In \emph{European Conference on Computer Vision}, pages 58--76.
  Springer, 2024{\natexlab{b}}.

\bibitem[Wang et~al.(2025{\natexlab{c}})Wang, Yu, Stengel-Eskin, Yoon, Cheng,
  Bertasius, and Bansal]{wang2025videotree}
Ziyang Wang, Shoubin Yu, Elias Stengel-Eskin, Jaehong Yoon, Feng Cheng, Gedas
  Bertasius, and Mohit Bansal.
\newblock Videotree: Adaptive tree-based video representation for llm reasoning
  on long videos.
\newblock In \emph{2025 IEEE/CVF Conference on Computer Vision and Pattern
  Recognition (CVPR)}, pages 3272--3282. IEEE, 2025{\natexlab{c}}.

\bibitem[Wei et~al.(2026)Wei, Wang, Gao, Hao, Li, Ding, Chen, He, and
  Yu]{wei2026seeing}
Zhaoyang Wei, Guoliang Wang, Guohua Gao, Yanchao Hao, Mingda Li, Wenchao Ding,
  Xi Chen, Shizhu He, and Xuehui Yu.
\newblock Seeing is believing: Grounding long-video understanding in
  spatio-temporal visual evidence.
\newblock In \emph{Proceedings of the AAAI Conference on Artificial
  Intelligence}, pages 10584--10592, 2026.

\bibitem[Wu et~al.(2026)Wu, Li, Shi, Duffy, Marques, Olson, Zhou, and
  Manocha]{wu2026co}
Xiyang Wu, Zongxia Li, Guangyao Shi, Alexander Duffy, Tyler Marques,
  Matthew~Lyle Olson, Tianyi Zhou, and Dinesh Manocha.
\newblock Co-evolving llm decision and skill bank agents for long-horizon
  tasks.
\newblock \emph{arXiv preprint arXiv:2604.20987}, 2026.

\bibitem[Xia et~al.(2026{\natexlab{a}})Xia, Chen, Wang, Liu, Zeng, Wang, Han,
  Zhou, Zhao, Chen, et~al.]{xia2026skillrl}
Peng Xia, Jianwen Chen, Hanyang Wang, Jiaqi Liu, Kaide Zeng, Yu Wang, Siwei
  Han, Yiyang Zhou, Xujiang Zhao, Haifeng Chen, et~al.
\newblock Skillrl: Evolving agents via recursive skill-augmented reinforcement
  learning.
\newblock \emph{arXiv preprint arXiv:2602.08234}, 2026{\natexlab{a}}.

\bibitem[Xia et~al.(2026{\natexlab{b}})Xia, Lai, Jiang, Tian, Zhou, Li, and
  Wang]{xia2026ser}
Sheng Xia, Zhengqin Lai, Tianxiang Jiang, Kanghui Tian, Shoujun Zhou, Bin Li,
  and Yi Wang.
\newblock Ser: Learning to ground video reasoning with semantic evidence
  rewards.
\newblock \emph{arXiv preprint arXiv:2606.24726}, 2026{\natexlab{b}}.

\bibitem[Xia et~al.(2026{\natexlab{c}})Xia, Hu, Sun, Xu, Xu, Wang, Xu, and
  Jiang]{xia2026grasp}
Tianle Xia, Lingxiang Hu, Yiding Sun, Ming Xu, Lan Xu, Siying Wang, Wei Xu, and
  Jie Jiang.
\newblock Grasp: Graph-structured skill compositions for llm agents.
\newblock \emph{arXiv preprint arXiv:2604.17870}, 2026{\natexlab{c}}.

\bibitem[Xiao et~al.(2024)Xiao, Yao, Li, and Chua]{xiao2024can}
Junbin Xiao, Angela Yao, Yicong Li, and Tat-Seng Chua.
\newblock Can i trust your answer? visually grounded video question answering.
\newblock In \emph{2024 IEEE/CVF Conference on Computer Vision and Pattern
  Recognition (CVPR)}, pages 13204--13214. IEEE, 2024.

\bibitem[Xie et~al.(2025)Xie, Chen, Ge, and Ni]{xie2025video}
Yuan Xie, Tianshui Chen, Zheng Ge, and Lionel Ni.
\newblock Video-mtr: Reinforced multi-turn reasoning for long video
  understanding.
\newblock \emph{arXiv preprint arXiv:2508.20478}, 2025.

\bibitem[Ye et~al.(2025)Ye, Wang, Sun, Chandrasegaran, Durante, Eyzaguirre,
  Bisk, Niebles, Adeli, Fei-Fei, et~al.]{ye2025re}
Jinhui Ye, Zihan Wang, Haosen Sun, Keshigeyan Chandrasegaran, Zane Durante,
  Cristobal Eyzaguirre, Yonatan Bisk, Juan~Carlos Niebles, Ehsan Adeli, Li
  Fei-Fei, et~al.
\newblock Re-thinking temporal search for long-form video understanding.
\newblock In \emph{2025 IEEE/CVF Conference on Computer Vision and Pattern
  Recognition (CVPR)}, pages 8579--8591. IEEE, 2025.

\bibitem[Yu et~al.(2025)Yu, Wu, Chu, Ren, Huang, Chu, Zhang, He, Li, Li,
  et~al.]{yu2025vrbench}
Jiashuo Yu, Yue Wu, Meng Chu, Zhifei Ren, Zizheng Huang, Pei Chu, Ruijie Zhang,
  Yinan He, Qirui Li, Songze Li, et~al.
\newblock Vrbench: A benchmark for multi-step reasoning in long narrative
  videos.
\newblock In \emph{2025 IEEE/CVF International Conference on Computer Vision
  (ICCV)}, pages 21655--21666. IEEE, 2025.

\bibitem[Yu et~al.(2026)Yu, Duan, and Zhang]{yu2026longvidsearch}
Rongyi Yu, Chenyuan Duan, and Wentao Zhang.
\newblock Longvidsearch: An agentic benchmark for multi-hop evidence retrieval
  planning in long videos.
\newblock \emph{arXiv preprint arXiv:2603.14468}, 2026.

\bibitem[Zhang et~al.(2024)Zhang, Lu, Islam, Wang, Yu, Bansal, and
  Bertasius]{zhang2024simple}
Ce Zhang, Taixi Lu, Md~Mohaiminul Islam, Ziyang Wang, Shoubin Yu, Mohit Bansal,
  and Gedas Bertasius.
\newblock A simple llm framework for long-range video question-answering.
\newblock In \emph{Proceedings of the 2024 Conference on Empirical Methods in
  Natural Language Processing}, pages 21715--21737, 2024.

\bibitem[Zhang et~al.(2026{\natexlab{a}})Zhang, Long, Bao, Feng, Zhang, Yue,
  and Wang]{zhang2026memskill}
Haozhen Zhang, Quanyu Long, Jianzhu Bao, Tao Feng, Weizhi Zhang, Haodong Yue,
  and Wenya Wang.
\newblock Memskill: Learning and evolving memory skills for self-evolving
  agents.
\newblock \emph{arXiv preprint arXiv:2602.02474}, 2026{\natexlab{a}}.

\bibitem[Zhang et~al.(2026{\natexlab{b}})Zhang, Shao, Li, Lin, Fu, Wang, Jiao,
  Lu, Liu, Zhang, et~al.]{zhang2026mmskills}
Kangning Zhang, Shuai Shao, Qingyao Li, Jianghao Lin, Lingyue Fu, Shijian Wang,
  Wenxiang Jiao, Yuan Lu, Weiwen Liu, Weinan Zhang, et~al.
\newblock Mmskills: Towards multimodal skills for general visual agents.
\newblock \emph{arXiv preprint arXiv:2605.13527}, 2026{\natexlab{b}}.

\bibitem[Zhang et~al.(2026{\natexlab{c}})Zhang, Pan, Kumar, Saini, Erdogan,
  Yang, Kang, Huang, Sato, and Kong]{zhang2026cast}
Mingfang Zhang, Jingjing Pan, Ashutosh Kumar, Rajat Saini, Mustafa Erdogan,
  Hsuan-Kung Yang, Caixin Kang, Yifei Huang, Yoichi Sato, and Quan Kong.
\newblock Cast-bench: Benchmarking causal chain-grounded spatio-temporal
  reasoning for video question answering.
\newblock \emph{arXiv preprint arXiv:2605.23216}, 2026{\natexlab{c}}.

\bibitem[Zhang and Qiu(2026)]{zhang2026spyce}
Ru Zhang and Weijie Qiu.
\newblock Spyce: Skill-policy co-evolution for multimodal agents.
\newblock \emph{arXiv preprint arXiv:2607.13854}, 2026.

\end{thebibliography}
